# Review of Mercury's dynamic magnetosphere: Post-MESSENGER era and comparative magnetospheres

Weijie SUN[1*], Ryan M. DEWEY[1], Sae AIZAWA[2,3], Jia HUANG[1], James A. SLAVIN[1], Suiyan FU[4], Yong WEI[5,6] & Charles F. BOWERS[1]

[1] *Department of Climate and Space Sciences and Engineering, University of Michigan, Ann Arbor, MI 48109, USA;*
[2] *Institut de Recherche en Astrophysique et Planétologie, CNRS-UPS-CNES, Toulouse 31400, France;*
[3] *Department of Geophysics, Graduate School of Science, Tohoku University, Sendai 980-8578, Japan;*
[4] *School of Earth and Space Sciences, Peking University, Beijing 100871, China;*
[5] *Key Laboratory of Earth and Planetary Physics, Institute of Geology and Geophysics, Chinese Academy of Sciences, Beijing 100029, China;*
[6] *School of Earth and Planetary Sciences, University of Chinese Academy of Sciences, Beijing 100864, China*

**Abstract**   This review paper summarizes the research of Mercury's magnetosphere in the Post-MESSENGER era and compares its dynamics to those in other planetary magnetospheres, especially to those in Earth's magnetosphere. This review starts by introducing the planet Mercury, including its interplanetary environment, magnetosphere, exosphere, and conducting core. The frequent and intense magnetic reconnection on the dayside magnetopause, which is represented by the flux transfer event "shower", is reviewed on how they depend on magnetosheath plasma $\beta$ and magnetic shear angle across the magnetopause, following by how it contributes to the flux circulation and magnetosphere-surface-exosphere coupling. In the next, Mercury's magnetosphere under extreme solar events, including the core induction and the reconnection erosion on the dayside magnetosphere, the responses of the nightside magnetosphere, are reviewed. Then, the dawn-dusk properties of the plasma sheet, including the features of the ions, the structure of the current sheet, and the dynamics of magnetic reconnection, are summarized. The last topic is devoted to the particle energization in Mercury's magnetosphere, which includes the energization of the Kelvin-Helmholtz waves on the magnetopause boundaries, reconnection-generated magnetic structures, and the cross-tail electric field. In each chapter, the last section discusses the open questions related to each topic, which can be considered by the simulations and the future spacecraft mission. We end this paper by summarizing the future BepiColombo opportunities, which is a joint mission of ESA and JAXA and is en route to Mercury.

## 1. Introduction

### 1.1 Planet Mercury in Solar System

Mercury is the smallest and the innermost planet in the Solar System. The mean diameter of Mercury is around 4880 km,

which is about 38% of Earth's diameter (~12,742 km). Mercury is even slightly smaller than moons such as Ganymede (~5262 km) and Titan (~5148 km, Zebker et al. (2009)). Ganymede is a Galilean moon (Jupiter III), and is the largest known satellite in the Solar System, while Titan is the largest satellite of Saturn. Interestingly, Ganymede contains significant amounts of water (Pilcher et al., 1972) and Titan contains a subsurface ocean (Grasset et al., 2000; Iess et al., 2012), while Mercury is a more massive rocky planet similar to Earth.

Due to its proximity to the Sun, Mercury is difficult to study with Earth's ground-based observatories. Figure 1 illustrates the orbits of Mercury and the Earth around the Sun. Mercury orbits the Sun in a period of ~88 Earth days with an orbital inclination angle of ~3.38° relative to the Sun's equator in the International Celestial Reference Frame (ICRF). Mercury's orbital eccentricity (~0.206) is the largest among the planets in the Solar System. The perihelion of Mercury's orbit is ~0.307 AU and the aphelion is ~0.467 AU. The distance from the Sun at perihelion is around two-thirds of the distance at aphelion. Here, AU is the astronomical unit (~1.496×10^{11} m), which is the average distance from Earth to the Sun. Taking Earth's orbit as a comparison, the perihelion of the Earth's orbit is ~0.983 AU and the aphelion is ~1.017 AU. The distance from the Sun at perihelion is ~96.7% of the distance at aphelion with an eccentricity of ~0.0167. Viewed from the Earth, Mercury's orbit around the Sun appears to swing back and forth, but never exceeds 28° of the angular distance. Hence, Mercury is always near the horizon. When Mercury is west of the Sun, it appears in the morning sky before sunrise but is soon covered by the sunlight. When Mercury is east of the Sun, it appears in the evening sky after sunset and sets soon after. Therefore, before the invention of the solar telescope, the best times to view Mercury were only short intervals before sunrise or after sunset on Earth.

Mercury is the least explored inner planet by spacecraft missions. Again, because of its proximity to the Sun, a spacecraft would require a relatively high speed to reach Mercury and would be difficult to insert into or stay at a stable orbit around the planet. Only two spacecraft have visited Mercury so far. The first is Mariner 10 (see, for example, Ness et al., 1974), which flew by Mercury three times in 1974 and 1975. See Russell et al. (1988) and Slavin (2004) for comprehensive reviews of Mercury's magnetosphere after Mariner 10. The second is MESSENGER (Solomon and Anderson, 2018), which flew by Mercury three times before it was inserted into a stable orbit on 18 March 2011. MESSENGER orbited Mercury over 4000 times in approximately four years before the spacecraft exhausted its fuel and crashed into Mercury's surface on 30 April 2015. MESSENGER provided continuous measurements of the magnetic field (Anderson et al., 2007) and plasma compositions (Andrews et al., 2007) around Mercury's magnetosphere. The Bepi-Colombo mission is currently en route to Mercury (e.g., Milillo et al., 2020), which was launched on 20 October 2018 and is planned to insert into Mercury's orbit in December 2025. The BepiColombo mission consists of two spacecraft, which are the Mercury Planetary Orbiter (MPO) and the Mercury Magnetospheric Orbiter (Mio). When BepiCo-lombo arrives at Mercury, one spacecraft will serve as a solar wind monitor to the other inside the magnetosphere.

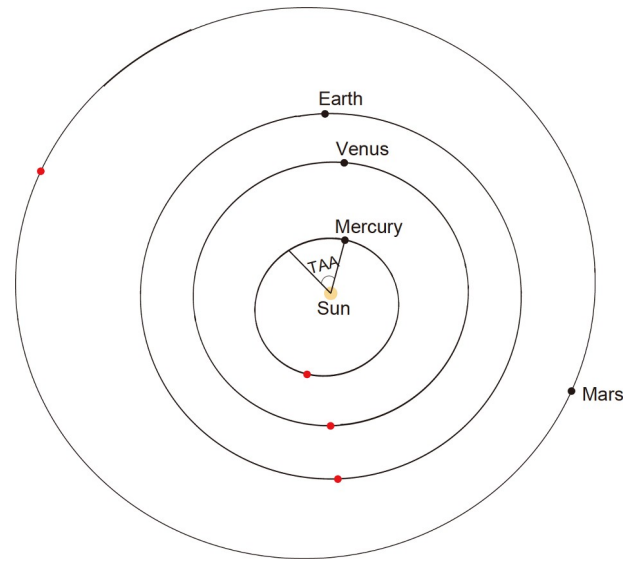

**Figure 1**  The orbits of Mercury, Venus, Earth, and Mars around the Sun. These orbits are projected onto the plane of Earth's orbit, which has an inclination angle of 7.155° to the Sun's equator. The black and red dots represent each planet's perihelion and aphelion, respectively. The true anomaly angle (TAA) of Mercury is the counterclockwise angle from the perihelion as viewed from the north celestial pole. The positions of the planets and the Sun are obtained through the International Celestial Reference Frame (ICRF), Jet Propulsion Laboratory Planetary and Lunar Ephemerides, see Park et al. (2021).

Mariner 10 discovered the intrinsic magnetic field of Mercury (Ness et al., 1974). Despite its small size and slow rotation (59 Earth days), Mercury has a global intrinsic magnetic field, which is approximately a magnetic dipole. Mercury's magnetic field is in a similar direction as that of the Earth's magnetic field but is much weaker (<1% of Earth's magnetic field) (Ness et al., 1976; Anderson et al., 2010). Mariner 10 also discovered a few traces of atoms in Mercury's surface bounded exosphere (Broadfoot et al., 1976). Later, the Earth-based telescopes remotely measured a group of planetary-originated atoms, such as Sodium (Na) (Potter and Morgan, 1985), Potassium (K) (Potter and Morgan, 1986), Calcium (Ca) (Bida et al., 2000), etc. MESSENGER provided much more comprehensive measurements of Mercury's magnetosphere and exosphere. The solar wind-magnetosphere-surface (exosphere) coupling has been extensively investigated and many meaningful conclusions have been achieved. For example, the low solar wind Alfvénic Mach number near Mercury's orbit significantly influences dayside magnetopause reconnection. Dayside magnetopause reconnection demonstrates clear magnetic shear angle dependency and plasma $\beta$ dependency. Moreover, Mercury's plasma sheet has a clear dawn-dusk asymmetry, with the dawnside plasma sheet being more dynamic than the duskside plasma sheet. Interestingly, the duskside plasma sheet is more dynamic in Earth's magnetosphere.

## 1.2 Purpose and structure of the review

Mercury's magnetosphere resembles Earth's magnetosphere to some extent since both planets possess a global magnetic field. However, Mercury is closer to the Sun, and therefore, it encounters different solar wind conditions than Earth's magnetosphere. Furthermore, Mercury does not have a dense atmosphere or ionosphere, but a surface bounded exosphere. In addition, Mercury has a large conducting core. All of these attributes make Mercury's magnetosphere act distinctly from Earth's magnetosphere in response to solar wind variations. This review aims to summarize recent progress in Mercury's magnetospheric dynamics and provide useful information for future investigations such as the BepiColombo mission and simulations. This review will avoid duplicating the previous reviews of Mercury's dynamic magnetosphere, such as the reviews of Sundberg and Slavin (2015), Slavin et al. (2018), Korth et al. (2018), Raines et al. (2015). Instead, the review focuses on the progress after 2015, that is, in the post-MESSENGER era. Note that Slavin et al. (2021), the most recent review, reviewed the progress on the Dungey cycle of Mercury's magnetosphere. Moreover, it is interesting to conduct comparative magnetosphere studies. Recently, Kepko et al. (2015) comparatively investigated dipolarizations of Earth's and Mercury's magnetospheres. In this review, we compare the dynamics of Mercury's magnetosphere more broadly to those in Earth's magnetosphere. In some places, we also compare Mercury's magnetosphere dynamics to those in magnetospheres of satellites, other terrestrial planets, and giant planets.

## 1.3 Mercury's interplanetary environment

Mercury is the closest planet to the Sun and as a result, experiences the strongest driving from the solar wind compared to other planets (see Slavin and Holzer, 1981; Gershman and DiBraccio, 2020). Previous studies have analyzed *in situ* measurements from Helios 1 and Helios 2 to investigate the solar wind of Mercury's orbital zone (from ~0.31 to 0.47 AU) (Russell et al., 1988; Burlaga, 2001; Sarantos et al., 2007). Helios 1 provided measurements from 1974 to 1986, and Helios 2 from 1976 to 1980. However, their data coverages were incomplete for many orbits. To avoid this issue, we employ the measurements from the latest Parker Solar Probe (PSP) to investigate the solar wind parameters of Mercury's orbital zone. For the solar wind parameters near Earth's orbit, the measurements are taken from the Wind spacecraft. We note that none of Helios 1, Helios 2, and the PSP provided measurements while MESSENGER orbited Mercury.

The PSP was launched on 12 August 2018. It orbits the Sun in the ecliptic plane, and the deepest perihelion is scheduled to be less than 10 solar radii (Fox et al., 2016). The Solar Wind Electrons, Alphas, and Protons (SWEAP) instrument suite (Kasper et al., 2016) and the FIELDS instrument suite (Bale et al., 2016) onboard provide plasma and magnetic field data, respectively. In this work, we use data taken from the first four orbits. The Wind spacecraft was launched on 1 November 1994. It had a complicated orbit around Earth, but has resided at the Lagrange 1 point since June 2004. Therefore, we select about 15 years' data since Wind has reached the Lagrange 1 point for this study. The SWE Faraday cups measure the reduced distribution functions of solar wind protons and helium ions (Ogilvie et al., 1995). The magnetic field data are measured by the MFI (Magnetic Field Investigation) (Lepping et al., 1995).

In Figure 2, the solar wind parameters, including speed, density, temperature, and the intensity of the interplanetary magnetic field (IMF), are shown as a function of heliocentric distances ($R$) from the Sun from 0.25 to 1.05 AU. In Figure 2a, the mean value of the solar wind speed is ~330 km s$^{-1}$ in Mercury's orbital zone, which is comparable to the value of ~390 km s$^{-1}$ at Earth's orbit. The mean solar wind speed is almost unchanged from Mercury's perihelion to Mercury's aphelion. The solar wind speed shows a small variation in Mercury's orbital zone from ~250 to ~450 km s$^{-1}$, while the variation of the solar wind speed near the Earth's orbit is broader, i.e., from ~300 to ~800 km s$^{-1}$.

The mean values of the solar wind density in Mercury's orbital zone ranges from ~30 to 120 cm$^{-3}$ (Figure 2b), which is six to thirty times the densities (~4.5 cm$^{-3}$) at Earth's orbit. The solar wind density at Mercury's aphelion (~40 cm$^{-3}$) is less than half of the density at perihelion (~100 cm$^{-3}$). In Figure 2c, the mean values of the solar wind temperature range from ~3×10$^4$ to ~1.5×10$^5$ K in Mercury's orbital zone, which is comparable to the mean temperature at Earth's orbit (~7.2×10$^4$ K). The solar wind temperature decreases from ~1.0×10$^5$ K near Mercury's perihelion to ~3×10$^4$ K near Mercury's aphelion. The magnetic field strength in Mercury's orbital zone varies from ~15 to 45 nT (Figure 2d), which is four to ten times the average magnetic field strength (~4 nT) at Earth's orbit. The magnetic field strength near Mercury's aphelion is less than half of the field strength near Mercury's perihelion.

Figure 3 shows several parameters of the solar wind that play important roles in controlling the dynamics of planetary magnetospheres. The first parameter is the convection electric field, $|E| = \left| -\mathbf{v}_p \times \mathbf{B} \right|$, in which $\mathbf{v}_p$ is the solar wind velocity and $\mathbf{B}$ is the IMF. In the study of Earth's magnetosphere, the solar wind convection electric field directly controls the energy flux that can be transported into the magnetosphere (Perreault and Akasofu, 1978; Akasofu, 1981), which comes as a result of magnetopause reconnection (also called the dayside merging) (Kan and Lee, 1979). Since the reconnected magnetic field lines connect to the polar cap, the cross-polar cap potential (CPCP) is also closely related to the solar wind convection electric field (Sonnerup, 1974; Kan and Lee, 1979). The magnetic reconnection in the nightside plasma sheet (Milan et al., 2006), and the ionospheric conductivity (Kivelson and Ridley, 2008) can both modulate the values of the CPCP. As a result, the solar wind convection electric field correlates with the large-scale electric field in the Earth's magnetosphere (Lei et al., 1981; Baumjohann and Haerendel, 1985; Lv and Liu, 2018), and, therefore, influences the particle motions/trajectories in the magnetosphere. Moreover, during the active geomagnetic periods induced by extreme solar wind events, the solar wind convection electric field is very strong and could even partially penetrate into Earth's low-altitude (<1000 km) and low-latitude ionosphere (Nishida, 1966; Fejer et al., 1979; Huang et al., 2005; Wei et al., 2009, 2015). This implies that a fraction of the Alfvén waves injected by the solar wind can overcome the shielding of the upper ionosphere and penetrate to the low-altitude and low-latitude ionosphere.

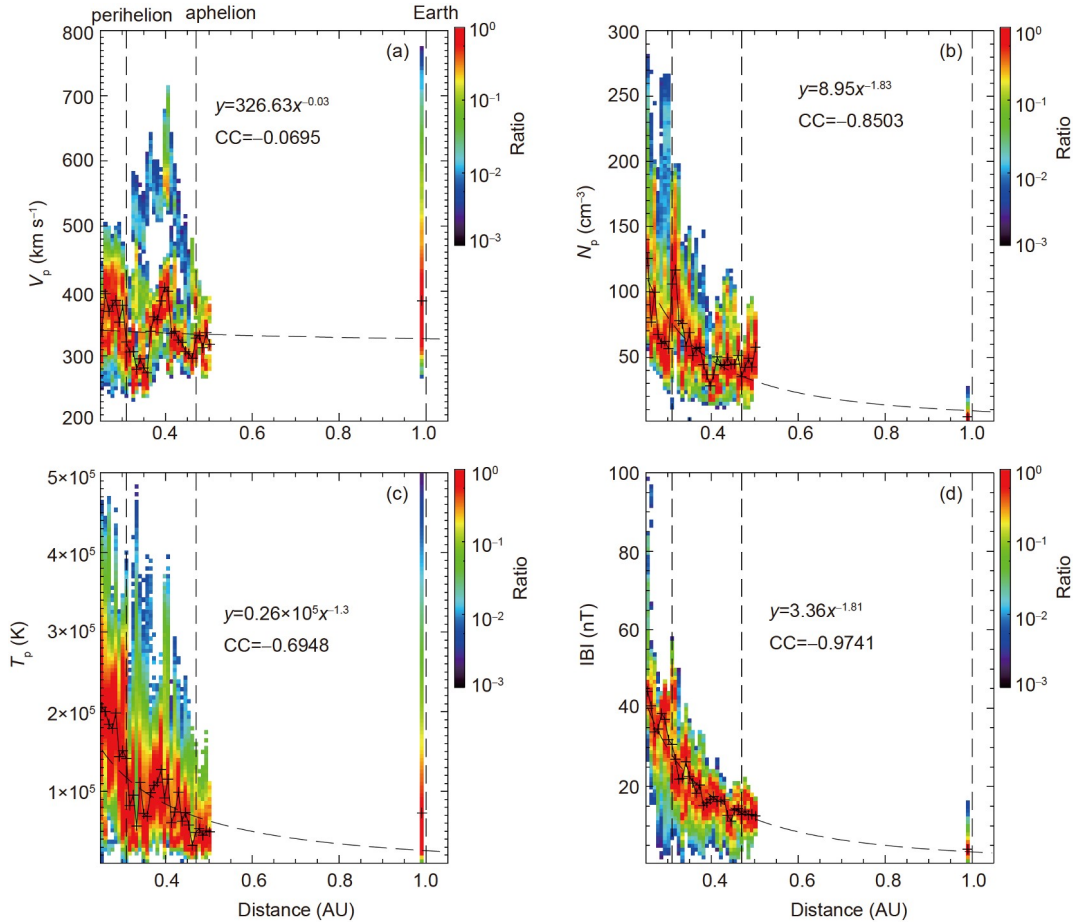

**Figure 2**   The heliocentric variations of (a) the solar wind bulk velocity, (b) the solar wind proton number density, (c) the solar wind proton temperature, and (d) the interplanetary magnetic field (IMF) intensity. The horizontal axis, Distance, represents the heliocentric distance from the center of the Sun. The measurements from 0.25 to 0.5 AU are from the Parker Solar Probe (PSP) and the measurements at near 1 AU are from the Wind spacecraft. The value of the color bar represents the normalization of the data point numbers in each bin to the maximum number among the bins in each vertical column. The vertical dashed lines represent Mercury's perihelion (~0.31 AU), Mercury's aphelion (~0.47 AU), and the average heliocentric distance of Earth (1 AU).

In Figure 3a, the convection electric field ranges from 2 to

10 mV m$^{-1}$ in Mercury's orbital zone, which is two to ten times the convection electric field at Earth's orbit (~1 mV m$^{-1}$). On the other hand, the convection electric field at Mercury's perihelion is ~6 to 10 mV m$^{-1}$, which is several times the convection electric field at Mercury's aphelion (~3 mV m$^{-1}$). The strong convection electric field could lead to fast convection in Mercury's magnetosphere.

The second parameter is the solar wind dynamic pressure, $P_d = n_{sw} m v_{sw}^2$, where $n_{sw}$ is the solar wind density, and $m$ is the mass of solar wind species. The solar wind dynamic pressure is also called the solar wind momentum flux density. The solar wind dynamic pressure could affect the shape and the location of magnetopause (Ferraro, 1960; Sibeck et al., 1991; Shue et al., 1998), and, therefore, the size of the magnetosphere. In addition to the dynamic pressure, the magnetopause reconnection can erode magnetic flux near the subsolar magnetopause (Coroniti and Kennel, 1972; Slavin and Holzer, 1979), which would result in the inward motion of the dayside magnetopause and the flaring of the nightside magnetopause. The reconnection effect is discussed in the following sections of solar wind plasma $\beta$ and Alfvénic Mach number.

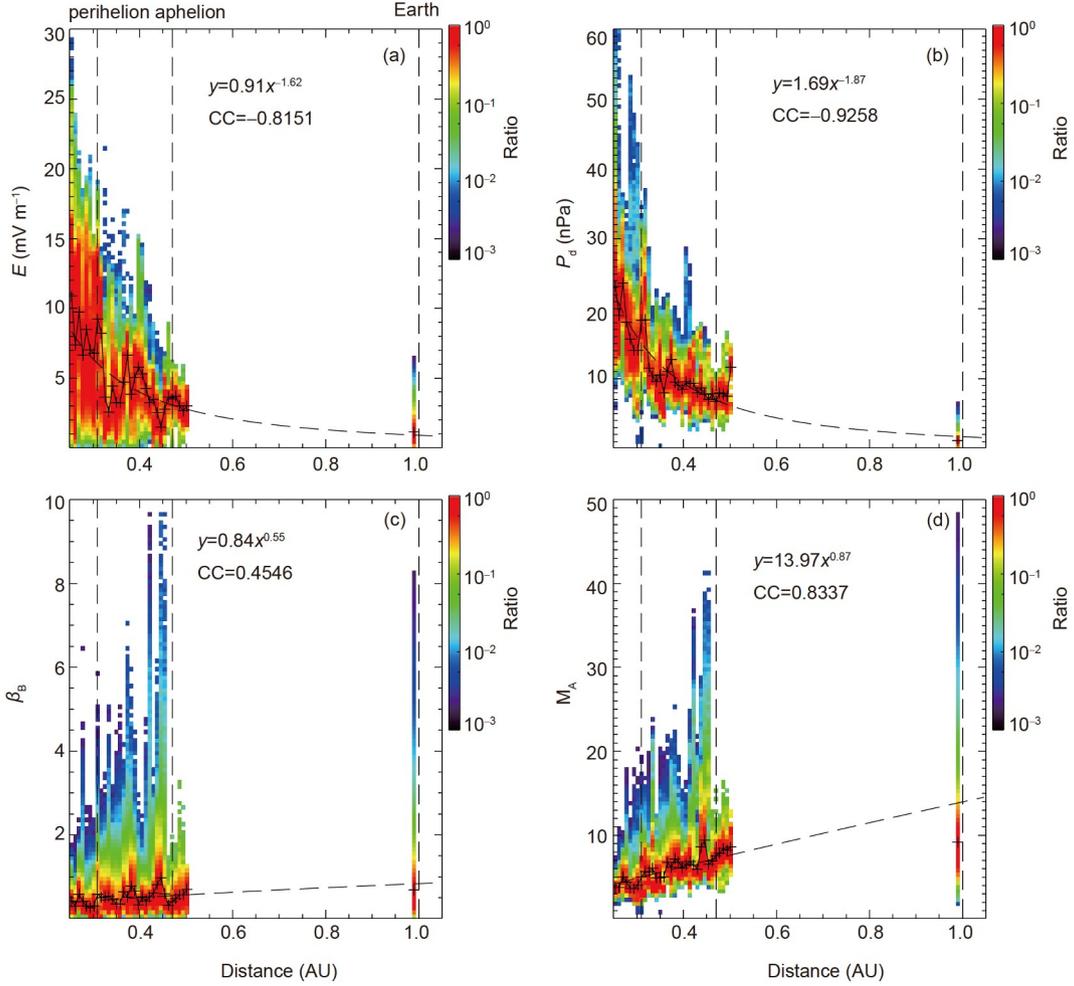

**Figure 3**   The heliocentric variations of (a) the solar wind convection electric field ($|E| = \left| -\mathbf{v}_p \times \mathbf{B} \right|$), (b) the solar wind dynamic pressure $\left( n_p m_p v_p^2 \right)$, (c) the solar wind plasma $\beta$, and (d) the solar wind Alfvénic Mach number $\left( V_{\text{Alfvén}} / v_p \right)$. Only the measurements of protons were considered for the parameters in this figure. This figure is in the same format as Figure 2.

In Figure 3b, the solar wind dynamic pressure in Mercury's orbital zone ranges from ~3.5 to ~9.0 nPa, which is around six to ten times the dynamic pressure at Earth's orbit (~0.55 nPa). Mercury's magnetosphere is under much stronger dynamic pressure than Earth's magnetosphere. Furthermore, the dynamic pressure at Mercury's perihelion (~9.0 nPa) is several times the dynamic pressure at Mercury's aphelion.

The other two parameters shown in Figure 3 are the solar wind plasma $\beta$ (Figure 3c), which is the ratio of the plasma thermal pressure to the magnetic pressure, and the solar wind Alfvénic Mach number (Figure 3d). The solar wind Alfvénic Mach number is the ratio of solar wind speed to the Alfvén speed, $V_{\text{sw}}/V_A$, where the Alfvén speed is $\mathbf{V}_A = \mathbf{B} / \sqrt{\mu_0 n_{\text{sw}} m_p}$. The plasma $\beta$ positively correlates with the Alfvénic Mach number in some extent, that is, the smaller the plasma $\beta$ the smaller the Alfvénic Mach number.

The Alfvénic Mach number negatively correlates with the Alfvén speed, that is, the smaller the Alfvénic Mach number the larger the Alfvén speed. The plasma $\beta$ and the Alfvén speed can influence the magnetic reconnection both in the matters of the reconnection rate and the occurrence of the magnetic reconnection (Sonnerup, 1979; Paschmann et al., 1986; Ding et al., 1992; Scurry et al., 1994). The outflows of magnetic reconnection transport magnetic flux at the rate of the upstream Alfvén speed (Sonnerup, 1979). Therefore, the higher the upstream solar wind Alfvén speed, the stronger the reconnection outflow speed, which corresponds to a larger reconnecting electric field ($E_{\text{rec}}$) and reconnection rates. The reconnection rate refers to the rate of annihilation of magnetic flux ($F_B$) during the magnetic reconnection ($\mathrm{d}F_B/\mathrm{d}t \sim E_{\text{rec}}$). This review also uses the dimensionless reconnection rate, which can be calculated through several equivalent methods (see, for example, Sonnerup et al., 1981). We often employ the ratio of $B_N/B_{\text{inflow}}$ to calculate the dimensionless reconnection rate with magnetic field mea-surements, where $B_N$ is the normal component of the magnetic field in the reconnecting current sheet and $B_{inflow}$ is the magnetic field intensity in the reconnection inflow region. In Figure 3c, the plasma $\beta$ in Mercury's orbital zone ranges from ~0.2 to ~1.0. The plasma $\beta$ at Earth's orbit is ~0.7.

In addition to the above discussions, the intensity of the solar wind number density, solar wind speed, and Alfvénic Mach number can influence viscous-like processes near the magnetopause, for example, the Kelvin-Helmholtz (K-H) instability. Despite the orientation of the IMF playing an important role in the occurrence of K-H instability on the magnetopause boundaries (e.g., Hasegawa et al., 2006), the increase of solar wind speed, solar wind number density, and Alfvénic Mach number enhance the occurrence rate of K-H waves near the Earth's magnetopause (Kavosi and Raeder, 2015).

### 1.4 Mercury's magnetic field and magnetosphere

Mercury's global intrinsic magnetic field has a dipole moment of around 190 nT•$R_M$3. The dipole center is offset approximately 479 km northward from the planet's center and the tilt of the magnetic pole relative to the planet's spin axis is less than 0.8° (Anderson et al., 2012). The polarity of Mercury's magnetic field is similar to that of Earth's dipole magnetic field. However, the magnetic field intensity near Mercury's magnetic equatorial plane is around 200 nT, which is less than 1% of the magnetic field strength of $3.05 \times 10^4$ nT at the Earth's equatorial plane. The weak magnetic field interacts with the solar wind forming a relatively small magnetosphere, although it is still able to separate the shocked solar wind above the planet's surface at an altitude of ~1200 km (e.g., Siscoe et al., 1975; Slavin et al., 2009). While structures of Mercury's magnetosphere resemble Earth's magnetosphere in many aspects, Mercury does not have a corotation region above the surface. Mercury spins very slowly at a period of ~59 Earth days. The corotation region, which is primarily controlled by the electric field due to the planet's spin, if it exists, would be beneath the surface of Mercury. Furthermore, whether a radiation belt, either long-survived or transient, can survive in Mercury's magnetosphere is still an open question.

Although the subsolar magnetopause ($R_{SS}$) of Mercury is normally above the surface, the small intrinsic magnetic field intensity, strong solar wind dynamic pressure, small solar wind plasma $\beta$, and slow spin of the planet together cause the subsolar magnetopause of Mercury to be located very close to the planet's surface.

We can apply the classic Chapman-Ferraro sixth-root relationship to estimate the $R_{SS}$ for planetary magnetospheres. The Chapman-Ferraro relationship assumes a balance of the solar wind dynamic pressure and the total pressure inside the planetary magnetospheres.

$$C\rho u^2 = nk_BT + B_{eq}^2 / 2\mu_0 R_{SS}^6. \tag{1}$$

On the left-hand side, the solar wind dynamic pressure $\rho u^2$ converts into plasma thermal pressure at the $R_{SS}$ with an efficiency of $C$. The $n$ and $T$ are the plasma density and temperature in the magnetosphere side at the $R_{SS}$. The $k_B$ is the Boltzmann constant. The planetary magnetic field intensity at the $R_{SS}$ is $B_{eq} / R_{SS}^3$, in which the $B_{eq}$ is the magnetic field intensity at the magnetic equator of the planet's surface. Therefore, the $R_{SS}$ can be calculated

$$R_{SS} = \left( \frac{B_{eq}^2}{2\mu_0 \left( C\rho u^2 - nk_BT \right)} \right)^{1/6}. \tag{2}$$

In the magnetospheres of Mercury and Earth, the thermal pressure, $nk_BT$, in the magnetosphere side at the $R_{SS}$, is much smaller than the magnetic pressure, and therefore, can be ignored. In Figure 2, the dynamic pressure ranges from 3.5 to 9 nPa near Mercury's orbit, the $R_{SS}$ is estimated to be from 0.14 to 0.34 $R_M$ above the planet's surface with the efficiency ($C$) of 0.8. In the case of Earth's magnetosphere, the $R_{SS}$ is estimated to be from ~9.7 $R_E$ above the planet's surface. In observations, the $R_{SS}$ is located at a distance of ~0.45 $R_M$ above Mercury's surface. In the Earth's magnetosphere, the subsolar magnetopause is around 10 $R_E$ above Earth's surface.

The fundamental flux circulation in the magnetospheres of Earth and Mercury is called the Dungey cycle (Dungey, 1961). As illustrated in Figure 4, the Dungey cycle starts at magnetic reconnection on the dayside magnetopause, where the IMF and the planetary magnetic field reconnects resulting in open field lines with one end connecting to the solar wind and the other end to the planetary magnetic field. The open field lines convect anti-sunward and transport magnetic flux to the nightside lobes (Figure 4a, the growth phase). The cross-tail electric field convects the open field lines in the lobes toward the plasma sheet in the magnetic equatorial plane, where magnetic reconnection closes the open field lines and convects the closed field lines to replenish the dayside magnetosphere (Figure 4b and 4c, the expansion phase and recovery phase). The entirety of this magnetic flux circulation constitutes the Dungey cycle.

The flux circulation of the Dungey cycle is revealed as the two cell convection patterns of the plasma flows in the ionosphere. The cell convection starts as plasma flows across the polar cap in the anti-sunward direction, which is driven by the reconnection on the dayside magnetopause. When the flow reaches the nightside auroral region, it returns to the dayside along the auroral ovals on both the dawnside and duskside, which is driven by the reconnection in the nightside plasma sheet (Siscoe and Huang, 1985; Crooker, 1992; Zhang et al., 2015). The investigations of plasma flows in the plasma sheet during the phases of the substorm can be found in Juusola et al. (2011a, 2011b) and Sun et al. (2017a), in which the sunward plasma flows are observed to replenish the dayside magnetosphere. However, the situation might vary in Mercury's magnetosphere. The slow rotation of Mercury and the relatively large volume of Mercury in the magnetosphere would cause a portion of plasma flows driven by the nightside reconnection to directly impact the planet's surface (see Dewey et al., 2020), which would constrain their ability in returning the magnetic flux to the dayside magnetosphere.

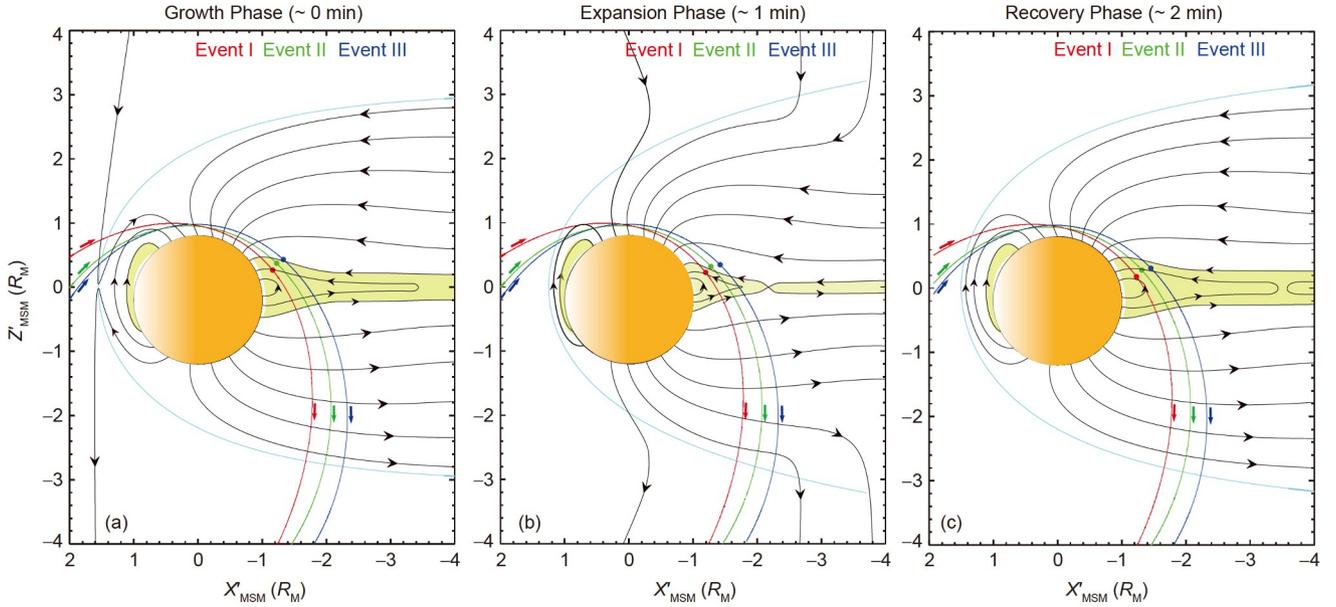

**Figure 4**  The evolution of Mercury's magnetosphere during a substorm. The location of the magnetopause, the occurrence of dayside magnetopause reconnection, the shape of the plasma sheet, and relative locations of three events observed by MESSENGER at the start of the substorm (a) growth phase, (b) expansion phase, and (c) recovery phase. The red dot represents Event I, during which MESSENGER was located in the plasma sheet during the substorm. The green dot represents Event II, during which MESSENGER moved from the plasma sheet into the lobe during the growth phase and moved back into the plasma sheet during the expansion phase. The blue dot represents Event III, during which MESSENGER was located in the lobe during the growth phase and entered the plasma sheet during the expansion phase. The whole process takes 1 to 3 min in Mercury's magnetosphere compared to 1 to 3 hours in Earth's magnetosphere. This figure is revised from Sun et al. (2015b).

The flux circulation corresponds to several magnetospheric response modes, including the substorm, the steady magnetospheric convection, and the sawtooth event. The substorm is the most well-known mode. The substorm is accompanied by many dynamics in the magnetosphere, and the magnetic flux loading-unloading in the lobes corresponds to the magnetic energy store and release. From Figure 4a to 4b, the open field lines accumulate in the lobe, which corresponds to the magnetic flux loading. From Figure 4b to 4c, the accumulated magnetic flux is released by magnetic reconnection in the plasma sheet, which corresponds to the magnetic flux unloading and produces structures known as dipolarizations, flux ropes and high speed flows.

The duration of the flux loading-unloading process can be roughly estimated from the loaded flux in the lobe ($\Delta\Psi$) divided by the CPCP ($\Phi$)

$$T_{\text{load}} = \Delta\Psi / \Phi. \tag{3}$$

The CPCP corresponds to the rate of magnetic flux transporting from the dayside magnetosphere to the nightside lobes. Here we take Mercury and Earth as examples to estimate $T_{\text{load}}$. At first, the magnetic flux in one lobe can be estimated from

$$\Psi = 0.5\pi R_{\text{Tail}}^2 B_{\text{Lobe}}. \tag{4}$$

At Mercury, $R_{\text{Tail}}\sim 2.5\ R_M$, $B_{\text{Lobe}}\sim 50$ nT, $\Psi\sim 3$ MWb. At Earth, $R_{\text{Tail}}\sim 20\ R_E$, $B_{\text{Lobe}}\sim 30$ nT, $\Psi\sim 760$ MWb. If the lobe flux was enhanced by 20%, the loaded magnetic flux ($\Delta\Psi$) would be $\sim 0.6$ MWb at Mercury and $\sim 150$ MWb at Earth. Secondly, we estimate the value of the CPCP. As discussed in Section 1.3, the CPCP is controlled by several factors. Here we simply estimate the CPCP according to the scale of the dayside magnetopause X-line. In Figure 3a, the solar wind convection electric field is $\sim 4$ mV m$^{-1}$ near Mercury's orbital zone and $\sim 1$ mV m$^{-1}$ near Earth's orbit, considering the extent of X-lines of $\sim 1.0\ R_M$ and 8 $R_E$, the CPCP is $\sim 10$ kV in Mercury's magnetosphere and $\sim 50$ kV in Earth's magnetosphere. Therefore, dividing $\Delta\Psi$ by $\Phi$, the loading period $T_{\text{load}}$ can be calculated to be 60 s in Mercury's magnetosphere and 50 min in Earth's magnetosphere. The period of the loading-unloading can be obtained by simply doubling the $T_{\text{load}}$, which is around two minutes in Mercury's magnetosphere and around two hours in Earth's magnetosphere.

In Mercury's magnetosphere, MESSENGER measurements have revealed that the magnetospheric substorm persists for one to three minutes (Slavin et al., 2010a). In Earth's magnetosphere, both Earth-based and spacecraft measurements have demonstrated that the magnetospheric substorm, which corresponds to flux loading-unloading in the lobe, could persist for one to three hours (Akasofu, 1964; Rostoker et al., 1980; Huang et al., 2003). The above estimations are consistent with these observations. In the magnetospheres of the giant planets, the flux loading-unloading driven by the solar wind-magnetosphere reconnection could persist for several Earth days.

## 1.5   Mercury's neutral exosphere

The Ultraviolet Spectrometers of Mariner 10 discovered Hydrogen (H) and Helium (He) in Mercury's surface bounded exosphere. The measurement of Oxygen (O) was described as "a tentative identification" due to a low signal-to-noise ratio (Broadfoot et al., 1976). Starting in the 1980s, the Earth-based solar telescopes began to take spectroscopy images of Mercury's exosphere, and the emissions of Sodium (Na) (Potter and Morgan, 1985), Potassium (K) (Potter and Morgan, 1986), and Calcium (Ca) (Bida et al., 2000) were discovered. The Mercury Atmospheric and Surface Composition Spectrometer (MASCS) (McClintock and Lankton, 2007) onboard MESSENGER detected several other species in Mercury's exosphere, including Magnesium (Mg) (McClintock et al., 2009), Manganese (Mn), and Aluminum (Al) (Vervack Jr et al., 2016). These measurements reveal that Mercury's exosphere is tenuous. The exospheric particles have long mean free paths and they more likely collide with the planet's surface rather than with each other.

Mercury's exosphere is supplied by both external and internal sources, which share many similarities with the Earth's Moon despite the Moon not having a global magnetic field. The external sources include solar wind, meteoroids, micrometeoroids, and comets. The internal sources are particles released from the planet's surface through a variety of processes, including thermal desorption, photon-stimulated desorption, electron-stimulated desorption, ion sputtering, micrometeoroid impact vaporization, and diffusion from the interior or the regolith.

Thermal desorption releases particles due to the high temperature of the surface, and the released particles then surround the surface since they contain very low energy (<eV). The photon-stimulated desorption corresponds to a process in which a bound electron in an atom absorbs solar ultraviolet emission. The atom becomes more energetic and escapes from the solid surface but is still in a bound state (Yakshinskiy and Madey, 1999). In electron-stimulated desorption (ESD), high-energy electrons (a few eV) neutralize surface ions, mostly Alkaid ions, which provide particles to the exosphere.

The solar wind sputtering corresponds to a process whereby the solar wind energetic particles (~1 keV) bombard the regolith and release particles. The sputtered particles gain relatively high energy (> a few eV up to 100 eV) through momentum transferred from the solar wind particles. Under most circumstances, the solar wind particles impact the high latitude cusp regions, which are controlled by both the planetary magnetic field and the solar wind convection electric field. The double-peaked sodium emissions on the northern and southern high-latitude regions imaged by the Earth-based telescopes are possibly due to the solar wind sputtering (Potter et al., 2006; Leblanc et al., 2009; Mangano et al., 2015; Orsini et al., 2018; Milillo et al., 2020) but still lack *in situ* evidence.

The micrometeoroid impact vaporization is similar to the solar wind sputtering in the sense of momentum transfer. However, the micrometeoroids are non-ionized and would not be influenced by the planetary magnetic field and the convection electric field. Therefore, the micrometeoroid impact could impact most places on the planet's surface. Since Mercury orbits the Sun at a speed of tens of kilometers per second, the moving forward portion on the dawnside has a higher chance of being bombarded with micrometeoroids. The bombard is expected to be significant toward the nightside when Mercury approaches the aphelion and the dayside when Mercury approaches the Sun (Pokorný et al., 2017). In addition, micrometeoroids are much heavier and larger in scale than solar wind particles. Many models suggest that the impact of micrometeoroids could release a component of neutrals with a much higher temperature (~3000 to 6000 K) (e.g., Killen et al., 2018). As a result, these atoms could reach a higher altitude (>1000 km) than those neutrals released during photon-stimulated desorption, electron-stimulated desorption, and thermal desorption.

We note that the above processes, including solar wind sputtering, micrometeoroid impact vaporization, photon-stimulated desorption, and thermal desorption, mostly eject neutrals from the regolith. Ions only take account for a small fraction (<10%) (Benninghoven, 1975; Hofer, 1991; Elphic et al., 1993). However, some experimental results suggest that the electron stimulated desorption releases particles from the regolith, particularly in the ionic form (~10%) (McLain et al., 2011) (Table 1).

For the notable gasses (Helium, He and Argon, Ar), the diffusion or degassing of the radiogenic He from the planet's interior can be an important source (Hodges, 1975; Goldstein et al., 1981). For example, Goldstein et al. (1981) estimated that the radiogenic supply of He could be >10% of the solar wind supply. The contribution from the diffusion to the Na and the K was still under debate. Sprague (1990) suggested that the diffusion occurring at the regolith is the dominant source for the exosphere. While other studies (Killen and Morgan, 1993) argued that the regolith diffusion was overestimated in contributing to the exosphere. However, several

**Table 1**    Processes contributing to Mercury's exosphere by internal sources

| Processes | Released energy | Ions percentage (%) | Impact regions |
|---|---|---|---|
| Thermal desorption | $\lesssim 0.1$ eV[a)] | <1 | Sunlit surface |
| Photon stimulated desorption | $\lesssim$ eV[b)] | <1 | Sunlit surface |
| Electron stimulated desorption | ~a few eV[c)] | ~10[d)] | Cusp regions/plasma sheet |
| Ion sputtering | Several eV, up to 100 eV[e)] | <10[f)] | Cusp regions/plasma sheet |
| Micrometeoroid impact vaporization | ~10 s to 100 s eV[g)] | <1 | More on downside surface |

a) Hunten et al. (1988); Yakshinskiy and Madey (2000); Leblanc and Johnson (2003). b) McGrath et al. (1986); Yakshinskiy and Madey (1999). c) Yakshinskiy and Madey (1999, 2004); Johnson et al. (2002). d) Yakshinskiy and Madey (2000); McLain et al. (2011). e) Sigmund (1969); Hofer (1991); Mura et al. (2007). f) Benninghoven (1975); Hofer (1991); Elphic et al. (1993). g) Morgan et al. (1988); Mangano et al. (2007). The percentages of ions generated by the thermal desorption, photon stimulated desorption, and micrometeoroid impact vaporization is set to be <1%. There are still lacking exact values on the portion of ions released by these processes.

studies suggest that the diffusion of atoms from the interior to the surface could enhance the effectiveness of thermal desorptions (see Salvail and Fanale, 1994; Killen et al., 2007).

Mercury's neutral exosphere, including Na, Ca, and K, exhibits seasonal repetitive variations in a period of Mercury orbital period, ~88 Earth days. The intensity of the neutral exosphere reaches a maximum when Mercury is located in the true anomaly angle (TAA) of 75° and 255° (Figure 1). This repetitive seasonal variation was induced by the thermal and photon stimulated desorption. In the TAA of 75° and 255°, the resonance wavelengths of the elements were shifted from the dips in the Fraunhofer lines of the solar irradiances to both edges due to the Doppler shift. The higher intensity of the solar irradiances near the edges of the Fraunhofer lines is responsible for the stronger emissions of Mercury's exospheric particles.

## 1.6    Ions in Mercury's magnetosphere

The measurements from the Fast Imaging Particle Spectrometer (FIPS) onboard MESSENGER have revealed the distribution of ion species in Mercury's magnetosphere (Zurbuchen et al., 2008, 2011). The distributions of several ion species are shown in Figures 5 and 6 (Raines et al., 2011, 2013). The $Na^+$-group with mass per charge (m q$^{-1}$) from 21 to 30, including $Na^+$, $Mg^+$, $Al^+$ and $Si^+$, was found to be the most abundant planetary ion population (Figure 5a to 5c). The $O^+$-group with m q$^{-1}$ from 14 to 20, including $O^+$ and water group ions, is the second most abundant planetary ion population (Figure 5d to 5f). The $Na^+$-group and $O^+$-group appeared throughout Mercury's magnetosphere and were enhanced in several regions, including the northern cusp, the pre-midnight (duskside) plasma sheet, the nightside southern high-latitude region, and the dawn terminator. The $Na^+$-group and $O^+$-group are mostly originated from the planet. The enhancements in the northern cusp and the downside terminator are consistent with the enhancements of neutrals in the exosphere. The enhancement in the pre-midnight plasma sheet can be evidence of the non-adiabatic accel-eration of the $Na^+$ and $O^+$ in the cross-tail direction (Ip, 1987; Delcourt et al., 2002; Gershman et al., 2014). At last, the enhancements of $Na^+$ and $O^+$ in the nightside southern high-latitude are similar to those energetic ion plumes that have been observed at Venus and Mars (Luhmann and Kozyra, 1991; Dubinin et al., 2011; Dong et al., 2015). However, this escaping channel has not been well studied in Mercury's magnetosphere. Moreover, Raines et al. (2013) showed that the average observed density of $Na^+$-group and $O^+$-group varied with the TAA.

As shown in Figure 6b and 6c, the $He^+$ is evenly distributed in the magnetosheath and the plasma sheet without notable dawn-dusk asymmetry. The $He^+$, mostly originating from the Sun (e.g., Skoug et al., 1999; Lepri and Zurbuchen, 2010; Gilbert et al., 2012), the planet (surface or exosphere), and even the picked-up $He^+$ (e.g., Möbius et al., 1985; Gloeckler et al., 1993), can contribute to the observed $He^+$. The $He^{++}$ is commonly observed in Mercury's magnetosphere (Figure 6d–6f). However, the densities in the magnetosphere are much lower than the densities in the magnetosheath, which could be because the $He^{++}$ are originated from the solar wind.

## 1.7    Mercury's conducting core and conductivity profile

Mercury has a large conducting core with a conductivity of $\gtrsim 10^6$ S (e.g., Hood and Schubert, 1979), which is composed of liquid/molten metal (Peale, 1976). Figure 7 shows the conducting core of Mercury and how the conducting core responds to the variations in the solar wind (Slavin et al., 2014; Jia et al., 2019). MESSENGER (Rivoldini and Van Hoolst, 2013) and Earth-based radar observations (Hauck II et al., 2013) of Mercury's gravity and rotation have revealed a liquid core radius of ~2000 km, 0.82 $R_M$. Measurements of the induced magnetic fields may also reveal the scales of the conducting liquid core, though the solid upper layer of the core would be included. MESSENGER measurements of the induced magnetic fields reveal a similar scale of the liquid core radius (~0.85 $R_M$), which varies across studies (Johnson et al., 2016; Wardinski et al., 2019; Katsura et al., 2021). Mercury's liquid core accounts for ~55% of the planet's volume, which is the largest among the planets in the Solar System. Recently, some studies have indicated that Mercury has a solid inner core with a radius of 30% to 70% of the liquid core (Genova et al., 2019). For comparison, Earth's liquid core has a radius of ~3480 km, which accounts for ~16% of the Earth's volume. The solid inner core of the Earth has a radius of ~1221 km, which accounts for ~35% of the entire liquid core (Alfè et al., 2007).

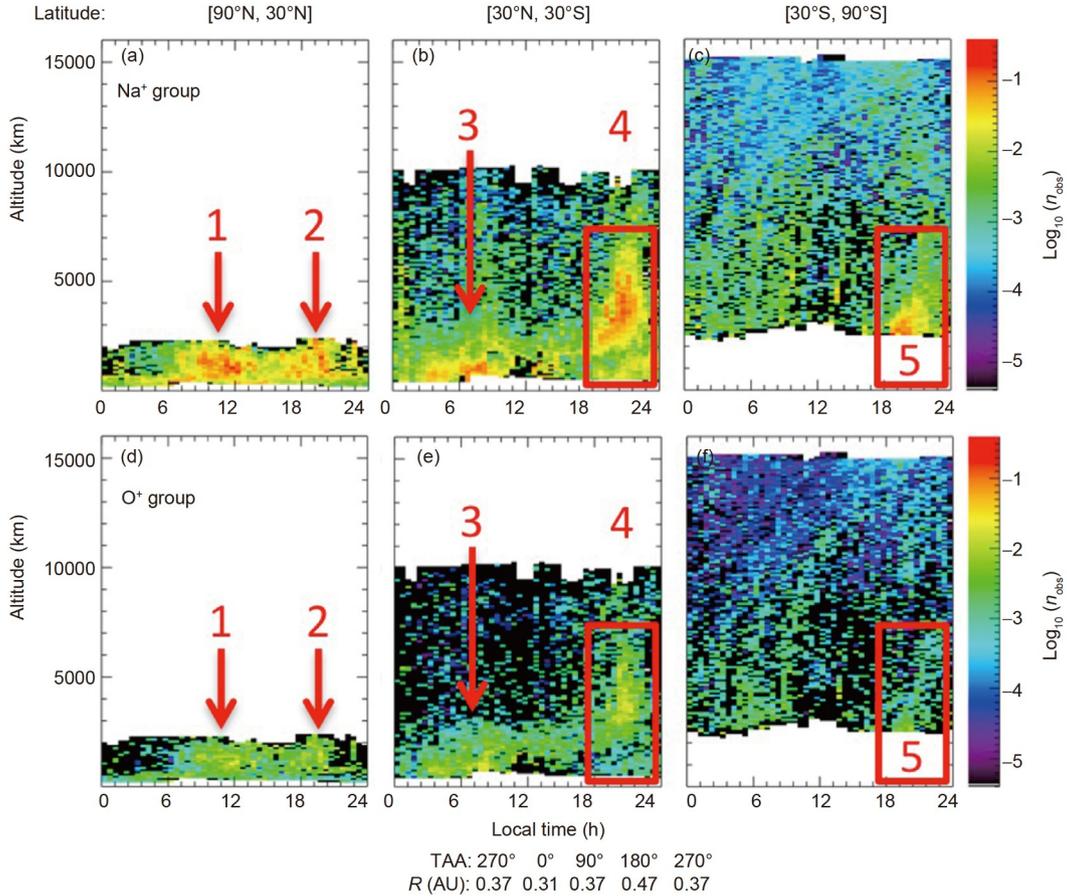

**Figure 5**   The distribution of Na$^+$-group ((a)–(c)) and O$^+$-group ((d)–(f)) heavy ions in Mercury's magnetosphere as functions of altitude (km) and local time (h). The Na$^+$-group includes ions with m q$^{-1}$ from 21 to 30, and the O$^+$-group with m q$^{-1}$ from 14 to 20. This figure shows the observed density of the ions (see Raines et al., 2011, 2013). This figure includes the data collected from 25 March 2011 to 31 December 2011, corresponding to 3.1 Mercury years. The numbers in the figure represent the enhancement features of the Na$^+$-group and O$^+$-group. The numbers 1 and 2 indicate the enhancements at high latitudes centered at local time ~10:30 hr and 19:00 hr, respectively. The numbers 3 and 4 indicate the enhancements around the dawn terminator and pre-midnight plasma sheet around the equator. The number 5 indicates an enhancement in an altitude of ~6000 km and high latitudes. This figure is adapted from Raines et al. (2013), and more information can be found in Raines et al. (2013).

The conductivity of Mercury's mantle and crust is discussed in several models. In Verhoeven et al. (2009), the conductivity of the mantle increases from ~10$^{-7}$ to ~10$^{-2}$ S m$^{-1}$ at depths of ~100 to ~500 km. The conductivity of the crust is ~10$^{-11}$ S m$^{-1}$ near the surface. However, the crustal conductivity could be significantly different in the model of metal-rich chondrite (Taylor and Scott, 2005).

The understanding of the core and the conductivity profile of the planets are important in many aspects. In the case of Mercury, the dynamo models on the generation mechanisms of Mercury's intrinsic magnetic field require accurate mea-surements of the core (Christensen, 2006; Takahashi and Matsushima, 2006). On the other hand, Mercury's interior has cooled more rapidly than the Earth's interior. The study of Mercury's interior can also help understand the evolution of the Earth's magnetic field. Mercury's liquid core is highly conducting with a conductivity of ≥10$^6$ S (Verhoeven et al., 2009). As a result, the conducting core responds to the external solar wind variations and, in turn, influences magnetospheric dynamics, which is illustrated in Figure 7 and is discussed in Glassmeier et al. (2007), Slavin et al. (2014, 2019), and Jia et al. (2019).

How field-aligned currents, such as Birkeland currents and the substorm current wedge, close without an ionosphere in Mercury's magnetosphere is an interesting topic. Anderson et al. (2014) proposed that the field-aligned current flows radially through the planet's low-conductivity crust until reaching the highly conducting core. The current then flows laterally from dawn to dusk through the more conductive material and flows upward through the low-conductivity

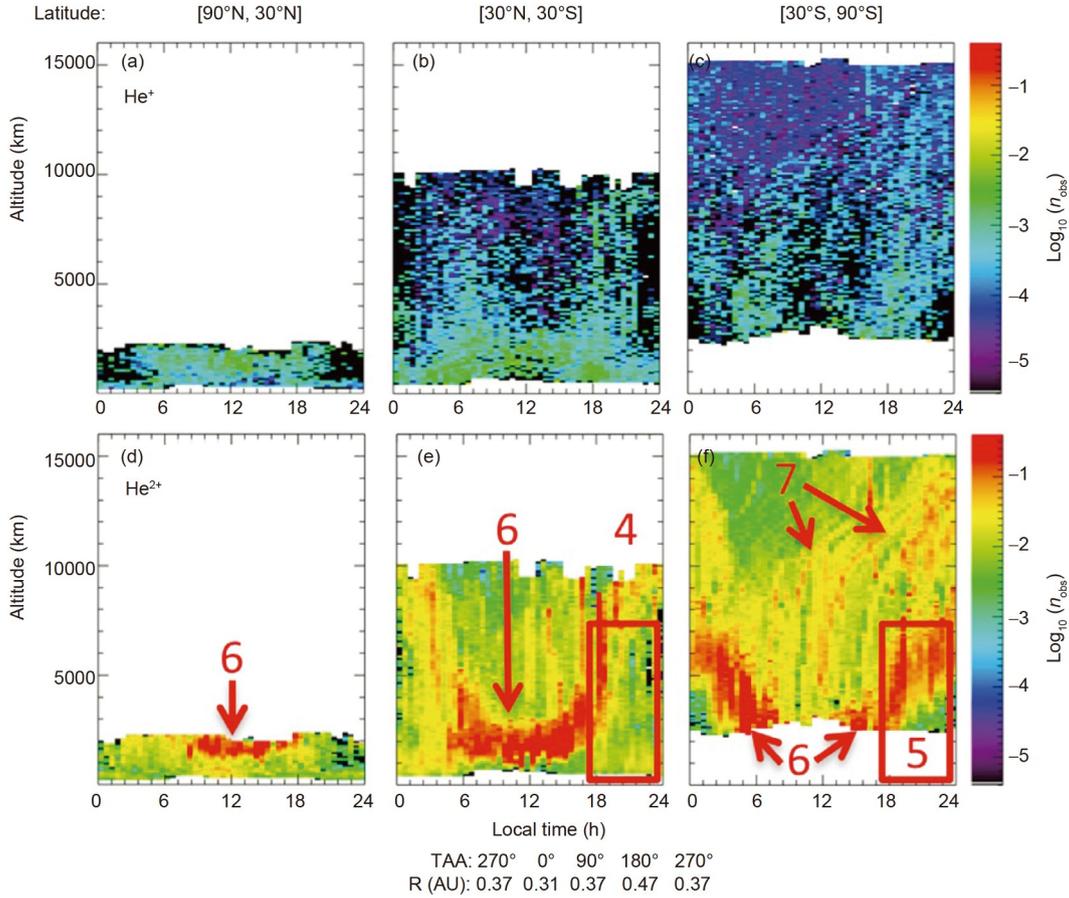

**Figure 6**   The distribution of He$^+$ ((a)–(c)) and He$^{++}$ ((d)–(f)) ions in Mercury's magnetosphere in the same format as Figure 5. The numbers 4 and 5 indicate the same enhancements of He$^{++}$ as the Na$^+$-group and O$^+$-group. The number 6 indicates the enhancement in the magnetosheath. The number 7 indicates the enhancements of He$^{++}$ in the orbits with an overall higher density. This figure is adapted from Raines et al. (2013), and more information can be found in Raines et al. (2013).

crust again, completing the closure of field-aligned current (see also, Janhunen and Kallio, 2004). The net electric conductance is estimated to be ~1 S. On the other hand, some studies indicate that due to the ionization of sodium or other atoms in Mercury's exosphere, a Pedersen conductivity can be defined since the newly created ions and electrons would be displaced a finite distance along the electric field and form an electric current. Note for the case in Mercury's magnetosphere; the electric field can be contributed by the convection electric field and the electric field associated with surface potential. This process is called pickup conductivity, which was first proposed for the partially ionized ionosphere of Io (Goertz, 1980; Ip and Axford, 1980). Cheng et al. (1987) discussed the pickup conductivity in Mercury's exosphere. In the pickup process, the equivalent integrated Pedersen conductivity $\Sigma_p$ can be estimated by,

$$\Sigma_p = N_0 m_i c^2 / \tau_i B^2, \qquad (5)$$

where $N_0$ is the column density of the neutral atom. $m_i$ is the mass of the atom. $c$ is the light speed. $\tau_i$ is the ionization time scale. $B$ is the magnetic field intensity. For sodium in the polar region, $N_0$ is on the order of ~$10^{11}$ atoms cm$^{-2}$. $\tau_i$ is

~$10^4$ s. $B$ is ~400 nT. We then can obtain a $\Sigma_p$ value of ~0.04 S. On the dayside equatorial region, the pickup Pedersen conductivity can reach a value of ~0.3 S. On the night side, the value should be much smaller than 0.04 S.

Considering that the Birkeland currents were distributed on the polar region, the pickup Pedersen conductivity might be able to close part of the field-aligned currents. The substorm current wedge is distributed on the nightside (Sun et al., 2015a; Poh et al., 2017a; Dewey et al., 2020). The pickup Pedersen conductivity can play a minor effect. However, some simulation studies indicate that under particular conditions, the dense sodium exosphere and the Pedersen and Hall currents might also close a large portion of the field-aligned current (Exner et al., 2020), which desires evidence from the observations. Another factor that still has not been carefully considered is how the electric field associated with surface potential influences the closure of the field-aligned current. The surface of Mercury can be charged by the solar radiation and plasma precipitation, similar to the situations at Moon (e. g., Manka, 1973; Freeman and Ibrahim, 1975; Halekas et al., 2008). However, modeling and observations of Mercury's surface potential and the electrostatic fields are still missing.

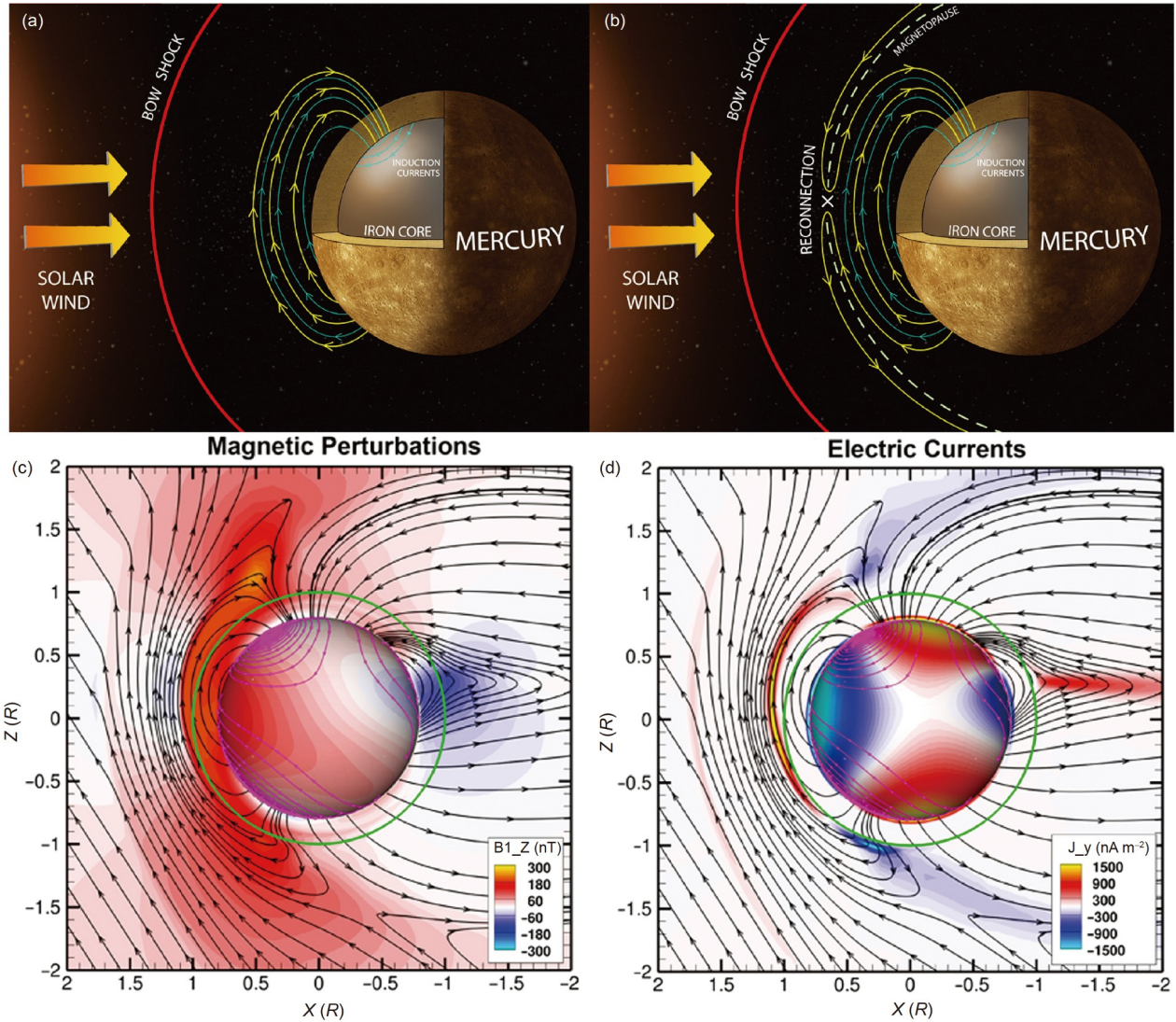

**Figure 7** The response of Mercury's large iron conducting core to the temporal variations of the solar wind. (a) and (b) are illustrations from Slavin et al. (2014). (c) and (d) are simulated magnetospheric configurations from Jia et al. (2019). An increase in solar wind pressure drives induction currents on the iron core (the cyan loops in (a) and (b), the magenta lines in (c) and (d)), which add magnetic flux to the intrinsic magnetic field (the cyan magnetic field lines in (a) and (b)). Dayside magnetopause reconnection can erode magnetic flux in the dayside magnetosphere, including the induction-created field lines, as shown in (b). In (c), the color indicates the $z$-component of magnetic field perturbations (in nT). In (d), the color indicates the current density in the $y$-direction (in nA m$^{-2}$). The panels (c) and (d) are from the global MHD model of Mercury's magnetosphere, in which the planetary conducting core is included and electromagnetically couples to the surrounding plasma environment (Jia et al., 2015, 2019). Dong et al. (2019) provide another version of Mercury's magnetosphere, which employs the ten-moment multifluid model and also couples to a conducting core.

## 2.   Flux transfer event "showers"

### 2.1   Flux transfer event and magnetic flux rope

Magnetic flux ropes are fundamental magnetic structures in space plasma physics and consist of magnetic field lines with helical topology. A magnetic flux rope is formed between neighboring reconnection X-lines as illustrated in Figure 8 and see also Lee and Fu (1985) and Raeder (2006), and thus, the generation of flux ropes frequently requires multiple reconnection X-line sites. However, in induced magnetospheres (Xie and Lee, 2019), or a region with velocity shear (see, for example, Nykyri and Otto, 2001), helical magnetic field lines can be generated without reconnection X-lines.

The magnetic flux rope is widely observed throughout the Solar System and possibly in galaxy clusters (e.g., Ruzmaikin et al., 1989), and plays an essential role in transporting flux and energy across different kinds of boundaries. For example, flux ropes are often observed optically in the Sun's corona, and interplanetary coronal mass ejections (ICMEs), one of the most well-known space weather events, are mostly formed on the solar surface as flux ropes and are subsequently ejected into interplanetary space (e.g., Cheng et al., 2017; Hu, 2017).

In planetary magnetospheres with global intrinsic magnetic fields, flux ropes have been frequently observed near the magnetopause and in the plasma sheet. In induced magnetospheres of Venus and Mars, flux ropes have been observed within the ionosphere (Russell and Elphic, 1979; Vignes et al., 2004; Bowers et al., 2021) and current sheets in the draped magnetic field (Eastwood et al., 2012; Zhang et al., 2012; Hara et al., 2017). Flux ropes near the magnetopause are known as the flux transfer events (FTEs), which were first observed on Earth's dayside magnetopause (Haerendel et al., 1978; Russell and Elphic, 1978), and then in the magnetospheres of Jupiter, Mercury, and Saturn (see Table 2 for more information). FTEs are generated during the magnetic reconnection between the IMF and the magnetospheric magnetic field. Inside FTEs, magnetic field lines have one end connected to the solar wind and the other to planetary magnetic fields (as illustrated in Figure 8). In spacecraft measurements, FTEs have bipolar signatures in the magnetic field component that is normal to the magnetopause surface, and enhancements in the magnetic field intensity towards the center of the structure.

FTEs play essential roles in planetary magnetospheres in many aspects. Here we discuss three important processes. First, FTEs are evidence of the occurrence of magnetic reconnection under most of the circumstances since they are believed to be generated by multiple reconnection X-lines. The direct measurements of the reconnection diffusion region require extremely high time resolution measurements of the fields and particles. For example, the NASA Magnetospheric Multiscale (MMS) mission (Burch et al., 2016) is designed to study magnetic reconnection around Earth's magnetosphere. However, most of the missions, especially planetary missions designed for other planets than the Earth, cannot provide such high-time and high-spatial resolution measurements. Since FTEs are consequences of magnetic reconnection, the study of FTEs can reveal the occurrence and property of magnetic reconnection. However, the occurrence of magnetic reconnection does not necessarily generate FTEs, which we need to keep in mind when applying FTEs to reveal the occurrence of magnetic reconnection.

Second, FTEs can transport magnetic flux from the dayside magnetosphere into the nightside and contribute to the flux circulation of the Dungey cycle in planetary magnetospheres (Section 1.4 introduced the Dungey cycle). Many studies have investigated the amount of magnetic flux transported by FTEs in the Dungey cycle. As illustrated in Figure 9, there are two flux transport models in the planetary magnetospheres. One is Dungey's initial single X-line reconnection (SXR) model (Figure 9a), the other is the multiple X-lines reconnection model (MXR) (Figure 9b). In the SXR, magnetic flux is transported by a single X-line reconnection, which continuously reconnects the magnetic field lines near the subsolar point (Dungey, 1961). While in MXR, magnetic flux can be transported by FTEs generated by the multiple X-lines. In the magnetospheres of Earth,

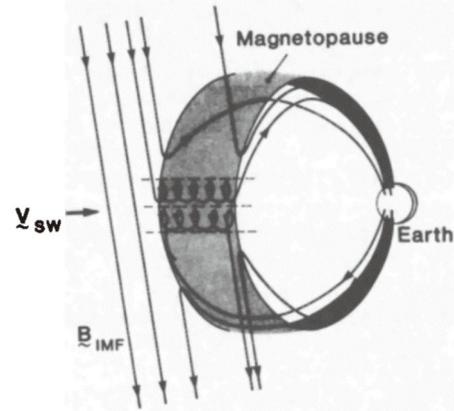

**Figure 8**   Multiple X-line reconnections generate flux transfer events (FTEs) on the dayside magnetopause. The FTE is formed between neighboring X-lines, in which reconnection occurs between the interplanetary magnetic field and the planetary magnetic field. The figure is adapted from Lee and Fu (1985).

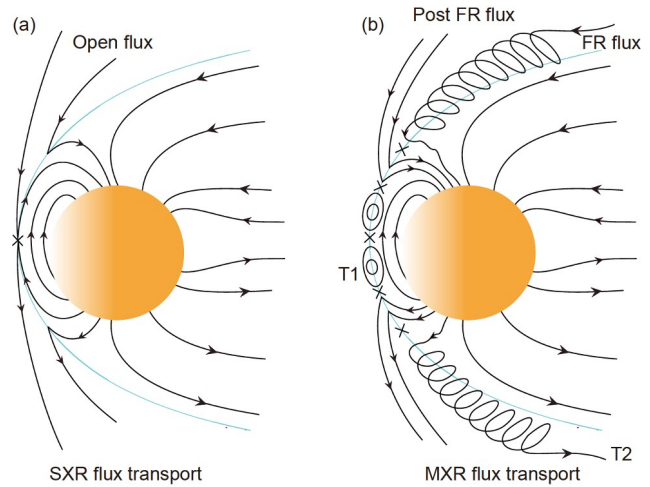

**Figure 9**   Schematic illustration of magnetic flux transport from (a) single X-line reconnection (SXR) model (Dungey, 1961) and from (b) multiple X-line reconnection (MXR) model (Lee and Fu, 1985). This figure is adapted from Sun et al. (2020b).

Jupiter, and Saturn, FTEs can transport only a small amount of the magnetic flux (<5%) compared to the flux circulation of the Dungey cycle (see Table 2 for detailed information). This might indicate that multiple X-line reconnections infrequently occur in these planetary magnetospheres.

Third, FTEs can exchange particles between the solar wind and the planetary magnetospheres. The solar wind particles can transport along with the open magnetic field lines inside the FTEs and enter into the magnetosphere, and the magnetospheric particles can leak into the magnetosheath along with the same flux tubes. Because the solar wind particles can influence the polar-cap regions, the precipitation of the solar wind particles attracts much attention from researchers. For example, the footprint of FTEs has been identified in the Earth's ionosphere, and FTEs likely enhance the local con-

**Table 2** Flux transfer events (FTEs) and their contributions to planetary magnetospheres

| Planet | Repetition | Flux content | Polar cap open flux | Percentage per FTE | Cross-magnetospheric potential drop | Contribution per FTE |
|---|---|---|---|---|---|---|
| Mercury | ~10 s[a)] | ~0.02 to 0.4 MWb[b)] | ~2.6 MWb[c)] | ~1% to 3% | ~20 kV[d)] | ~6 to 8 kV (~40%) |
| Earth | ~5 min[e)] | ~0.4 to 4 MWb[f)] | ~600 MWb[g)] | ~0.1% to 1% | ~80 to 100 kV[h)] | ~1 to 10 kV (~1%) |
| Saturn | ~5 min[i)] | ~0.1 to 0.8 MWb[j)] | ~10 to 50 GWb[k)] | $\sim 10^{-3}$ to $10^{-4}$ | ~50 to 200 kV[l)] | ~2 to 9 kV (~1%) |
| Jupiter | ~10 min[m)] | ~0.3 MWb[n)] | ~300 to 500 GWb[o)] | $\sim 10^{-6}$ | ~100 to 500 kV[p)] | ~1 kV (<1%) |

a) Russell and Walker (1985); Slavin et al. (2010b, 2012b, 2014, 2019); Sun et al. (2020b). b) Slavin et al. (2010b, 2019); Imber et al. (2014); Sun et al. (2020b). c) Slavin et al. (2009, 2010a); Imber and Slavin (2017). d) Slavin et al. (2009); DiBraccio et al. (2015a); Jasinski et al. (2017); Dewey et al. (2018). e) Rijnbeek et al. (1984); Lockwood et al. (1995). f) Rijnbeek et al. (1984); Hasegawa et al. (2006); Zhang et al. (2008); Fear et al. (2017). g) Milan et al. (2004). h) Wygant et al. (1983); Boyle et al. (1997). i) and j) Jasinski et al. (2016, 2021). k) Jackman et al. (2004); Badman et al. (2014). l) Jackman and Arridge (2011); Masters (2015). m) and n) Walker and Russell (1985). o) Nichols et al. (2006); Badman and Cowley (2007). p) Nichols et al. (2006); Jackman and Arridge (2011) vection (van Eyken et al., 1984; Goertz et al., 1985; Provan et al., 1998). The precipitation rates of solar wind particles depend on the solar wind particle flux and the loss cones of the injected solar wind particles. As shown in Figures 2 and 3, the solar wind particle flux is highest at Mercury compared to the other planets in the Solar System. The loss cones rely on the magnetic field intensities of the planets. Poh et al. (2018) have demonstrated that the loss cone of the particles is largest in Mercury's magnetosphere (~20° to 30° or even larger) (Korth et al., 2014; Winslow et al., 2014) compared to other planetary magnetospheres (< a few degrees or smaller). Therefore, a study on how solar wind particles influence Mercury's magnetosphere and exosphere could be useful. We note that the magnetic field topologies inside FTEs are more complex than what has been discussed here. Some studies in Earth's magnetosphere have shown that both ends of the magnetic field lines could be closed to the planets or connected to the solar wind (Fu et al., 1990; Pu et al., 2013).

The remaining sections of this chapter discuss the recent progress on these three questions. However, there are several other important questions related to FTEs. For example, how were the FTEs formed? Are they formed by the simultaneous X-line reconnections (Lee and Fu, 1985) or the sequential X-line reconnections (Raeder, 2006)? How do the FTEs evolve? Can the FTEs energize particles, etc. Section 2.6 briefly discusses a few of these questions.

### 2.2 Cases of flux transfer event showers at Mercury

In spacecraft observations, FTE signatures last approximately one minute and are separated by tens of minutes in the magnetospheres of Earth (Lockwood et al., 1995), Jupiter (Walker and Russell, 1985), and Saturn (Jasinski et al., 2016, 2021). Meanwhile, the formation of FTEs at intrinsically magnetized planets normally requires magnetic field lines on the two sides of the dayside magnetopause to be nearly anti-parallel, i.e., the magnetic shear angle is approximately 180° (e.g., Kuo et al., 1995). However, in Mercury's magnetosphere, FTE signatures only last around one second and are separated by a few seconds in spacecraft observations (see

Figure 10 and Russell and Walker, 1985; Slavin et al., 2009). In particular, FTEs often appear in large numbers (>10) in a few minutes, which is known as the FTE shower (Slavin et al., 2012b). Although the occurrence of FTEs is higher as the magnetic shear angle across the dayside magnetopause becomes larger (Leyser et al., 2017), FTE showers have been observed under the northward IMF on Mercury's dayside magnetopause (Figure 10; Slavin et al., 2014; Sun et al., 2020b).

Figure 10a shows an FTE shower that was observed during the southward IMF with a magnetic shear angle of ~110°. The magnetosheath plasma $\beta$ adjacent to the dayside magnetopause was estimated to be ~0.05. In the calculation of plasma $\beta$, the magnetosheath plasma pressure is obtained through the assumption that pressure is balanced on the dayside magnetopause and the thermal pressure is negligible compared to the magnetic pressure in the dayside magnetosphere. Figure 10b shows an FTE shower that occurred during the northward IMF during an ICME impact event. This event has been analyzed by Slavin et al. (2014). This FTE shower corresponded to a magnetic shear angle of ~60° and magnetosheath plasma $\beta$ of ~0.06 (Slavin et al., 2014). It can be seen that the magnetosheath plasma $\beta$ was small in both FTE showers, which implies that the magnetic reconnection is nearly symmetric on Mercury's dayside magnetopause and the magnetosheath plasma $\beta$ could play an important role in the occurrence of FTE showers.

### 2.3 Dayside magnetopause reconnection

The investigation of the occurrence of FTEs on the magnetopause could reveal the properties of magnetic reconnection because of their close relationship that the FTE is an outcome of magnetic reconnection. Most of the observations from Mariner 10 and MESSENGER on Mercury's dayside magnetosphere, for instance, Russell and Walker (1985); Slavin et al. (2009, 2010b); and Imber et al. (2014), were made during the southward IMF, while only a few were during northward IMF conditions, for example, Slavin et al. (2014). Leyser et al. (2017) have done a statistical study of the FTEs on Mercury's dayside magnetopause. They have shown that the FTEs are prevalent during the southward IMF, which corresponds to the magnetic shear angle larger than 90°. Leyser et al. (2017) have also shown that the FTEs are observed preferentially on the pre-noon sector of the dayside magnetopause, which is likely due to the dawnward component ($-B_y$) bias of the Parker spiral.

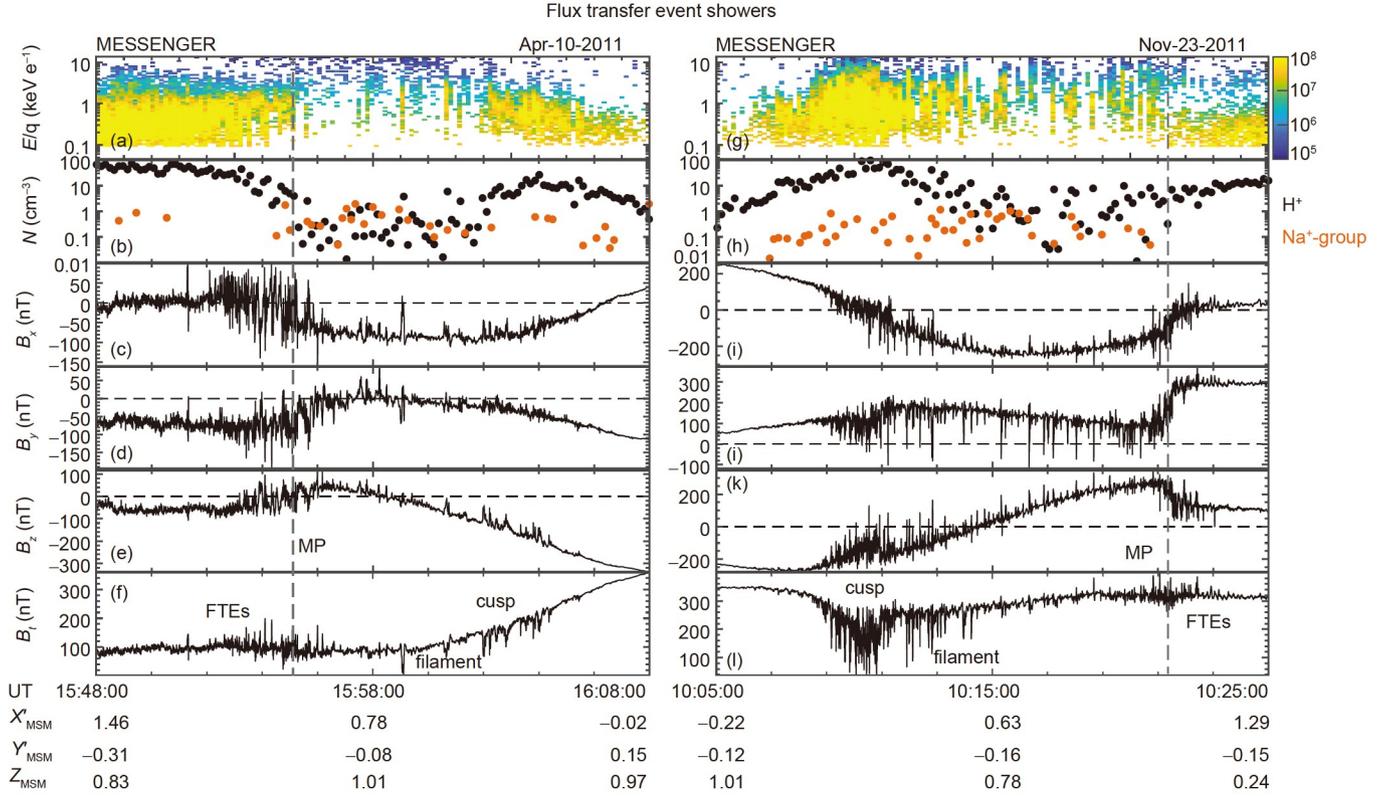

**Figure 10**   Flux transfer event showers under southward IMF (left panel, 10 April 2011) and northward IMF (right panel, 23 November 2011, adapted from Slavin et al. (2014)). (a) and (g), proton differential particle flux, unit is (cm$^{-2}$ s$^{-1}$ sr$^{-1}$ keV$^{-1}$ e); (b) and (h), the observed densities (Raines et al., 2011, 2013) of proton (H$^+$, black dots) and sodium group ions (Na$^+$-group, gold dots); (c) and (i), magnetic field $x$-component, $B_x$; (d) and (j) $B_y$; (e) and (k) $B_z$; (f) and (l) magnetic field intensity, $B_t$.

Sun et al. (2020b) have surveyed the entire MESSENGER database from 11 March 2011 to 30 April 2015, which have identified a total number of 3748 dayside magnetopause crossings. They have employed an established automatic flux rope detection technique (Smith et al., 2017a, 2017b) to identify flux ropes about the dayside magnetopause crossings. They require flux ropes to contain bipolar deflections in the magnetic field component normal to the magnetopause, which is coincident with enhancements in either of the other two field components and the magnetic field intensity. The bipolar magnetic fields correspond to magnetic field rotations and the enhancement in magnetic field intensity to the core field. Detailed descriptions can be found in Sun et al. (2020b). In that study, they investigated the properties of FTE showers, which are defined as the magnetopause crossings accompanied by at least ten flux ropes.

The occurrence of FTE showers on Mercury's dayside magnetopause crossings is shown as a function of the magnetic shear angle and the magnetosheath plasma $\beta$ in Figure 11. The FTE showers can be observed with magnetic shear angles from 0° to 180°, and magnetosheath plasma $\beta$ from 0.1 to 10. Occurrences of the FTE showers increase with increasing magnetic shear angles (Figure 11a–11c), and decreasing plasma $\beta$ (Figure 11a, 11d, and 11e), with the highest occurrence of the FTE showers corresponding to a value of ≥85%, which is located in the region with a magnetic shear angle exceeding 150° and a plasma $\beta$ value less than 0.5. We note that the occurrence rates are larger than 50%, even for a small magnetic shear angle of ~70°. The magnetic shear angle dependency of the occurrence of FTEs in Mercury's magnetosphere is similar to other planetary magnetospheres. However, the occurrence of the FTEs also displays clear plasma $\beta$ dependency at Mercury.

The curved line in Figure 11a indicates a theoretical relation of magnetic reconnection between the magnetic shear angle and the plasma $\beta$ difference on the two sides of the reconnecting current sheet (Swisdak et al., 2010),

$$\Delta\beta = 2\frac{L_{cs}}{\lambda_i}\tan\left(\frac{\theta}{2}\right), \tag{6}$$

where $L_{cs}$ is the thickness of the current sheet and $\lambda_i$ is the ion inertial length. The curve corresponds to $L_{cs}/\lambda_i=1$. The theory predicts that the region below the curve favors magnetic reconnection, and the region above the curve suppresses magnetic reconnection. The occurrence of the FTE showers in Figure 11a does show higher values below the curve than above the curve on the region of large magnetic shear angle (>110°). However, the occurrence below the curve on the region of small magnetic shear angle region (<60°) is not higher than the value above the curve.

The intensity of the IMF is largest (Figure 2d) and the solar wind plasma $\beta$ (Figure 3c) is lowest near Mercury's orbit compared to other planets in the Solar System. As a result, the IMF can be easily draped ahead of the dayside magnetopause and forms a plasma depletion layer (PDL) (see Gershman et al., 2013). The PDL corresponds to low plasma $\beta$ (<0.1) and, therefore, causes the occurrence of magnetic reconnection on the dayside magnetopause to be less dependent on the magnetic shear angle. Furthermore, the PDL can contain a higher Alfvén speed, and the dayside magnetopause reconnection would correspond to a higher reconnection rate. The results from global magnetohydrodynamics (MHD) simulation have indicated that the solar wind $M_A$ can influence the occurrence rates of the FTEs, i.e., the higher the $M_A$ is, the lower the occurrence rates of the FTEs (Chen C et al., 2019), which is somewhat consistent with the observations.

## 2.4    Transport of flux from dayside to nightside

The axial magnetic flux of a single FTE-type flux rope on Mercury's dayside magnetopause ranges from 0.02 to 0.05 MWb (Slavin et al., 2010b; Sun et al., 2020b). The loaded magnetic flux in the lobe has an upper limit value of 1.07 MWb in Mercury's loading-unloading events (0.69±0.38 MWb) (Imber and Slavin, 2017). In order to estimate the amount of magnetic flux transported by flux ropes, i.e., the MXR, during Mercury's Dungey cycle, we estimate the number of FTEs during the loading interval ($T_{loading}/T_{spacing}$), and then multiply it by the mean magnetic flux of FTEs.

$$\Phi_{Flux} = \Phi_{FTE} T_{loading} / T_{spacing}. \qquad (7)$$

The $T_{loading}$ is 115 s, which is the average period of the loading phase obtained by Imber and Slavin (2017). Figure 12 shows the results of the accumulated magnetic flux by FTEs as a function of the magnetic shear angle and magnetosheath plasma $\beta$. The horizontal dashed line represents the loaded magnetic flux in the lobe (1.07 MWb) in Mercury's loading-unloading events (Imber and Slavin, 2017). The FTE's transported magnetic flux increases with magnetic shear angle (0.65 to 0.9 MWb) and does not depend on magnetosheath plasma $\beta$ (0.8 MWb). The FTEs can transport 60% to 85% of the magnetic flux that is required by the flux loading-unloading of Mercury's Dungey cycle. Therefore, the results from Figure 12 strongly suggest that the MXR is the primary magnetic reconnection process in Mercury's dayside magnetopause during the FTE shower periods, which is significantly different from the reconnection model in the magnetospheres of Earth, Jupiter, and Saturn (see Table 2 for detail). We note that Fear et al. (2019) argue that the post-flux rope flux (see Figure 9) can be several times the flux content inside the flux rope in some cases. Although Fear et al. (2019) conclude that the magnetic flux related to FTE-type flux ropes contributes most of the flux transport, it is the SXR process that transports the majority of the magnetic flux in their study.

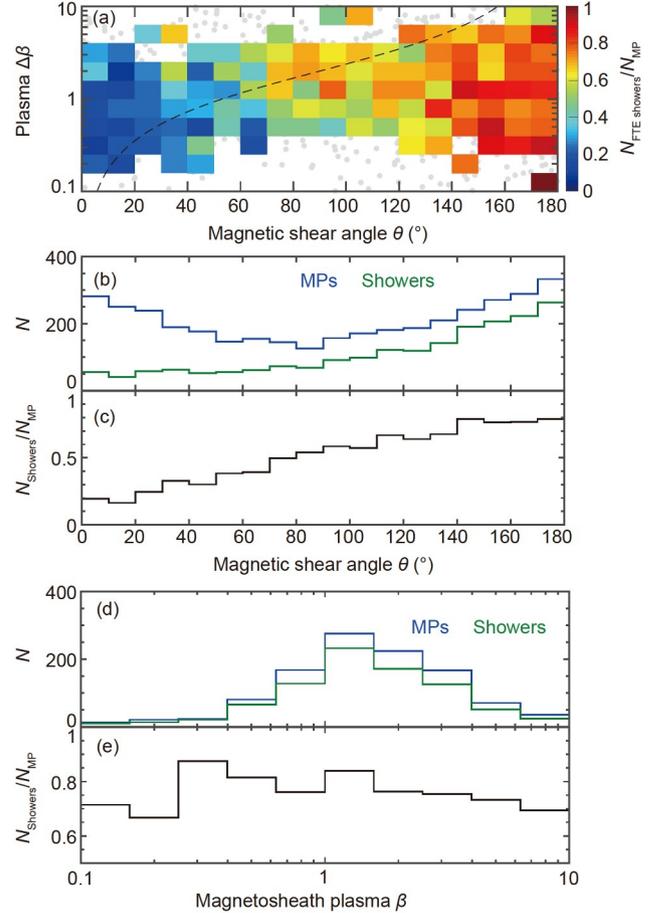

**Figure 11**  The occurrence of FTE showers in Mercury's dayside magnetopause crossings as functions of magnetic shear angle ($\theta$) across the magnetopause and magnetosheath plasma $\beta$. (a) The fraction of dayside magnetopause crossings that contain FTE showers as a function of both $\theta$ and plasma $\beta$. The dashed curve in (a) represents a theoretical relation of $\theta$ and plasma $\beta$ (Swisdak et al., 2010), above which reconnection is expected to be suppressed, and below which reconnection is expected to be favored. (b) Marginal distributions of the number of magnetopause crossings (blue), FTE showers (green), and (c) occurrence of FTE on $\theta$. (d) Marginal distributions of the number of magnetopause crossings (blue), FTE showers (green), and (e) occurrence of FTE on plasma $\beta$. This figure is adapted from Sun et al. (2020b).

## 2.5    Transport particles into magnetosphere

As introduced in Section 2.1, magnetic field lines inside the FTEs have one end connected to the solar wind while the other connected to the planetary magnetosphere. The solar wind particles travel along the open magnetic field lines and enter into the magnetosphere. On the other hand, magnetospheric particles transport into the magnetosheath along these open field lines. The footprints of the FTEs have been identified and investigated in the Earth's ionosphere. The footprints of the FTEs are associated with ionospheric flows, which are thought to be driven by the enhanced solar wind precipitations (e.g., Lockwood et al., 1990; Fear et al., 2007).

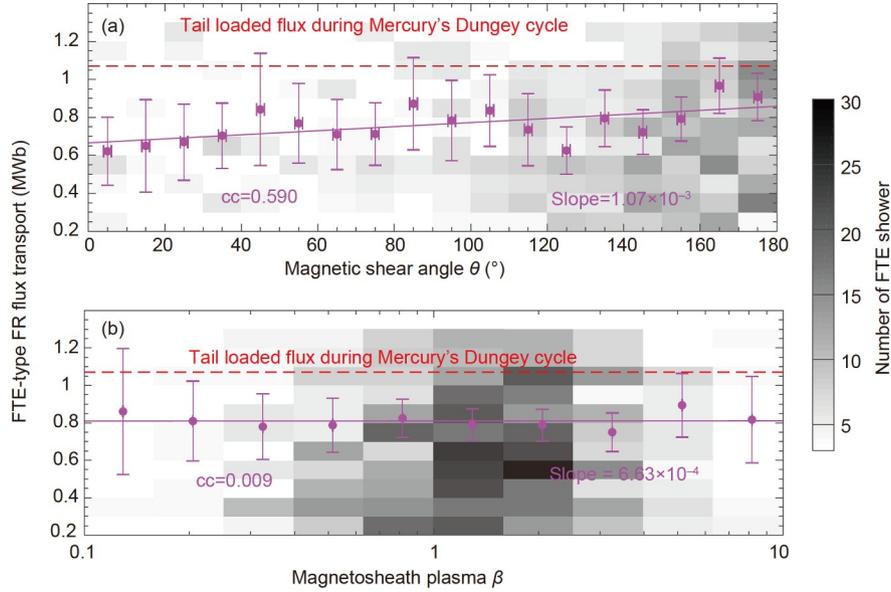

**Figure 12**   The amount of magnetic flux transported by FTE-type flux ropes during a nominal loading phase of Mercury's substorms as functions of (a) magnetic shear angle and (b) magnetosheath plasma $\beta$. This figure is adapted from Sun et al. (2020b).

The investigation of the FTEs related particle transport is important within Mercury's magnetosphere. First, FTEs are important magnetic structures on Mercury's dayside magnetopause, as discussed in Section 2.4. Second, FTEs can appear in extremely high frequency and are only separated by a few seconds. Third, the loss cone of injected solar wind particles is much higher at Mercury than at Earth, Jupiter, and Saturn. Fourth, Mercury does not have an intense atmosphere, and the solar wind particles with loss cones precipitate into the surface underneath Mercury's cusp and cause sputtering. Since sputtering is an effective way to eject particles out of the planet's surface, it could influence the exospheric dynamics at Mercury in a short time interval (~10 minutes).

In Figure 10b, Slavin et al. (2014) reported the cusp plasma filaments at Mercury, which are discrete magnetic field decreases that last a few seconds in the spacecraft's measurements and contain magnetosheath plasma. These magnetic decreases are termed cusp plasma filaments since they appear near or in the cusp region. Poh et al. (2016) investigated the cusp plasma filaments in a larger dataset. They found that the cusp plasma filaments have scales larger than the gyro-radii of the background protons, and the filaments contain plasmas that have similar properties as the magnetosheath plasma. They even identified several flux ropes among those plasma filaments. As shown in Figure 13, Poh et al. (2016) estimated a precipitation rate on the surface beneath the cusp from the cusp filaments of ~2.7×10²⁵ s⁻¹, which is around an order of magnitude larger than the average precipitation rates obtained from averaging over the entire cusp region (Winslow et al., 2012, 2013).

The investigation of Poh et al. (2016) implies that the cusp plasma filaments map to FTEs at the magnetopause. The high precipitation rate associated with the cusp filaments has important implications for surface sputtering and space

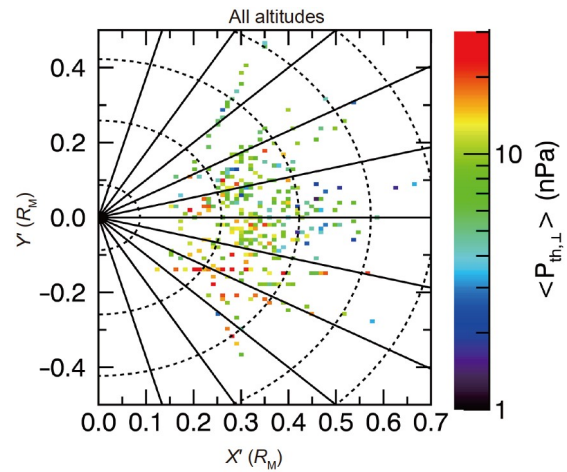

**Figure 13**   The spatial distribution of the perpendicular thermal pressure inside cusp plasma filaments. The perpendicular thermal pressure is derived from the magnetic depression inside the filaments, which are assumed to be pressure balanced with the surrounding magnetic field. This figure is adapted from Poh et al. (2016).

weathering at Mercury.

### 2.6  Open questions regarding FTE showers

The reviewed works in this subsection are the starts of the studies of FTE showers in Mercury's magnetosphere. This phenomenon desires further investigations, including the formation and the influence on Mercury's exosphere and magnetosphere, etc. This is partly limited by the spacecraft measurements. In this section, we discuss several open questions of the FTE showers in Mercury's magnetosphere.

The first question is related to the FTE showers on nightside high latitude magnetopause.

An FTE shower was first reported by Slavin et al. (2012b) on the nightside high latitude magnetopause during a northward IMF (Figure 14). However, FTE showers on the nightside high latitude magnetopause have not been well investigated thus far. Slavin et al. (2012b) showed that the large number of FTEs analyzed in their study were likely formed during the high latitude magnetopause reconnection that is tailward of Mercury's southern cusp (also called the cusp reconnection). The origin of the shower is in agreement with the Cooling model (Cooling et al., 2001). However, whether FTE showers formed on the dayside magnetopause can be transported to the nightside high latitude magnetopause is still unknown.

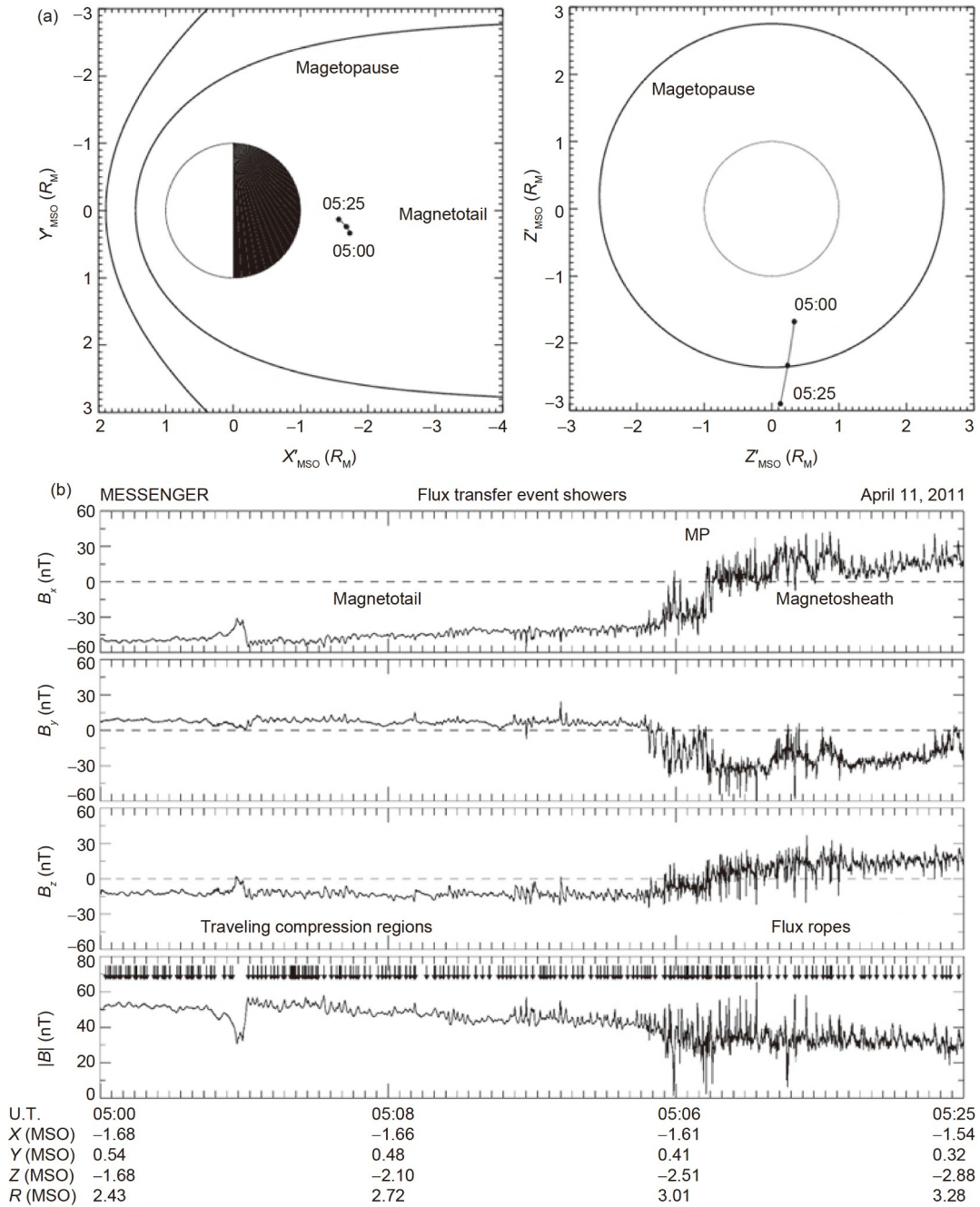

**Figure 14**  An FTE shower on Mercury's nightside high-latitude magnetopause. (a) The location of MESSENGER in the $X'_{MSO}$-$Y'_{MSO}$ and $Y'_{MSO}$-$Z_{MSO}$ planes, respectively. (b) The magnetic field measurements of the FTE shower. This figure is adapted from Slavin et al. (2012b).

These high latitude FTEs containing open field lines are expected to transfer magnetosheath plasma into the high latitude magnetotail. This transported magnetosheath plasma can form the plasma mantle, which drifts towards the magnetic equatorial plane driven by the cross-tail electric field and eventually supplies the plasma sheet. Moreover, the injected magnetosheath plasma can directly precipitate into the polar region. In studies at the Earth, the high-latitude magnetopause reconnection can occur continuously for many hours and inject and accelerate solar wind particles into the magnetosphere, creating the proton auroral spot (Frey et al., 2003). In the case of Mercury, the injected and accelerated solar wind particles can directly precipitate into the planet's surface and cause sputtering. Whether this process is an important source for the neutral sputtering near Mercury's polar region is an interesting topic that is yet to be investigated.

The second open question is the coalescence of FTEs. In studies of Earth's magnetosphere, it is suggested that the FTEs start out being small-scale (electron-scale or ion-scale) and are generated by multiple X-lines in the magnetopause current layer (e.g., Lee and Fu, 1985; Daughton et al., 2006). As the FTEs travel along the magnetopause driven by magnetosheath flow or reconnection outflows, they grow in scale. Several processes are proposed for this growth. For example, multiple X-lines may continuously occur as the FTEs travel along the magnetopause (Paschmann et al., 1982; Akhavan-Tafti et al., 2020), or within a chain of FTEs, neighboring FTEs are forced to merge into larger FTEs, in a process called coalescence (Biskamp and Welter, 1980; Dorelli and Bhattacharjee, 2009; Fermo et al., 2011; Hoilijoki et al., 2017).

FTE showers at Mercury contain extremely frequent flux ropes separated by only a few seconds, which makes Mercury's magnetopause an ideal environment to investigate the merging of flux ropes. Figure 15 shows examples of coalescence on Mercury's magnetopause (see also, Zhong et al., 2020b). In coalescence, multiple flux ropes merge into a single larger flux rope. In Figure 15, several large-scale FTEs contain at least three small-scale flux ropes, which are likely to merge. In this case, secondary reconnection would occur between neighboring flux ropes, which is a complicated scenario that might require three-dimensional analysis. This process would be very consequential as particles would be energized, accompanying the formation of a larger scale flux rope.

The third open question is on the formation mechanism of flux ropes along Mercury's dayside magnetopause. Lee and Fu (1985) proposed the simultaneous X-line reconnection, in which a magnetic flux rope forms in between two X-lines that occur simultaneously. Raeder (2006) proposed a scenario of sequential X-line reconnections. In this model, one X-line occurs and convects to high latitude due to the magnetosheath shear flow, and subsequently, another X-line occurs in the initial place of the first X-line resulting in a flux rope between the two X-lines. In the latter model, the non-zero dipole tilt is crucial since it requires that the stagnation point of the magnetosheath flow does not overlap with the reconnection X-line, i.e., the largest magnetic shear angle between the IMF and the planetary magnetic field. Studies in Earth's magnetosphere have shown that sequential X-lines could generate flux ropes near the dayside magnetopause (Hasegawa et al., 2010; Zhong et al., 2013). However, the dipole almost aligns with the rotation axis at Mercury. Therefore, we would expect sequential X-line reconnections to be limited in Mercury's magnetosphere.

We can also apply the theory of tearing instability on Mercury's dayside magnetopause. The tearing instability grows fast and can form a chain of secondary islands, i.e., ion-scale or electron-scale flux ropes. Loureiro et al. (2007) showed that the wavenumber scales $S^{3/8}$, where $S$ is the Lundquist number. Here we make a simple estimation: the Lundquist number $S$ is approximately $\sim(L/d_i)^2$, in which $L$ is the extent of the thin current sheet, $d_i$ is the ion inertial length. On Mercury's dayside magnetopause, we assume $L\sim1$ $R_M$, ion density $\sim50$ cm$^{-3}$. The Lundquist number, therefore, is $S\sim4\times10^3$, and if the tearing instability occurred, the wavelength of the secondary island generated by the tearing instability is $2\pi L/S^{3/8}$, which corresponds to a length of $\sim600$ km. The scale of flux ropes along Mercury's dayside magnetopause is $\sim500$ to 1000 km, which is a similar scale of the wavelength derived from the tearing instability. This consistency could imply that the tearing instability can indeed occur on Mercury's dayside magnetopause, which indicates that simultaneous X-line reconnection can account for the formation of the large-number and high-frequency of FTEs during the FTE shower intervals. However, this certainly deserves a more detailed analysis.

## 3. Magnetosphere under extreme solar events

### 3.1 Core induction and reconnection erosion

In chapter 1, we introduced the large metallic core of Mercury, which is highly conducting and can react to the variations in planetary magnetic fields that are caused either by the motion of magnetospheric boundaries or the changes of the large-scale current systems (e.g., Glassmeier et al., 2007; Heyner et al., 2016). For example, when the solar wind dynamic pressure increases, the Chapman-Ferraro current system on the magnetopause is enhanced, and the magnetopause moves closer to the surface in response to the out-ward pressure enhancement. However, the conducting core responds to the changes in the magnetic field, i.e., the perturbed magnetic field associated with the current enhanced system, and resists the inward motion of the dayside magnetopause. As a result, electric currents appear at the top of the conducting core and produce a magnetic field in the same sense as a dipole magnetic field which resists the inward motion of the dayside magnetopause (Slavin et al., 2014, 2019; Jia et al., 2015, 2019). At times, the induced fields can be significant compared to the intrinsic magnetic fields and have significant effects on the global magnetospheric configurations (e.g., Hood and Schubert, 1979; Suess and Goldstein, 1979; Glassmeier et al., 2007; Jia et al., 2019; Slavin et al., 2019). The induced fields during the period of extreme solar events can prevent the motion of the dayside magnetic field into the nightside or underneath the surface of the planet.

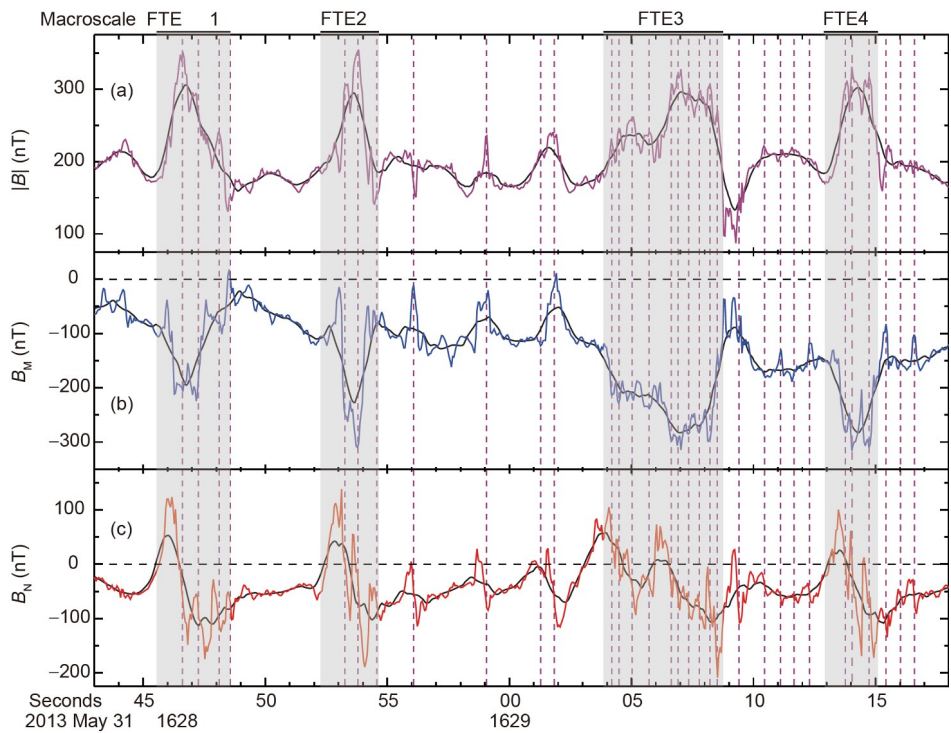

**Figure 15** A coalescence event of the FTEs in local magnetopause coordinates. (a) The magnetic field intensity, |**B**|; (b) The duskward component, $B_M$; (c) The normal component, $B_N$. This figure is adapted from Zhong et al. (2020b).

However, reconnection erosion due to magnetic reconnection plays the opposite role as induction on the dayside magnetopause (e.g., Slavin and Holzer, 1979). Reconnection erosion results in the transfer of magnetic flux from the dayside magnetosphere into the nightside magnetotail. As a consequence, reconnection erosion can erode the dayside closed magnetic flux, and the dayside magnetopause moves inward.

The induced magnetic field can be investigated through several approaches. One approach is to investigate long-term variation (Zhong et al., 2015b; Johnson et al., 2016). Due to the eccentricity of Mercury's orbit, the solar wind dynamic pressure varies significantly in a timescale of Mercury year (~88 Earth days). The variations of the dipole moment due to induction are estimated to be ~4% of the planet's dipole (Johnson et al., 2016). However, reconnection erosion is likely an important factor in controlling the standoff distance of Mercury's dayside magnetopause, and the subsolar magnetopause can reach the sunlit planetary surface ~1.5% to 4% of the time (Zhong et al., 2015b; Johnson et al., 2016).

Another approach is to investigate the short-term variations associated with the transient events in the solar wind. The solar wind dynamic pressure transient events include several types, including interplanetary shocks, solar wind discontinuities, ICMEs, and high-speed streams (HSS). Interplanetary shocks and discontinuities associated with pressure events are often small in amplitude and correspond to durations of a few minutes, which provides challenges for spacecraft without solar wind monitors, such as MESSENGER, to determine the arrival time of these events. In the following section, we introduce the transient effects of two types of extreme solar events: ICMEs and high-speed streams (HSSs).

### 3.2 Dayside magnetosphere under extreme solar events

Slavin et al. (2014) and Jia et al. (2019) have investigated a group of (eight events) highly compressed magnetosphere (HCM) events. In the HCM events, the magnetic field in-tensity adjacent to the dayside magnetopause within the magnetosphere was greater than 300 nT, which was much larger than the average value ~100 nT. This implies a strong dynamic pressure in the solar wind and high compression of the magnetosphere. Figure 10 on the right-hand side shows an ICME event, which was caused by an ICME on 23 November 2011, and Figure 7c and 7d show the magnetospheric configuration obtained from a global magnetohydrodynamic (MHD) simulation, including the conducting core (Jia et al., 2019) for the same event. The ICME brought a high dynamic pressure of ~50 nPa and a very low Alfvénic Mach number, although these conditions varied during the ICME (Exner et al., 2018). In this event, the magnetic field intensity on the two sides of the dayside magnetopause was almost constant, indicating a plasma depletion layer (Figure 10). Although the magnetic shear angle across the magnetopause was only ~60°, magnetic reconnection occurs at high rates, with a dimensionless reconnection rate of ~0.16 and high-frequent FTEs. The deep cusp and abundant cusp filaments (Figure 10) confirmed the high reconnection occurrence in this low magnetic shear angle event. In this event, the subsolar magnetopause was located at ~1.16 $R_M$ (see a point in Figure 16), which was closer to the planet than expected by the model including induction current (see Glassmeier et al., 2007). If there were no reconnection erosion, the solar wind dynamic pressure would have to be ~90 nPa, almost double the 50 nPa, for the dayside magnetopause to reach such a low altitude. Slavin et al. (2014) pointed out that a high dimensionless reconnection rate mitigates the effects of induction currents and causes the standoff distance of the dayside magnetopause to be closer to the planet (see also Jia et al., 2019).

Figure 16 displays the distribution of the HCM events as functions of $R_{SS}$ and solar wind dynamic pressure, which is adapted from Slavin et al. (2019). Each square represents either an HCM event or a disappearing dayside magnetosphere (DDM) event, with the scale of the square proportional to the dimensionless reconnection rate of the dayside magnetopause. The DDM events in this figure are discussed in the following. There are two curves in Figure 16, which represent the locations of $R_{SS}$ of the classic Chapman-Ferraro sixth root of the solar wind dynamic pressure (on the lower) and considering the effect of induction of Mercury's interior (Glassmeier et al., 2007) (on the upper). Neither curve includes the effects of reconnection erosion. In Figure 16, HCM events with lower dimensionless reconnection rates are closer to the upper curve including induction effects, while HCM events with higher dimensionless reconnection rates are closer to the lower curve without considering the induction effect.

Moreover, Jia et al. (2019) employ the global MHD model of Mercury's magnetosphere, as introduced in Figure 7. The green stars in Figure 16 represent the runs they performed that include the induction effect from the core, but the IMF is set to zero, so the model does not consider reconnection erosion on the dayside magnetopause. The green stars follow the upper curve in Figure 16, which is consistent with the induction effect theory. In the runs represented by red stars, both the induction effect and reconnection erosion effect, i.e., large magnetic shear across the dayside magnetopause, are considered. The red stars follow the lower curve, which is in good agreement with MESSENGER observations corresponding to a large dimensionless reconnection rate.

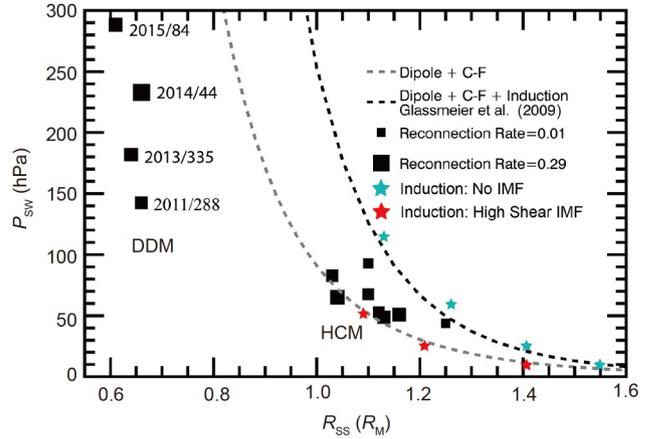

**Figure 16** The distribution of eight highly compressed magnetosphere (HCM) events and four disappearing dayside magnetosphere (DDM) events as functions of the subsolar magnetopause distance ($R_{SS}$) and the solar wind dynamic pressure ($P_{sw}$). Each event is represented by a square with the size indicating the dimensionless reconnection rate. The stars represent the locations of $R_{SS}$ from simulations (Jia et al., 2019). The green stars correspond to simulations that include induction effects but no IMF, and the red stars correspond to simulations that include induction effects and high-shear IMF. The lower dashed line is the curve of Chapman-Ferraro (C-F)'s sixth root of $P_{sw}$ with the mean subsolar distance determined by Winslow et al. (2013). The upper dashed line is a theoretical model that includes the effects of induction in Mercury's interior (Glassmeier et al., 2007). Note the $R_{SS}$ for the DDM and HCM events are not directly measured by MESSENGER, but are inferred from MESSENGER's observations by fitting to the empirical equations obtained by Winslow et al. (2013). This figure is adapted from Slavin et al. (2019) with the green and red stars from Jia et al. (2019).

Under extreme solar events, researchers have also analyzed an event in which MESSENGER did not encounter the dayside magnetosphere until the tailward of the northern cusp during its dayside crossing (Raines et al., 2015; Zhong et al., 2015a; Slavin et al., 2018, 2019; Winslow et al., 2020). Slavin et al. (2019) termed this kind of event a DDM event. Figure 17 shows one of the DDM events, which occurred on 15 October 2011. In this event, MESSENGER had a periapsis altitude of ~300 km on the dayside. In the magnetosheath, MESSENGER constantly observed magnetosheath proton flux, the southward IMF, and numerous FTEs. The FTEs are evidence of magnetic reconnection on the dayside magnetosphere, which suggests strong reconnection erosion in this event. Furthermore, the bow shock was located at an unusually low altitude, which strongly suggested that a large portion of the planet on the sunlit hemisphere was directly exposed to the solar wind. If this were the case, then the solar wind sputtering would be occurring at a maximum rate and would significantly influence the exosphere.

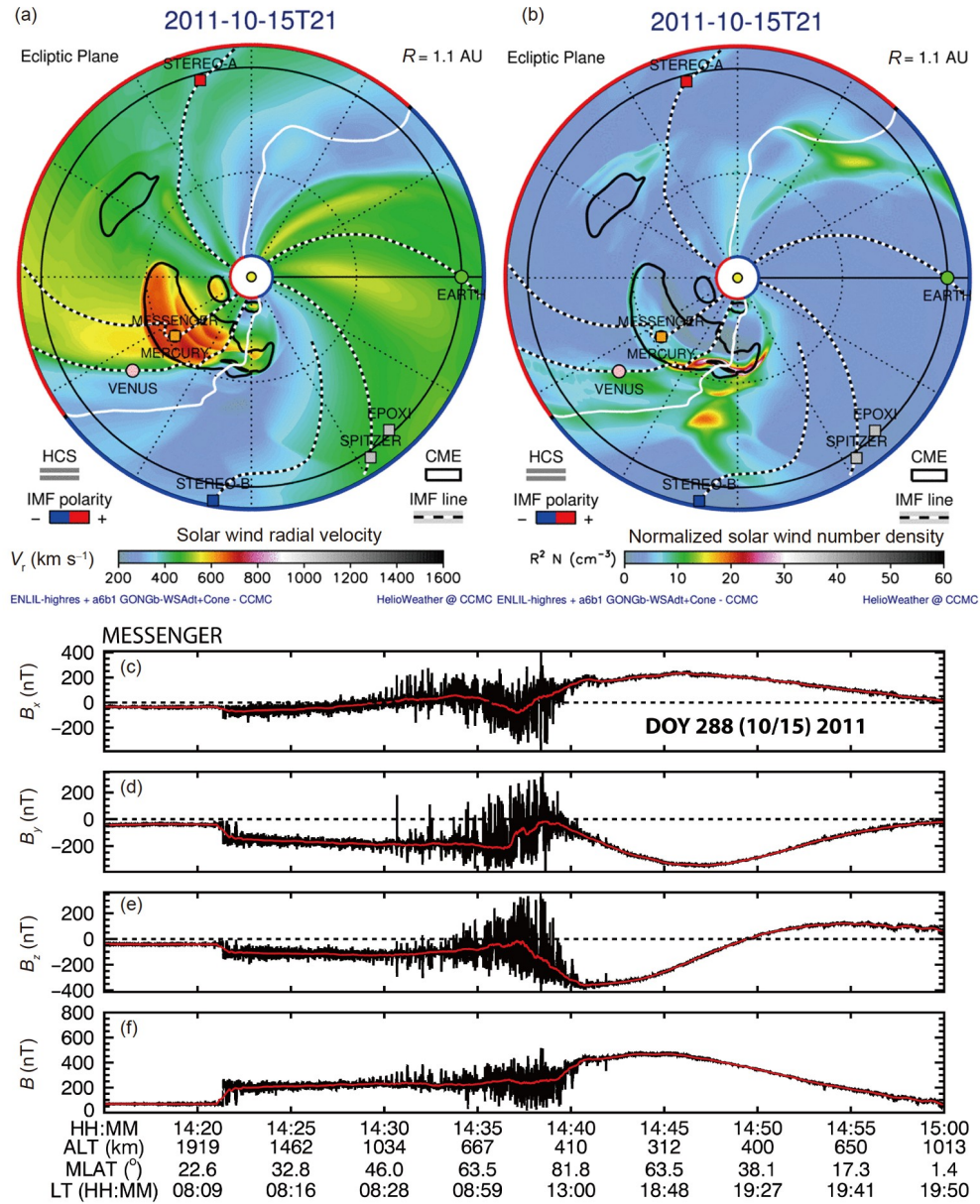

**Figure 17**  An example of a disappearing dayside magnetosphere (DDM) event. (a) Solar wind radial speed ($V_r$) and (b) the normalized solar wind number density (nr$^2$) as modeled from ENLIL-WSA+Cone (Tóth and Odstrčil, 1996; Odstrcil et al., 2004) from the Sun to 1.1 AU on 15 October 2011 at 21:00 UTC. The locations of Earth, Venus, and Mercury are indicated by different colored dots. Coronal mass ejections are enclosed by solid black lines. (c) $B_x$, (d) $B_y$, (e) $B_z$, (f) magnetic field intensity. Black lines are magnetic field measurements at native instrument resolution (20 Hz); the red lines are smoothed magnetic field measurements. This figure is adapted from Slavin et al. (2019).

In Figure 16, four DDM events are included in the left upper corner. In these DDM events, the smallest solar wind dynamic pressure was ~140 nPa. This value is much smaller than the value predicted by the upper curve, including the induction, which suggests that the reconnection erosion plays an important role in moving the magnetopause planetward. Note that the Chapman-Ferraro current system on the dayside would disappear along with the absorption of the solar wind particles by the surface (see also, Slavin et al., 2019).

### 3.3  The nightside magnetosphere response

The previous two sections introduce the progress of the dayside magnetosphere response to extreme solar events. This section will focus on the response of the nightside magnetosphere. Figure 18 shows an overview of the plasma and magnetic field measurements of Mercury's nightside magnetosphere during the ICME event on 23 November 2011. Exner et al. (2018) modeled the large-scale magnetic fields of this event and Sun et al. (2020a) have analyzed the

large- and small-scale plasma and magnetic features. First, both studies found that the magnetic field intensity in the lobes was quite high (~100 nT, Figure 18g), which is almost double the average intensity of the lobe field. The open magnetic flux in the lobe took more than half (~58%) of Mercury's available magnetic flux. This open magnetic flux is much larger than the maximum open lobe magnetic flux observed during Mercury's Dungey cycle (~42%). Second, Sun et al. (2020a) obtained a cross-magnetosphere potential drop of ~45 kV, which is triple the average value in Mercury's magnetosphere. The cross-magnetosphere potential drop was derived from the energy dispersion of the plasma mantle in Figure 18a. Whether the cross-magnetosphere potential drop can be saturated in Mercury's magnetosphere similar to that in Earth's magnetosphere is still an open question, which we will address in the next section.

Third, no magnetic flux loading-unloading events were observed in the tail lobes during a period of at least ten Mercury's Dungey cycles (~40 min, Figure 18g). This suggests that the amount of magnetic flux into and out of the magnetotail was similar, and Mercury's magnetosphere was under quasi-steady convection, which is similar to the steady magnetospheric convection (SMC) in Earth's magnetosphere. However, the SMC events require steady solar wind with average intensities of the velocity and IMF intensity (see O'Brien et al., 2002), which is significantly different from the extreme solar condition of this event. Slavin et al. (2012a) reported a quasi-steady convection event in Mercury's magnetosphere which occurred during average solar wind conditions. Together with the event in Figure 18, it suggests the quasi-steady convection events could occur both during average and extreme solar wind conditions in Mercury's magnetosphere.

Fourth, magnetic reconnection in the plasma sheet had a strong guide field during this event ($B_{guide}/B_{Lobe}$~0.29), which produced a distorted Hall magnetic field pattern. The dimensionless reconnection rate was estimated to be ~0.093, and the dawn-dusk extent of the X-line took ~20% of the tail width. Magnetic reconnection in the plasma sheet produced many flux ropes at a high frequency, which appeared as quasi-periodic flux rope groups (Figure 18), and see Figures 4 and 5 in Sun et al. (2020a). The flux rope group has a period of ~70 s, and, in each flux rope group, larger-scale flux ropes at the leading edge of the group were followed by the smaller-scale flux ropes. Zhong et al. (2020a) have investigated these flux rope groups. In their study, they proposed that the larger-scale flux ropes were generated by the interaction and coalescence of multiple smaller-scale flux ropes.

Sun et al. (2020a) have also investigated the response of the nightside magnetosphere to a high-speed stream (HSS) in the solar wind. The responses shared many similarities with the above ICME event. For example, the open magnetic field took nearly half (~44%) of Mercury's available magnetic flux. The magnetic flux into and out of the magnetotail was similar, indicating that Mercury's magnetosphere was under quasi-steady convection. However, in the HSS event, MESSENGER crossed the plasma sheet in the distance much closer to the planet, which was planetward of the near Mercury neutral line. High-frequent and large numbers of dipolarization fronts were observed, which also implied that intense magnetic reconnection occurred in the plasma sheet. Moreover, the occurrence rate of the dipolarization fronts is around two orders of magnitude higher than the occurrence rate averaged over all plasma sheet observations (Sun et al., 2016).

## 3.4  Polar cap potential saturations

Studies in Earth's magnetosphere have revealed that the CPCP, also called the cross-magnetosphere potential drop, was linearly related to the solar wind convection electric field, but saturated, i.e., reached an upper limit, during the periods of extreme solar events, i.e., ICMEs and Co-rotating Interaction Regions (CIRs), when the convection electric field became much larger ($\gtrsim$5 mV m$^{-1}$) (Reiff et al., 1981; Wygant et al., 1983). There are several theories on how the potential drop in the magnetosphere was saturated. Here, we employ a theory proposed by Kivelson and Ridley (2008) (hereinafter referred to as KR2008) based on the solar wind-magnetosphere interactions. In the KR2008 model, the potential drop was contributed by solar wind electric potential but controlled by the conductance difference between the solar wind and the polar ionosphere, which would be the planet's regolith at Mercury. In the KR2008, the potential drop is proportional to $2\Sigma_A/(\Sigma_A+\Sigma_P)$. The $\Sigma_A$ is Alfvén conductance of the solar wind, which equals $1/\mu_0 v_A$, and $v_A$ is Alfvén speed in upstream solar wind, and $\mu_0$ is the magnetic permeability in free space. The $\Sigma_P$ is the Pedersen conductance of the ionosphere. When the $\Sigma_A$ is much larger than the $\Sigma_P$, the Alfvén waves incident from the solar wind are not able to be reflected from the polar ionosphere. Therefore, the potential drop is linearly related to the solar wind electric field. However, when the $\Sigma_A$ is comparable to the $\Sigma_P$, the Alfvén wave is largely reflected, resulting in the saturation of the potential drop.

In Earth's polar ionosphere, $\Sigma_P$ is ~5 to 10 S. The solar wind $\Sigma_A$ at 1 AU is much larger than the $\Sigma_P$ in Earth's ionosphere under average conditions ($\gtrsim$50). As a result, the Alfvén waves can hardly be reflected, and the potential drop is linearly related to the solar wind convection electric field. However, when solar wind Alfvén Mach number is low (<5), i.e., during the periods of ICMEs and CIRs, the $\Sigma_A$ at 1 AU becomes small and is comparable to the $\Sigma_P$. The Alfvén waves can be reflected and cause potential saturation. Mercury corresponds to a low solar wind Alfvén Mach number (<5) and, therefore, a smaller $\Sigma_A$. However, $\Sigma_P$ at Mercury is also small (~1 S). Therefore, it would be interesting to investigate the CPCP saturation of Mercury's magnetosphere.

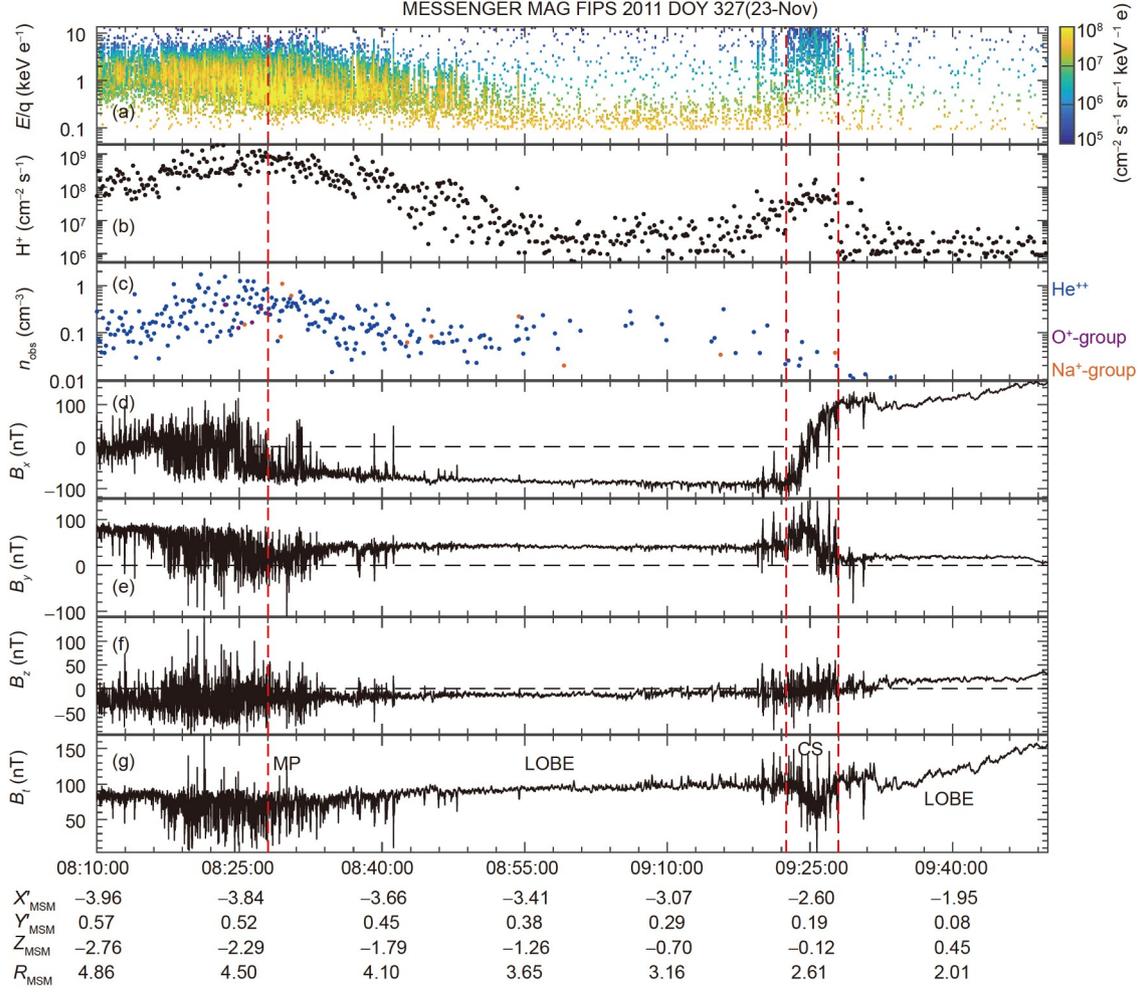

**Figure 18** Overview of MESSENGER plasma and magnetic field measurements during the spacecraft's nightside magnetospheric crossing on 23 November 2011. (a) Differential proton particle flux; (b) integrated proton particle flux; (c) the observed densities of He$^{++}$ (in blue), O$^+$-group (purple), and Na$^+$-group (gold); (d) $B_x$; (e) $B_y$; and (g) magnetic field intensity ($B_t$). The vertical dashed lines indicate, from left to right, the average magnetopause location, the southern plasma sheet boundary, and the northern plasma sheet boundary, respectively. The magnetopause (MP), lobe (LOBE), and current sheet (CS) are labeled. This figure is adapted from Sun et al. (2020a).

The calculation of the potential drop in KR2008 (eq. (13) in Kivelson and Ridley (2008)) is

$$\text{Potential} = 10^{-7}u_x^2 \\ + 0.1\pi R_{sd}B_{sw,yz}u_x\left(\frac{\theta}{2}\right)\frac{2\Sigma_A}{\Sigma_A + \Sigma_P}, \qquad (8)$$

where the $u_x$ is $x$ component of solar wind speed, the $R_{sd}$ is the subsolar standoff distance of the magnetopause, the $B_{sw,yz}$ is IMF component in the $y$-$z$ plane, the $\theta$ is the IMF clock angle. The $10^{-7}u_x^2$ is a viscous interaction term (Boyle et al., 1997), $0.1\pi R_{sd}$ refers to the length of the dayside magnetopause reconnection line, $B_{sw,yz}u_x(\theta/2)$ corresponds to the reconnection electric field (Sonnerup, 1974).

Figure 19 shows the potential drop estimated from the

KR2008, in which the subsolar standoff distance of Mercury's magnetopause ($R_{sd}$) was ~1.13 $R_{M}$, and the solar wind velocity was ~450 km s$^{-1}$. In Figure 19, the black line represents the magnitude of the potential drop from the KR2008 when $\Sigma_P$=0, which indicates that the incident Alfvén waves from the solar wind would not be reflected at all. The black line represents a linear relationship between the potential drop and the solar wind convection electric field, though the $x$-axis is on a logarithmic scale. The blue line represents the situation when $\Sigma_P$=1 S, i.e., Mercury. The saturation of the potential drop would be very small (<1%) during the average IMF condition in Mercury's magnetosphere. Here the saturation refers to the differences between the blue and the black curves. Even under the impact of the ICMEs where the IMF intensity reaches 100 nT, the saturation is still small (<10%). For comparison, the red line in Figure 19 represents the potential drop when $\Sigma_P$ is 10 S, which is much larger than the conductivity at Mercury. The potential drop is largely saturated and starts to reach a constant value when the solar wind magnetic intensity is ~30 nT.

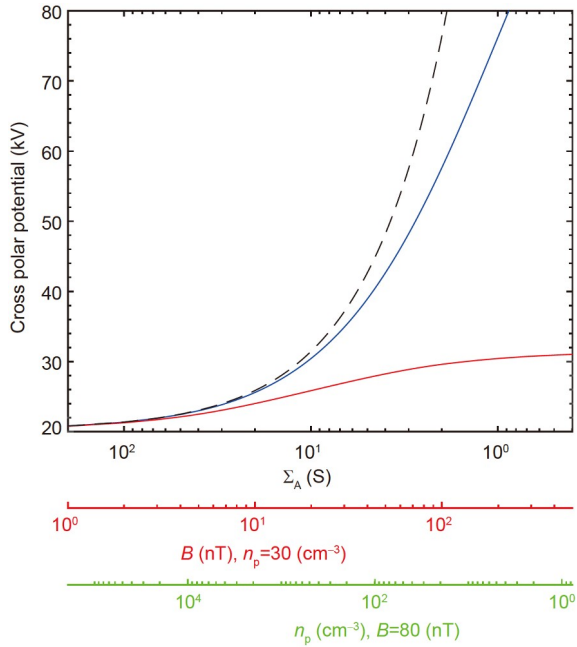

### 3.5  Open questions regarding magnetosphere under extreme solar event

Many important questions regarding Mercury's magnetosphere under extreme solar event remain to be answered. In the previous chapter, the intense solar wind sputtering caused by FTE showers is discussed, which can induce short-term variations in Mercury's exosphere. We note that solar wind sputtering is expected to be much more intense under the impact of the ICMEs. For example, Slavin et al. (2014) showed a cusp with multiple short and deep magnetic depressions. The variations of the exosphere were observed by ground-based imaging (Orsini et al., 2018). In particular, solar wind sputtering is substantially enhanced in the DDM events (Slavin et al., 2019; Winslow et al., 2020). During the intervals of DDM events, most of the sunlit hemisphere of Mercury is exposed to the solar wind directly. This makes the planetary environment analogous to that at the Moon when it is outside of Earth's magnetotail, as well as other airless bodies, such as Ceres, Io, and the asteroids, in the Solar System.

Extreme solar wind sputtering can dramatically enhance the density of species of the exosphere in a short period, which might provide a good opportunity to detect the minor species in the exosphere, and might also help to investigate the surface elements. On the other hand, the prevalence of energetic components of the exospheric species may be due to micro-meteoroid impact vaporizations on the planet's surface (e.g., Killen et al., 1999; Domingue et al., 2007; Wurz et al., 2010). Solar wind sputtering has been observed capable of launching atoms and some ions from the surface of the Mercury with large energies, which would allow many of them to escape (e.g., Schmidt et al., 2012). Many simulation works have proposed that solar wind sputtering can be a source for the energetic component of the exosphere (Potter et al., 2006; Mura et al., 2007; Orsini et al., 2018). DDM intervals would be a favorable period to make such investigations.

Also, the study of Earth's magnetosphere has shown that solar wind-magnetosphere coupling produces several response modes of the magnetosphere, which includes the substorms, SMC, and sawtooth oscillations. In the SMC events, magnetic reconnection continuously occurs in the magnetotail, and the magnetic flux transferred into and out of the magnetotail is balanced. In Earth's magnetosphere, SMC events are driven by the average and steady solar wind, which usually continues more than several hours corresponding to the duration of several substorms (O'Brien et al., 2002; Partamies et al., 2009). Sawtooth oscillations consist of quasi-periodic substorms. Magnetic flux in the lobe exhibits quasi-periodic loading-unloading, and energetic particle injections occur quasi-periodically on geosynchronous orbit. Sawtooth oscillations are detected during more intense solar wind conditions than substorms, like those during ICMEs (Huang et al., 2003; Henderson et al., 2006).

Under the impact of an ICME and an HSS, MESSENGER did not observe the quasi-periodic flux loading-unloading in Mercury's magnetotail but did observe a steady magnetic field in the lobe persisting more than 10 Mercury's Dungey cycle, which seems to indicate steady convection. Several studies in Earth's magnetosphere have proposed that the ionosphere, i.e., the ionospheric outflows, could modulate the reconnection rate and tail reconnection. As a result, tail reconnection cannot reach a steady state under extreme solar events (see Brambles et al., 2011; Zhang et al., 2020). Although planetary ions are frequently observed in Mercury's magnetotail, the significance of these planetary ions in Mercury's magnetospheric dynamics is still an open question.

Furthermore, we still do not know whether a ring current can be formed in Mercury's magnetosphere and whether there are magnetic storms when Mercury's magnetosphere is influenced by extreme solar events. There are many questions regarding the formation of a ring current:

(1) Can particles accomplish a complete drift around the planet? Are there significant particles accomplishing the

drift? In Mercury's magnetosphere, particles can be easily lost into the planet's surface because of a large loss cone, and particles can encounter the dayside magnetopause and leak into the magnetosheath easily in a process called magnetopause shadowing. The variations of the solar wind dynamic pressure can magnify the effects of magnetopause shadowing in Mercury's magnetosphere. Despite all of these effects, MESSENGER observations have shown possible quasi-trapped, low energy ion populations (Schriver et al., 2011) and possible drifting echoes of energetic electrons associated with a dipolarization (Baker et al., 2016). These echoes are evidence of energetic electrons completing drifts around the planet. Statistical investigations have shown that energetic electrons can appear throughout the dayside magnetosphere, which suggests that energetic electrons can accomplish a complete drift around the planet (Lawrence et al., 2015; Baker et al., 2016; Ho et al., 2016; Dewey et al., 2017). Simulation studies have suggested that both ions and electrons can accomplish a complete orbit around the planet (Delcourt et al., 2007; Trávníček et al., 2010; Walsh et al., 2013; Yagi et al., 2017). Walsh et al. (2013) further noted that the drift path of particles could split into two (northern and southern) trajectories due to the local maximum of the magnetic field near the magnetic equator in a process called the Shabansky orbit (Shabansky, 1971). However, we emphasize that several questions relating to the drift around the planet remain unanswered. For example, there is no observational evidence so far of the Shabansky orbit, and we still do not know whether ions can accomplish a complete orbit around the planet.

(2) Are there sufficient energetic particles for the formation of a significant ring current? How efficient is the energization of particles during the active magnetospheric periods? Substorm dipolarizations can effectively energize particles, which is one of the most important processes during the active magnetospheric period. MESSENGER observations of proton energizations can be found in Sun et al. (2015b, 2017b, 2018) and Dewey et al. (2017), and electron energization can be found in Dewey et al. (2017). Simulation works can be found in Ip (1987) and Delcourt et al. (2007). In the periods when Mercury's magnetosphere was under extreme solar events, flux ropes and dipolarization fronts also appear in extreme frequencies (Sun et al., 2020b; Zhong et al., 2020a), which might provide plenty of energetic particles for the formation of a ring current. However, as pointed out by Dewey et al. (2020), if the dipolarization fronts directly impact the planet's surface without significant braking, then there might not be large amounts of energetic particles generated. The next two chapters contain more discussions on particle energization and dipolarizations in Mercury's magnetosphere.

(3) As discussed by Ip (1987) and Schriver et al. (2011), and others, due to the low altitude of Mercury's dayside magnetopause, only ions with energy lower than super-thermal ($\sim$10 keV) can be possibly quasi-trapped. Slavin et al. (2014, 2019), Jia et al. (2019), and Winslow et al. (2020) have shown that the subsolar magnetopause standoff distance is much closer to the planet's surface (only a few hundred kilometers above the planet's surface). Occasionally, the dayside magnetosphere completely disappears, under which conditions, all of the orbiting particles would likely leak out of the magnetosphere and enter the magnetosheath. Therefore, the strong compression that occurs during extreme solar events is an important factor and should be considered in determining the formation of a ring current at Mercury.

## 4.    Dawn-dusk properties of the plasma sheet

The post-MESSENGER era has highlighted the extent and mystery of dawn-dusk asymmetries in Mercury's magnetotail. Asymmetries have been observed in planetary magnetotails throughout the Solar System and are inferred from differences between the pre-midnight (dusk or $+Y$) and post-midnight (dawn or $-Y$) magnetotail regions. In the absence of multipoint measurements, these asymmetries are usually characterized by statistical studies of local magnetic fields and plasma measurements that sacrifice temporal resolution for spatial coverage. As such, investigation of these asymmetries at Mercury has flourished in the post-MESSENGER mission era as studies can leverage the most statistically complete set of *in situ* observations. These studies have revealed dawn-dusk asymmetries in plasma and magnetic field properties, large-scale structure, and reconnection dynamics of Mercury's magnetotail, as summarized in Table 3. While some of these asymmetries are similar to those observed at Earth, others are curiously different. These descriptions of Mercury's magnetotail are statistical in nature, however, so they are subject to several considerations.

### 4.1    Magnetotail ion distributions

Protons, the most abundant ion in Mercury's magnetotail, exhibit strong cross-tail asymmetries in their density, temperature, and spectra. Zhao et al. (2020) present the most comprehensive analysis of proton asymmetries to date, shown in Figure 20. Within each spatial bin, proton 1D velocity distribution functions are averaged together and fit with Kappa and Gaussian functions (e.g., Sun et al., 2017b). Proton density and pressure are enhanced post-midnight compared to pre-midnight, while proton temperature displays no apparent cross-tail asymmetry. The proton spectral index ($\kappa$) is greater in the post-midnight magnetotail, indicating a steeper distribution function (i.e., less super-thermal flux) than the pre-midnight magnetotail. Flux above

**Table 3**  Dawn-dusk asymmetries in the magnetotail of Mercury and Earth

| Process/property | Asymmetry preference (higher side) | |
|---|---|---|
| | Earth | Mercury |
| Occurrence of reconnection | Dusk[a] | Dawn[b] |
| Hot component (warm flux) | Dusk[c] | Dawn[d] |
| Cold component (thermal flux) | Dawn[e] | Dawn[f] |
| Proton density | Dawn[g] | Dawn[h] |
| Northward (z) component of magnetic field | Dawn[i] | Dawn[j] |
| Kappa value | Dusk[k] | Dusk[l] |

a) Slavin et al. (2005); Imber et al. (2011); Nagai et al. (2013). b) Sun et al. (2016); Smith et al. (2017b). c) Wang et al. (2006); Keesee et al. (2011). d), f) and l) Zhao et al. (2020). e) and g) Wing et al. (2005). h) Korth et al. (2014); Zhao et al. (2020). i) Fairfield (1986). j) Poh et al. (2017a). k) Espinoza et al. (2018)

the average proton plasma sheet energy (i.e., thermal flux) is higher on the dawnside, while flux below the average energy (i.e., warm flux) is organized by the magnetotail flanks.

Additional studies generally support these proton asymmetry observations. Sun et al. (2017b) examined proton suprathermal flux (>4.7 keV) and proton temperature as a function of location in the plasma sheet for different levels of magnetospheric activity, inferred from the thickness of the current sheet. Neither suprathermal flux nor temperature display dawn-dusk asymmetries during magnetospherically quiet intervals (i.e., thick current sheets). During active intervals (i.e., thin current sheets), both suprathermal flux and temperature increase preferentially post-midnight, resulting in cross-tail asymmetries. Although the energy thresholds are different between Zhao et al. (2020) and Sun et al. (2017b), and Zhao et al. (2020) do not differentiate the plasma sheet crossings by magnetic activity, both studies agree on a dawn-dusk asymmetry in the absolute proton flux at above-average energies. However, these studies disagree on cross-tail proton temperature trends. Finally, distributions of mean proton flux observed by FIPS and plasma pressure inferred from magnetic field deficits (Korth et al., 2014) display similar enhancements of dawnside proton flux and pressure.

Several mechanisms have been proposed to account for the asymmetry in proton properties across Mercury's magnetotail. Foremost, magnetic reconnection has been discussed with regards to proton temperature and super/suprathermal flux distributions. Magnetic reconnection in Mercury's magnetotail occurs more frequently in the post-midnight magnetotail than the pre-midnight magnetotail (see Section 4.3). Mercury's magnetotail reconnection and its byproducts (e.g., dipolarizations) can heat and energize protons (Dewey et al., 2017; Sun et al., 2018), so the dawnside enhancements in proton temperature and superthermal flux have been usually attributed to reconnection (Sun et al., 2017b; Zhao et al., 2020). Reconnection has also been suggested to be responsible for the dawnside enhancement of proton density since it opens the plasma sheet to plasma convected from the dayside (Zhao et al., 2020). However, observations of the proton plasma mantle have identified a clear enhancement in plasma content in the pre-midnight magnetotail (Jasinski et al., 2017). Although the plasma mantle is not observed in every orbit, plasma mantle observations analyzed by Jasinski et al. (2017) are expected to be representative of typical magnetospheric conditions (DiBraccio et al., 2015b; Dewey et al., 2018). It is currently unknown how efficient reconnection in Mercury's magnetotail is at entraining mantle plasma on newly-closed plasma sheet field lines. If this efficiency is much greater than in non-reconnecting regions of the magnetotail, then it may be possible for reconnection to account for the dawnside enhancement in proton density. In addition to magnetic reconnection, low latitude dynamics near the flanks are also discussed regarding cross-tail proton asymmetries. Velocity shear between the magnetosheath and the magnetosphere can result in Kelvin-Helmholtz vorticities that transport magnetosheath plasma into the magnetosphere, particularly contributing to the distribution of warm plasma flux. Kelvin-Helmholtz waves are more frequently observed in Mercury's duskside magnetotail flank (see Slavin et al. (2018) for a recent review). Enhancement in warm flux on the dawnside flank in Figure 20f may be due to averaging over the downside's thicker, denser low-latitude boundary layer (Liljeblad et al., 2015).

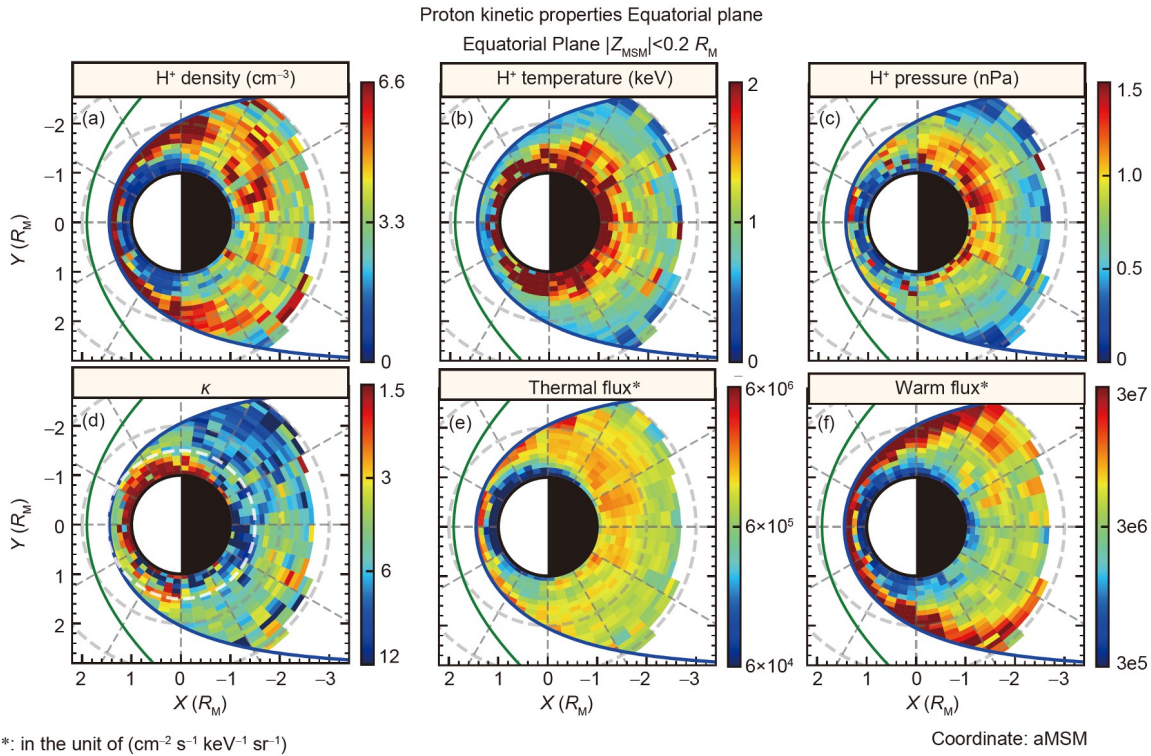

**Figure 20**  Proton properties near the magnetic equator from FIPS measurements: (a) density, (b) temperature, (c) pressure, (d) spectral index $\kappa$, (e) flux <0.83 keV, and (f) flux >0.83 keV. Adapted from Zhao et al. (2020).

Generally, these cross-tail trends in proton properties are opposite to those observed in Earth's magnetotail but may be consistent given the differences in magnetotail reconnection. Cold, hot, and superthermal ion components in Earth's plasma sheet each possess dawn-dusk asymmetries that are generally dependent on the interplanetary magnetic field (see Walsh et al., 2014). The hot component ions, for example, have higher temperatures towards dusk during magnetospherically quiet intervals, attributed to gradient-curvature drift (Spence and Kivelson, 1993; Keesee et al., 2011). During northward IMF, Wang et al. (2006) found greater total ion density and equal pressures at dawn and dusk. With increasing geomagnetic activity, the density asymmetry weakens, resulting in greater ion pressures at dusk. Fluxes of hot and superthermal ions are greater at dusk, with the cross-tail assymmetry increasing for increasing magnetic activity and higher energies. These terrestrial cross-tail asymmetries were derived from Geotail measurements and are expected to represent primarily protons. Although these trends in proton density, temperature, and pressure are generally opposite to that at Mercury, the source of these trends is consistent if magnetic reconnection is the primary mechanism. Magnetotail reconnection at Earth occurs more frequently pre-midnight and increases in frequency with increasing geomagnetic activity (e.g., Genestreti et al., 2014). The trends in enhanced duskside proton pressure, flux, and temperature during active intervals at Earth parallel the enhanced dawnside proton pressure, flux, and temperature observed at Mercury.

Finally, minor ions in Mercury's magnetotail also exhibit strong dawn-dusk asymmetries. In particular, heavy ions of planetary origin are more abundant in the pre-midnight magnetotail (Raines et al., 2013). The dominant planetary ions in the pre-midnight magnetotail are Na$^+$-group ions (mass-per-charge 21–30 amu e$^{-1}$), which have number densities typically ~10% that of H$^+$ (Gershman et al., 2014). While the number density of these ions may be small, their large mass and slightly greater temperatures result in them contributing ~50% to the mass density and ~15% to the thermal plasma pressure. The dawn-dusk asymmetry of planetary ion abundance is opposite to that of protons and is a result of non-adiabatic effects. Planetary ions have gyroradii much larger than protons that result in less adiabatic behavior in the magnetotail and subsequently stronger centrifugal effects that transport them duskward (Delcourt, 2013).

## 4.2 Magnetotail current sheet structure

Dawn-dusk asymmetries in ion distributions combined with asymmetries in the magnetic field produce differences between the pre-midnight and post-midnight magnetotail structure at Mercury. Poh et al. (2017a, 2017b) and Rong et al. (2018) identified similar dawn-dusk differences in the magnetic field using independent methods. Poh et al. (2017a) analyzed 319 central plasma sheet crossings and fit them with a 1D Harris current sheet. The magnetic field and Harris current sheet properties of these crossings revealed a substantial, significant dawn-dusk asymmetry in the plasma sheet magnetic field strength: $B_z$ is enhanced at dawn. A linear fit to $B_z(Y)$ for $-1.7<X<-1.4\ R_M$ yielded a slope of $-2.77\pm0.37$ nT/$R_M$ and a local midnight value of ~10 nT. Dividing the two for a more dimensionless expression gives $-0.28\pm0.04\ R_M^{-1}$. This dawn-dusk asymmetry becomes less substantial and less significant with downtail distance: the dimensionless trend is $-0.09\pm0.07\ R_M^{-1}$ in the range $-2.3<X<-2.0\ R_M$. In addition to the stronger dawnside magnetic field, the Harris current sheet fits the revealed additional cross-tail asymmetries. While the current sheet is thinnest at local midnight, it is thicker on the dawnside than on the duskside, consistent with the magnetic field strength distribution. The cross-tail current density peaks near midnight but is greater at dusk than dawn. Finally, the proton plasma $\beta$ shows an approximately linear trend across the magnetotail and is systematically greater at dusk than dawn.

Rong et al. (2018) examined all MAG observations within the magnetotail and found similar cross-tail differences in Mercury's magnetotail structure. This study corroborated the dawn-dusk asymmetry in $B_z$ and its dependence on downtail distance. From spatially binned magnetic field observations, the study found that maximum current density peaks near midnight but is greater at dusk than at dawn, while the minimum radius of curvature is near midnight and is generally smaller at dusk than at dawn. These two asymmetries are stronger closer to the planet ($-2.0 < X < -1.5$ $R_M$) than further downtail ($-2.5 < X < -2.0$ $R_M$), similar to the asymmetry in $B_z$. Finally, this study identified that magnetic field lines flare away from local midnight more at dawn than dusk.

While Poh et al. (2017a) and Rong et al. (2018) find similar dawn-dusk differences in magnetic field structure, Poh et al. (2017b) and Rong et al. (2018) disagree on the planetward edge of the current sheet. Poh et al. (2017b) fit a semi-infinite current sheet slab model to dipole-subtracted magnetic field measurements to estimate that the most likely location of the current sheet inner edge is $X \approx -1.22$ $R_M$. Rong et al. (2018) calculate the statistical duskward current density in the $X$-$Z$ plane from dipole-subtracted field measurements. The study identifies the inner edge to be near $X = -1.5$ $R_M$ planetward of which the current density begins decreasing. A recent analysis of fast flow braking places the inner edge of the current sheet between $-1.3 < X < -1.1$ $R_M$, coincident where the magnetotail proton plasma $\beta$ reaches unity (Dewey et al., 2020). This boundary is not parallel to the terminator but rather bows about the planetary dipole, as shown in Figure 21. Although both proton pressure and magnetic field pressure are stronger at dawn than at dusk, the strengths of these asymmetries are not equal; $\beta$ exhibits a dawn-dusk asymmetry. Proton plasma $\beta$ is greater at dusk than at dawn (Poh et al., 2017b), causing the inner current sheet boundary to be shifted systematically closer to the terminator at dawn (Dewey et al., 2020) and indicating that the dawn-dusk asymmetry in magnetic pressure is stronger than the asymmetry in proton pressure.

The origins of Mercury's dawn-dusk magnetic field asymmetries remain unclear. Earth's magnetotail also possesses asymmetries in current sheet thickness and current density, but unlike trends in protons (Section 4.1) and magnetic reconnection (Section 4.3), they are in the same direction as Mercury's. Similar to Mercury, the current sheet thickness and current density are greater at dawn in Earth's magnetotail than at dusk (e.g., Artemyev et al., 2011; Rong et al., 2011). Thinner current sheets are more likely to reconnect, so this dawn-dusk asymmetry at Earth is consistent with the magnetotail's reconnection asymmetry (e.g., Imber et al., 2011). At Mercury, the thicker dawnside is associated with more frequency reconnection. Poh et al. (2017a) propose that the thinner duskside current sheet may be a result of mass loading from heavy planetary ions more abundant there. The presence of heavy ions at dusk may also produce a weaker local magnetic field via diamagnetism (Rong et al., 2018). Alternatively, the thicker dawnside current sheet has been suggested to be a statistical result from averaging over magnetic-field-enhanced products from reconnection (Poh et al., 2017a). Finally, numerical simulations (e.g., Chen Y et al., 2019) that reproduce the dawn-dusk asymmetry in current sheet thickness suggest it be related to the Hall effect and external driving conditions.

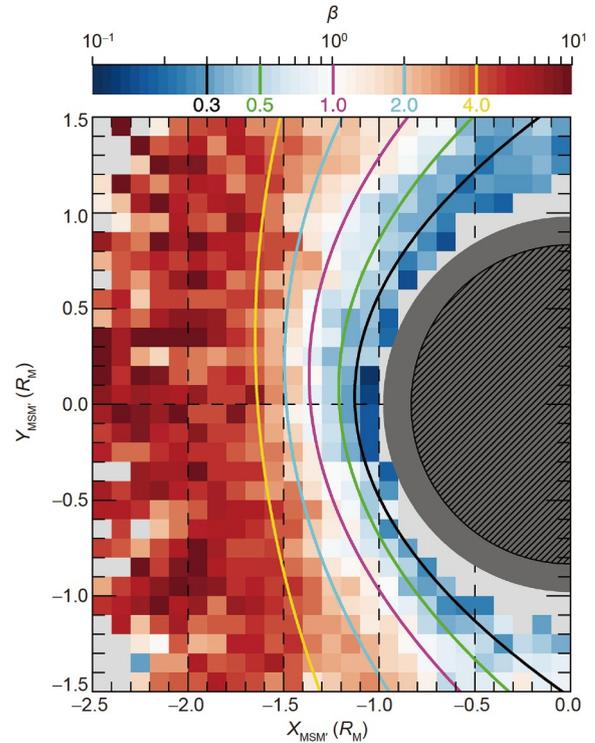

**Figure 21**    Proton plasma $\beta$ as a function of spatial location near Mercury's magnetic equator. Colored polynomials indicate specific $\beta$ contours. Adapted from Dewey et al. (2020).

### 4.3    Magnetotail reconnection dynamics

Magnetotail dynamics, particularly those related to magnetic reconnection, exhibit substantial dawn-dusk asymmetries. Mercury possesses terrestrial-like substorms (Sun et al., 2015b), signatures of which are observed more frequently at dawn than dusk. In the magnetotail lobes, more instances of magnetotail loading-unloading have been recorded post-midnight than pre-midnight (Imber and Slavin, 2017). Within the plasma sheet, dipolarizations (Sun et al., 2016; Dewey et al., 2020) and flux ropes (Sun et al., 2016; Smith et al., 2017b, 2018b; Zhao et al., 2019) are more abundant at dawn as measured either by event frequency or by the number of MESSENGER orbits that observed these events. As magnetotail reconnection produces these events (e.g., Slavin et al., 2012a; DiBraccio et al., 2015a; Dewey et al., 2020), their cross-tail asymmetry is indicative of the dawnside preference in Mercury's magnetotail reconnection (Sun et al., 2016; Smith et al., 2018b).

Furthermore, as dipolarizations and flux ropes interact with the surrounding plasma environment, they generate additional asymmetries in the magnetotail, including en-ergetic electron injections, fast sunward flows, magnetic pileup, and substorm current wedge formation. Dipolarizations, for example, are a dominant source of energetic electron acceleration within Mercury's magnetosphere (Dewey et al., 2017), resulting in energetic electron injections to be more frequent at dawn than at dusk (Baker et al., 2016), independent of electrons' eastward drift about the planet (Walsh et al., 2013; Nikoukar et al., 2018; Dong et al., 2019). A fraction of these energetic electrons precipitate to the planet, and in the absence of a thick atmosphere, collide with the planet's surface to produce X-ray fluorescence and space weathering (Starr et al., 2012). The asymmetry in dipolarizations (and subsequently in energetic electrons) results in a dawnside preference in this auroral-like emission (Lindsay et al., 2016), as shown in Figure 22. This asymmetry in energetic electrons and their precipitation has been replicated well by global simulations (Chen Y et al., 2019; Dong et al., 2019).

Dawn-dusk asymmetries associated with magnetic reconnection at Mercury are opposite to those observed at Earth. At Earth, magnetic reconnection occurs more frequently in the pre-midnight magnetotail as verified by different spacecraft and various direct and indirect signatures (e.g., Slavin et al., 2005; Imber et al., 2011; Liu et al., 2013; Gabrielse et al., 2014; Genestreti et al., 2014; Walsh et al., 2014). Recent local and global simulations suggest that differences in Mercury's and Earth's dawn-dusk magnetotail reconnection asymmetries are related to kinetic effects.

Global magnetohydrodynamic simulations with embedded particle-in-cell (PIC) regions (i.e., MHD-EPIC models) of Mercury's magnetosphere have reproduced many of the dawn-dusk asymmetries at Mercury, including that of magnetotail reconnection. Chen Y et al. (2019) performed MHD-EPIC simulations of Mercury with the PIC region covering the magnetotail. These numerical simulations reproduce: the thicker dawnside current sheet, the enhanced dawnside proton plasma density and pressure, dawnside preference for dipolarizations and energetic electron injections, and dawnside stronger magnetic field. The results suggest that the dawn-dusk reconnection asymmetry depends on the strength of external solar wind driving. During moderate conditions, reconnection is slightly preferred in the pre-midnight region as the dawnside current sheet is significantly thicker than the duskside current sheet. However, during strong solar wind driving, the dawnside current sheet thins to the same thickness as the duskside sheet, and reconnection shifts dawnward. Kinetic Hall effects result in electron and magnetic structures leaving the magnetotail reconnection site(s) to travel dawnward, increasing the apparent dawn-dusk asymmetry near the planet.

These global MHD-EPIC simulation results are similar to localized PIC simulations. Liu et al. (2019) demonstrated that a two-region structure develops within an active reconnection site in 3D PIC simulations of thin magnetotail current sheets. Within the ion diffusion region, electrons remain magnetized (Hall effect) and preferentially transport

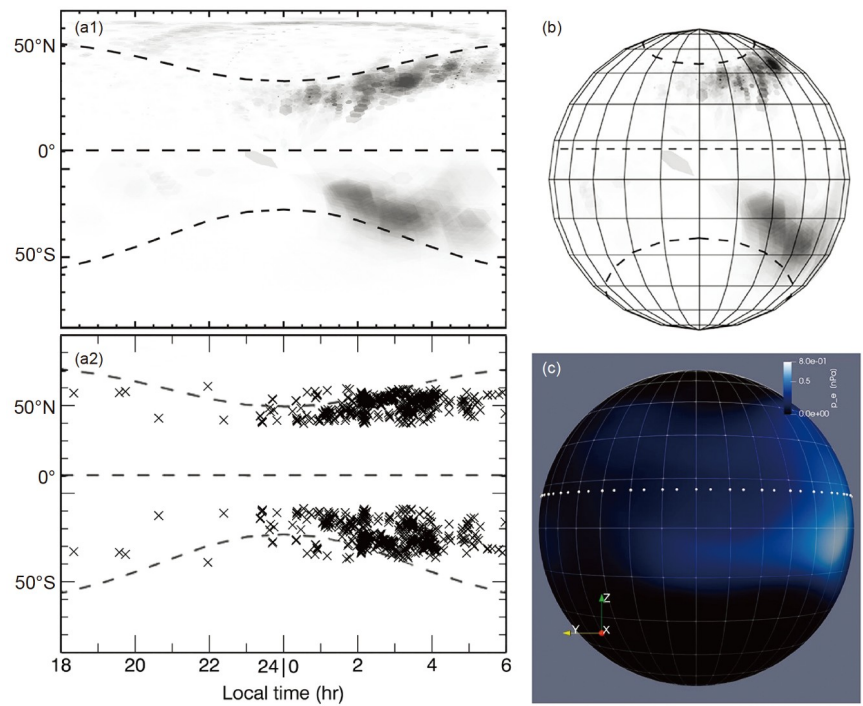

**Figure 22** (a1) Observed X-ray fluorescence of Mercury's nightside surface from precipitating energetic electrons from Lindsay et al. (2016). (a2) Predicted locations of fluorescence from magnetotail dipolarization events from Dewey et al. (2017). (b) Spherical projection of (a1). (c) Electron pressure at Mercury's surface simulated by Dong et al. (2019).

reconnected magnetic flux in the direction of the electron drift. The electron-drift side continues to thin while the ion-drift side experiences a relative suppression in the magnetic reconnection rate. The suppression region develops with a cross-tail with ~10 ion inertial lengths ($d_i$), implying a minimum cross-tail width for magnetic reconnection sites. Liu et al. (2019) apply these PIC simulation results to address the opposite dawn-dusk asymmetry in magnetotail reconnection at Earth and Mercury, shown in Figure 23. Earth's magnetotail is hundreds of $d_i$ wide compared to Mercury's of tens of $d_i$; Earth's magnetotail can accommodate more reconnection sites than Mercury. As both planetary magnetotails possess thinner duskside current sheets, reconnection initiates pre-midnight for both magnetotails. As reconnection proceeds, the suppression region develops on the local duskside of the reconnection site. In the Earth's relatively wide magnetotail, this local dawnward shift in reconnection does not affect the global location of reconnection. However, in Mercury's relatively narrow magnetotail, this local dawnward shift results in post-midnight reconnection. This hypothesis suggests that structures from magnetic reconnection in Mercury's magnetotail should be on the order of a few $d_i$. With plasma densities ~3–6 cm$^{-3}$ in Mercury's plasma sheet (see Figure 20), the ion inertial length is ~0.04–0.06 $R_M$. Dewey et al. (2020) measured the average width of dipolarizations in Mercury's magnetotail to be 0.30 $R_M$, implying the non-suppressed width of the reconnection site to be ~6–8 $d_i$ and is consistent with that predicted by Liu et al. (2019). Smith et al. (2018b) utilized Monte Carlo simulations to estimate that the typical reconnection site width is 2.16 $R_M \approx$40–60 $d_i$, approximately half the width of Mercury's magnetotail (Slavin et al., 2012a) and the full width of the thin current sheet (Poh et al., 2017a; Rong et al., 2018). The difference between these two estimates may originate from the Monte Carlo simulation assumption. In the Monte Carlo simulations, Smith et al. (2018b) assumed that there was at most a single-static reconnection site that persisted for the duration of the MESSENGER's crossing of the plasma sheet. Allowing for multiple reconnection sites that last for variable durations may result in closer estimates.

## 4.4 Substorm current wedge and current closure

Asymmetries in Mercury's magnetotail reconnection affect the coupling between the plasma sheet and the planet's conducting core. In addition to hosting induction currents (see Section 3), Mercury's core provides current closure to static and/or large-scale field-aligned currents. Mercury's ion pick-up conductivity near the nightside polar regions is insufficient to close currents (see Section 1.6); instead, field-aligned currents are expected to flow radially through the low-conductance regolith and mantle to close over the surface of the highly conducting planetary core (~2,000 km radius) (Janhunen and Kallio, 2004; Anderson et al., 2014). While the first observations of Birkeland currents were reported from Mariner 10 (Slavin et al., 1997), MESSENGER observations have provided additional evidence for Birkeland currents and this current closure model. Anderson et al. (2014) identified statistical perturbations over Mercury's northern pole consistent with Region-1 field-aligned currents. These currents flow towards the planet on the dawnside and away from the planet on the duskside with typical intensities between 20–40 kA. The study argues that current closure via the core is most likely and that the height-integrated conductance above the core is ~1 S. In addition, this study shows that there is no clear signature of the region-2 field-aligned current, which is expected to flow towards the planet on the duskside and away from the planet on the dawnside, i.e., in the opposite sense of region-1 field-aligned

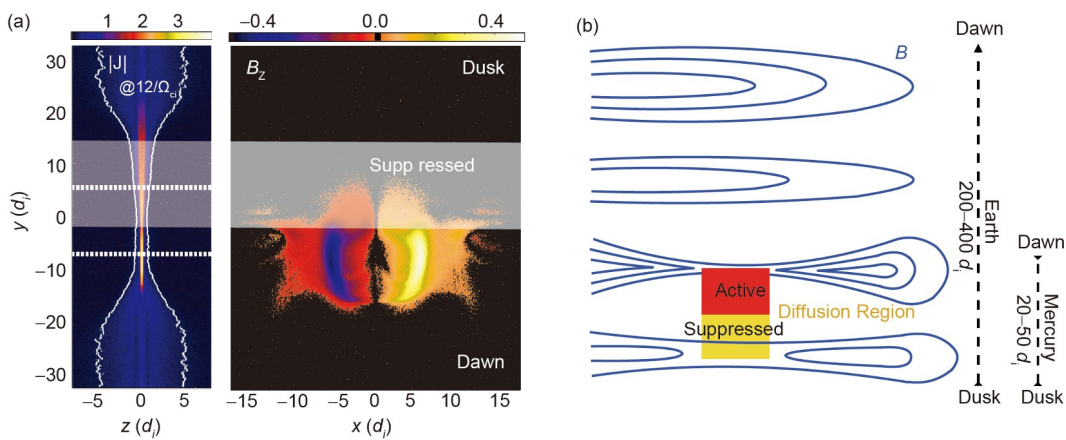

**Figure 23** (a) 3-D Particle-in-Cell (PIC) simulation of magnetic reconnection. The grey shaded region denotes where reconnection is locally suppressed along $y$. The scale is comparable to the cross-tail scale of Mercury's magnetotail. The scale of X-line in $y$ is ~31 $d_i$, where $d_i$ is the ion inertial length. "Dawn" and "Dusk" correspond to the respective sides of Mercury's magnetotail. (b) Schematic of dawn-dusk asymmetry in magnetotail reconnection at Earth and Mercury resulting from local suppression in reconnection. The figure is adapted from Liu et al. (2019). A global MHD-embedded PIC model results in a similar feature of magnetic reconnection in Mercury's magnetotail can be found in Chen Y et al. (2019).

current.

In planetary magnetospheres, the substorm current wedge is another important source for field-aligned currents. Similar to Region-1 field-aligned currents, the substorm current wedge is expected to close via Mercury's core. This current wedge, schematically shown in Figure 24, forms during magnetospherically active periods. Intense magnetotail reconnection generates fast plasma flows that carry newly-closed magnetic flux tubes (i.e., dipolarizations). As the flows encounter the planetary dipole field, the increased magnetic pressure brake and deflect the flows. As a result, the magnetic flux carried in the dipolarizations accumulates near Mercury (~700–1,200 km above the planet's nightside surface). The braking and flux pileup produce Alfvén waves that form the substorm current wedge (Sun et al., 2015a; Kepko et al., 2015; Dewey et al., 2020), which can be described as a stable standing Alfvén wave (Glassmeier, 2000). At an estimated Alfvén speed of ~1,000 km s$^{-1}$, the skin depth of these waves (~750–960 km) is larger than the depth of the regolith/mantle layer (~400 km). Although these waves can reach and be partially reflected by the core, their magnitude dampens exponentially. It is unlikely that a single Alfvén wave injection can form a stable substorm current wedge. Rather, the formation of a stable current wedge at Mercury requires a series of dipolarizations as sequential events can supply new Alfvén waves to replace damped ones (Sun et al., 2015a; Dewey et al., 2020).

Asymmetries in Mercury's magnetotail reconnection manifest in the substorm current wedge. Dewey et al. (2020) examined statistical signatures of magnetic flux pileup associated with dipolarizations in the near magnetotail. Pileup is more commonly observed post-midnight, while the average pileup strength is stronger pre-midnight. Dewey et al. (2020) interpret this signature as being related to Mercury's magnetotail asymmetry in dipolarizations: the more frequent post-midnight dipolarizations initiate pileup on the dawnside of the magnetotail, and after sufficient accumulation, the pileup region can expand duskward. The substorm current wedge at Mercury is therefore expected to form post-midnight and expand duskward, opposite to that at Earth.

Several studies have estimated the intensity of Mercury's substorm current wedge, as summarized in Table 4. Sun et al. (2015b) examined several individual substorms and estimated a current wedge intensity of ~60 kA by applying the ~1 S conductance (Anderson et al., 2014) to the magnetic energy dissipated during each of those substorm expansion phases. Poh et al. (2017a) examined a subset of central plasma sheet crossings (see Section 4.2) and applied a line-current model to a local statistical enhancement in the magnetic field near midnight. They found that the enhancement is consistent with a substorm current wedge of ~11 kA and a closure conductance of ~1.2 S. Dewey et al. (2020) examined the near-planet braking and flux pileup of dipolarizations to estimate a current wedge intensity of ~14.6 kA

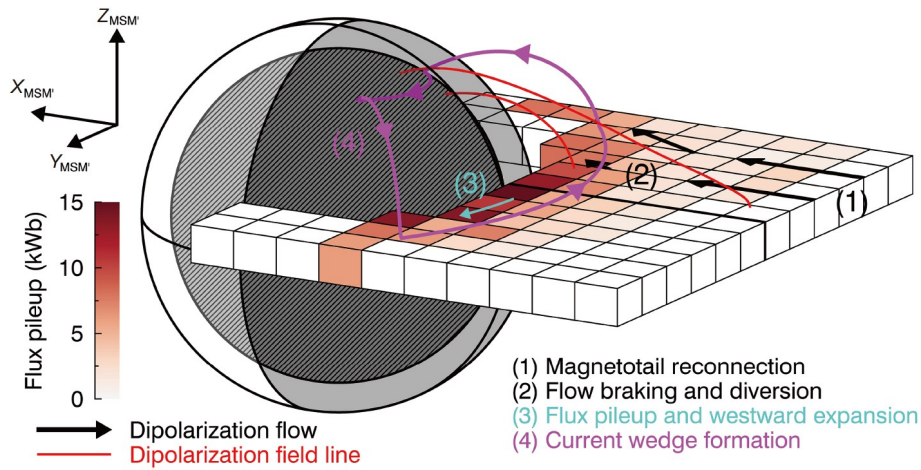

**Figure 24**   A schematic figure of magnetotail reconnection, plasma flow, flow braking and diversion, flux pileup, and current wedge formation in Mercury's magnetotail. This figure is adapted from Dewey et al. (2020).

**Table 4**   The intensity of the field-aligned currents in substorm current wedge and the height-integrated electrical conductance at Mercury from different studies

| Studies | Intensity of SCW (kA) | Height-integrated electrical conductance (S) | How to estimate |
|---|---|---|---|
| Sun et al. (2015b) | ~60 | ~1[a] | Dissipation of magnetic energy |
| Poh et al. (2017a) | ~11 | ~1.2 | Flux Pileup |
| Dewey et al. (2020) | ~14.6±5.0 | ~0.8±0.4 | Dipolarization related Flux Pileup |

a) Anderson et al. (2014)

and a closure conductance of ~0.8 S. Poh et al. (2017a) and Dewey et al. (2020) obtained similar estimates, but both found intensities much weaker than those reported by Sun et al. (2015b). The discrepancy may be due to sample size, as Sun et al. (2015b) investigated a small subset of the most intense substorms. Additionally, Dewey et al. (2020) suggest that a portion of dipolarizations may reach the planet's nightside surface. Such events would not contribute to flux pileup. Sun et al. (2015b) assumed that all dipolarizations would stop before reaching Mercury's surface, which would result in an overestimation of typical substorm current wedge intensities.

### 4.5   Open questions regarding plasma sheet dynamics

MESSENGER returned a rich set of measurements that have equipped the community to investigate the dynamics and dawn-dusk asymmetries of Mercury's magnetotail. However, open questions on the origin, character, and extent of these dynamics and asymmetries remain to be answered.

First, the origin of Mercury's post-midnight bias in magnetotail reconnection remains unresolved. Numerical simulations (e.g., Chen Y et al., 2019; Liu et al., 2019) have reproduced this dawn-dusk asymmetry and proposed hypotheses related to Hall effects. However, the precise mechanisms that produce the asymmetry differ. While both simulations are consistent with published MESSENGER observations, further comparison with observations is required to refine the origin of this cross-tail reconnection asymmetry and its difference from that at Earth. Dawn-dusk differences are also observed in the magnetotail, with the typical cross-tail width of reconnection sites still under debate. While Dewey et al. (2020) find the widths are small (~0.3 $R_M$), Smith et al. (2018a) identify a much greater extent (~2 $R_M$). MESSENGER's polar orbit limits the extent to which we can constrain cross-tail widths of individual structures. Therefore, numerical simulations will prove particularly valuable in characterizing Mercury's magnetotail reconnection sites.

Second, the frequency and character of Mercury's substorm current wedge warrant further study. While recent studies have demonstrated that a substorm current wedge can form in Mercury's magnetotail, the frequency and duration of the structure remain unknown. Based on the number and frequency of observed dipolarizations, Dewey et al. (2020) estimated ~6% of MESSENGER plasma sheet crossings could form a current wedge. This fraction is a lower bound since multiple magnetotail reconnection sites could produce a dipolarization that would not be observed by a single spacecraft. As discussed in Section 4.4, the intensity of the substorm current wedge is also undetermined. The difference in estimates from Sun et al. (2015b) compared to Poh et al. (2017a) and Dewey et al. (2020) demonstrates that the in-tensity of such a current structure may span an order of magnitude depending on the driver of the system. Additionally, while Dewey et al. (2020) provide a detailed estimation of substorm current wedge formation, several details require further attention, such as the mechanism and threshold for duskward magnetic flux pileup expansion.

Third, mass entry and transport in the plasma sheet for different ion species would benefit from further refinement. Theoretical and numerical studies (e.g., Delcourt, 2013) have advanced our understanding of dawn-dusk distributions of some ion species. However, the different mechanisms of mass entry and their relative contributions have not been quantified. For example, protons dominate the plasma sheet by number density, and several mechanisms have been proposed to describe their average dawn-dusk distributions. Zhao et al. (2020) suggest that entry from the low latitude boundary layer produces the enhancement of warm flux near the magnetotail flanks, while the dawnside preference in magnetic reconnection controls the dawnside enhancement in thermal flux. The efficiencies and precise details of these mechanisms require further analysis, particularly as they relate to proton distributions in the magnetosheath and plasma mantle (Jasinski et al., 2017). Energization (see Section 5), transport (e.g., Poh et al., 2018), and loss of plasma in the plasma sheet constitute additional open topics that connect entry mechanisms to the observed magnetotail distributions in proton density and temperature.

Finally, the statistical nature of analyzing dawn-dusk asymmetries from MESSENGER observations alone introduces difficulties in understanding how these cross-tail trends depend on solar wind drivers and Mercury's eccentric orbit.

(1) Mercury's cross-tail trends are expected to depend on upstream solar wind driving (Sun et al., 2017b; Chen Y et al., 2019). However, MESSENGER's elliptical orbit introduces long delays between plasma sheet crossings and solar wind observations. The nearest magnetopause crossing is ~1–2 h from the spacecraft's pass through the plasma sheet, during which solar wind and interplanetary magnetic field conditions are expected to change (James et al., 2017). These long delays prevent accurate pairing of MESSENGER magnetotail observations and external driving conditions. Some studies (e.g., Sun et al., 2017b) have used proxies within the magnetosphere to infer magnetospheric activity levels in order to mitigate this limitation. The dependence of cross-tail trends on upstream conditions deserves further dedicated attention.

(2) It is unknown if Mercury's orbit around the Sun influences dawn-dusk asymmetries in the magnetotail. MESSENGER's orbital plane is fixed in inertial space, so the spacecraft samples different local times of the magnetosphere as Mercury's position changes with respect to the Sun. As a result, the local time of MESSENGER observa-tions is fixed to Mercury's orbital position. For example, MESSENGER's orbit always samples the dawn-dusk terminator at perihelion. Mercury possesses the most eccentric planetary orbit within the Solar System, and as a result, the solar wind conditions change substantially over the course of its orbit (see Section 1). Without correcting with seasonal effects in MESSENGER observations, examining cross-tail properties convolutes seasonal trends, seasonal coverage, and dawn-dusk asymmetries. Some studies (e.g., Korth et al., 2014) have made assumptions on seasonal effects (e.g., in proton density) to mitigate these uncertainties; however, such effects and their influence on dawn-dusk asymmetries remain observationally unconstrained.

## 5. Particle energization in Mercury's magnetosphere

In this chapter, we summarize particle energization by various processes in Mercury's magnetosphere through comparisons with other magnetospheres in the Solar System. Mercury's relatively small magnetosphere results in the temporal and spatial scale of the electric and magnetic field variations to be comparable to those of plasma motion at the planet. Under such circumstances, plasma motion and field variation strongly affect each other, resulting in a possible non-adiabatic energization of plasmas which leads to significant plasma heating (e.g., Delcourt et al., 2002, 2007, 2017; Zelenyi et al., 2007). In the following subchapters, we focus on the different possible mechanisms of particle energization expected in Mercury's magnetosphere.

### 5.1 Kelvin-Helmholtz waves energizations

The K-H instability is an important process in which solar wind plasma penetrates to the magnetosphere when the IMF field lines are parallel to the planet's magnetospheric field lines. K-H waves have been observed by various spacecraft at Earth (e.g., Hasegawa et al., 2004; Hwang et al., 2011; Yan et al., 2014) and by Cassini at Saturn (Burkholder et al., 2020). At Mercury, MESSENGER had detected the K-H vortices several times in the magnetospheric flanks (Slavin et al., 2008; Sundberg et al., 2011, 2012; Liljeblad et al., 2014; Gershman et al., 2015). In general, the rolled-up K-H vortex leads to the transport of mass and momentum into the magnetosphere, and thus mixes two adjacent plasmas. Furthermore, magnetic reconnection inside the vortex enhances the plasma mixing within the vortex (e.g., Nykyri and Otto, 2001; Eriksson et al., 2016). At Earth, ion heating is discussed by several authors (e.g., Nykyri et al., 2006; Moore et al., 2017; Masson and Nykyri, 2018), and the particle energization associated with the K-H instability is mainly caused by secondary mechanisms such as magnetic re-connection, ion scale waves (Moore et al., 2017), kinetic Alfvén waves (Johnson and Cheng, 1997, 2001), and fast magnetosonic waves (Moore et al., 2016, 2017). At Saturn, Delamere et al. (2018) discussed that the magnetic fluctuations due to the growth of the K-H instability are turbulent, and thus, proton heating is expected.

In the magnetospheric flanks of Mercury, similar mechanisms of particle energization are expected. However, because of the limited particle instrument onboard MESSENGER, ion heating associated with the secondary mechanisms of the K-H instability cannot be analyzed in detail, i.e., magnetic reconnection inside a vortex would result from a thin current sheet compared to that at the dayside magnetopause. This would result in current sheet signatures with very short time intervals in the MESSENGER data, and thus, the ion heating cannot be addressed using the data from the FIPS instrument with a time resolution of around 10 s. Particle energization associated with the secondary mechanisms during the development of the K-H instability will be necessarily discussed by kinetic modeling and the new data from the BepiColombo mission in the future. On the other hand, due to Mercury's relatively small magnetosphere, the temporal and spatial field variation during the development of the K-H instability could be comparable to those of plasma motion. The estimated wavelength of the K-H waves observed by MESSENGER is about $1.5\ R_M$ (Mercury radius, 2440 km) (Gershman et al., 2015), and the heavy ions of planetary origin such as $Na^+$ observed in the vicinity of the K-H waves have a few keV of energy, meaning that particles have about a gyroradius of a few hundreds of kilometers. In this case, particles can undergo E-burst, in which a particle accelerates due to the sudden variation of an electric field within one gyroperiod.

Aizawa et al. (2018) conducted the particle tracing technique in the magnetic vector field obtained by ideal MHD simulations and found that the electric field variation during the development of K-H instability at Mercury can non-adiabatically energize planetary ions. The energization depends on the direction of particle motion and the electric field that the particles experience during one gyroperiod. The energy that particles can gain is controlled by the energy of the field, that is, the $\mathbf{E}\times\mathbf{B}$ drift speed (see Figure 25). Therefore, if the particle velocity vector and the electric field vector point in the same direction, and the particle has relatively small energy compared to the field that the particle passes through, its gyro motion is modified, and the particle gains a large boost of energy (encircled by red in Figure 25). However, if the particle is moving against the electric field and has comparable energy to that of the field, the particle can be decelerated (encircled by blue). If the particle has large energy compared to the $\mathbf{E}\times\mathbf{B}$ field, the particle does not gain energy (encircled by green).

Aizawa et al. (2020a) conducted a statistical analysis of the extensive work of Aizawa et al. (2018) to understand the particle energization and transport on both dawn and dusk configurations. They found that particle energization occurs on both the dawn and dusk side of the planet under north and southward IMF, but transport is more likely controlled by the convection electric field in the magnetosheath. For example, while the K-H instability is equally present on both the dawn and dusk sides under the northward IMF, particles are more energized and transport across the magnetopause on the dawnside, while less energization and transport of plasmas are observed on the duskside. Non-adiabatically energized ions are expected to sputter and produce secondary particles from the exosphere and surface. Contrary to simulation studies, Aizawa et al. (2020b) analyzed particle data obtained by the FIPS instrument. They focused on the sodium ion group and compared the phase space density distribution of the particle's path without K-H waves. Although there are some instrumental constraints, such as the limited field of view and energy range, the phase space distribution between K-H and non-K-H events exhibit significant differences when the spacecraft is inside the magnetosphere adjacent to the region where K-H waves are observed. The different distributions indicate that planetary ions are likely decelerated when K-H waves are present. K-H waves are frequently observed by MESSENGER on the dusk-nightside magnetopause, where FIPS detected the population of sodium ions with a few keV (Raines et al., 2011; Zurbuchen et al., 2011). According to the mechanisms suggested by Aizawa et al. (2018), it seems that they observed ions that have a large enough energy compared to the **E**×**B** energy, and move against the electric field, corresponding to the circled group by blue in Figure 25.

## 5.2 Dipolarization fronts and flux rope energization

Dipolarization fronts are also key phenomena that lead to particle energization in the magnetotail. There have been several mechanisms suggested for the formation of these dipolarization fronts, such as flow braking (e.g., Shiokawa et al., 1997; Birn et al., 2011; Fu et al., 2011; Pan et al., 2016), transient magnetic reconnection (e.g., Sitnov et al., 2009; Fu et al., 2013), and plasma instabilities (e.g., Pritchett and Coroniti, 2010; Runov et al., 2012; Pritchett and Coroniti, 2013; Pan et al., 2018). At Earth, previous studies using both simulations and observations have revealed that both ions and electrons are significantly accelerated during these dipolarizations (see Fu et al. (2020) and references therein). Particle acceleration in Earth's magnetotail contains both adiabatic (Fermi and betatron) and non-adiabatic (wave-particle interaction, resonance) acceleration. Moreover, flux ropes may also energize particles by reflecting particles between the two ends inside the rope, i.e., Fermi-like acceleration, either by the reconnection electric field (see Drake et al., 2006) or via the contraction of the flux rope (e.g., Dahlin et al., 2014; Sun et al., 2019). Earthward traveling flux ropes could interact with the dipole magnetic field and re-reconnect (e.g., Slavin et al., 2003; Lu et al., 2015; Poh et al., 2019), which may also energize particles. At the very large magnetosphere of Jupiter, recent observations from the Juno spacecraft show that flux increase and particle acceleration are likely due to adiabatic acceleration (Artemyev et al., 2020).

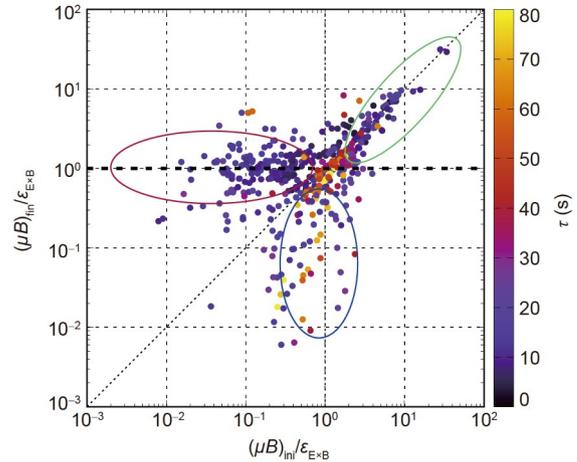

**Figure 25** Energization of planetary ions associated with K-H instability. Final perpendicular energy $(\mu B)_{\text{fin}}$, normalized to the maximum **E**×**B** drift energy $(\varepsilon_{E\times B})$ of $\text{Na}^+$ ions as a function of normalized initial perpendicular energy, $(\mu B)_{\text{ini}}/\varepsilon_{E\times B}$. The magnetic moment $\mu$ is the perpendicular energy of particles divided by magnetic field intensity $(B)$. The time scale of the corresponding E burst is coded according to the color scale at the right. The figure is adapted from Aizawa et al. (2018).

On the other hand, at Mercury, data from Mariner-10 first revealed energetic electron bursts associated with dipolarization events (Eraker and Simpson, 1986; Christon, 1987). In the MESSENGER era, the energetic particle spectrometer (EPS), the X-ray spectrometer, and the Gamma-ray and neutron spectrometer (GRNS) onboard MESSENGER have been used to identify energetic electrons, and the Fast Imaging Plasma Spectrometer (FIPS) has been used to examine ion properties. Baker et al. (2016) first reported energetic electron injections in conjunction with magnetic field dipolarizations using GRNS. Later, Dewey et al. (2017) identified 538 dipolarizations with energetic electron injections and conducted a statistical analysis of these events (see Figure 26 for dipolarization related energetic electrons). Their results suggest that electrons are likely energized by betatron or Fermi acceleration, and are likely not accelerated directly by magnetic reconnection, but by the subsequent dipolarization. Ion heating during dipolarizations at Mercury has also been investigated by Sun et al. (2017b) and Dewey et al. (2017), with a comparative study at the Earth conducted by Sun et al. (2018). Sun et al. (2017b) reported that proton suprathermal flux and temperature in the plasma sheet are attributed to the different levels of magnetospheric activity, primarily the different thickness of the current sheet. Ion properties before and after the dipolarization are fitted by Maxwellian and kappa distributions, and they concluded that thermal protons energized to suprathermal energies appeared as a significant kappa value in the distribution (see Figure 27). Proton temperature increases were also observed accompanying the dipolarization. However, the statistical analysis conducted by Dewey et al. (2017) showed that the typical temperature of thermal protons in dipolarization events is 38.0 MK compared to the 35.2 MK of ambient thermal plasma. This acceleration seems to be betatron acceleration, and they find that the average factor of plasma heating is about 1.1 at Mercury, while the average plasma heating factor at Earth is about 1.3. Two cases described in Sun et al. (2017b) indicate significant proton acceleration events (heating factor of 2) are extreme cases.

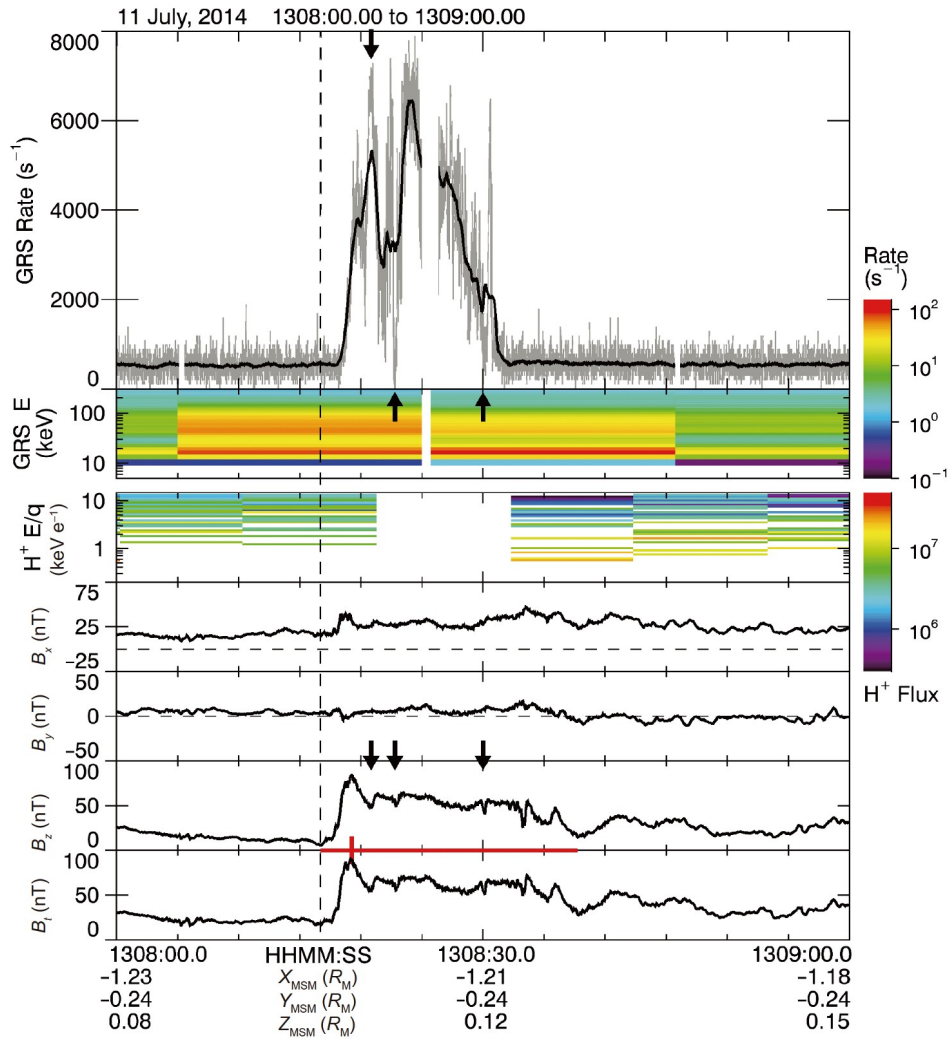

**Figure 26** A dipolarization is associated with several energetic electron injections. Each panel from top to bottom: GRS count rate; GRS accumulated spectra; FIPS H$^+$ spectrum; magnetic field x component ($B_x$); magnetic field y component ($B_y$); magnetic field z component ($B_z$); and magnetic field intensity ($B_t$). This figure is adapted from Dewey et al. (2017).

Since the timescale of dipolarizaions observed in Mercury's magnetotail is about 5–10 s, which is comparable with the gyroperiod of protons in the vicinity, particle acceleration is expected to be non-adiabatic (Delcourt et al., 2007). Sun et al. (2018) made a comparative study of the properties of magnetic dipolarizations at Earth and Mercury. The ion density in the plasma sheet is an order of magnitude higher at Mercury (~3–10 cm$^{-3}$) than at Earth (~0.1–0.6 cm$^{-3}$), while the temperature is several times lower at Mercury (~ 1–5 keV) than at Earth (~3–10 keV). The kappa value at Mercury has a broader range (~2 to ~60) than that at Earth (~5 to 20), and its variability during the dipolarization is larger (>60%) than that at Earth (<20%). This large variation in kappa value at Mercury indicates energy-dependent energy increments, while the small kappa variation at Earth indicates betatron acceleration under the conservation of the magnetic moment. Two possible mechanisms described the energy-dependent energy increments were discussed: non-adiabatic cross-tail particle motion associated with thin current sheets and wave-particle interactions.

Although particle acceleration, especially electron acceleration directly by magnetic reconnection, has not been well investigated in Mercury's magnetosphere, a few simulations have suggested that turbulent plasmoid reconnection can effectively accelerate electrons and form a power-law spectrum (Büchner et al., 2018; Zhou et al., 2018). These studies further pointed out that these accelerations occurred in the region around a reconnecting X-line. MESSENGER observations of a high-frequency chain of flux ropes in the magnetotail (e.g., Sun et al., 2016, 2020b; Zhong et al., 2020a), and on the magnetopause (e.g., Slavin et al., 2012b, 2014; Sun et al., 2020b) support the idea of plasmoid reconnection. However, their efficiency in energizing particles requires further studies.

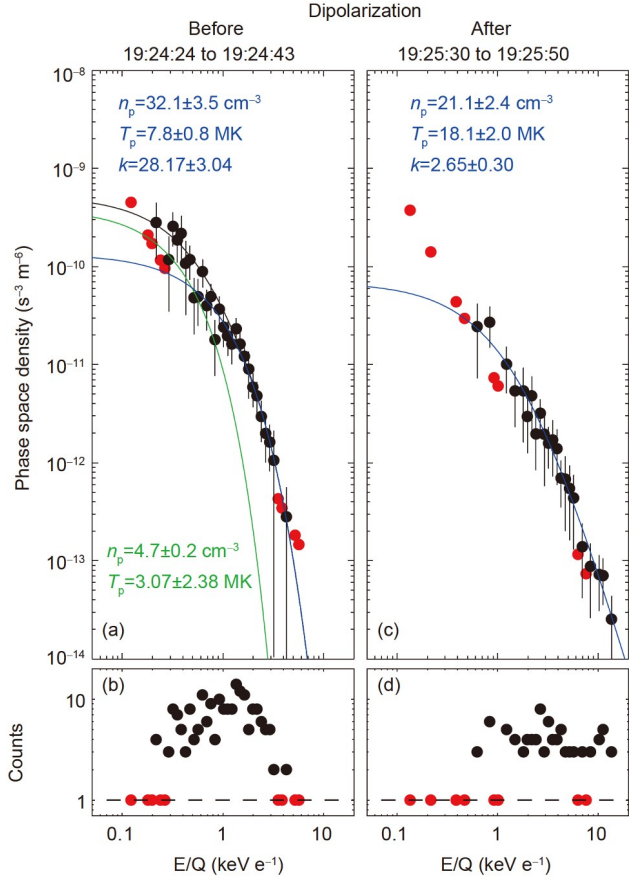

**Figure 27**  Proton phase space density (PSD) and counts versus E/Q before and after a dipolarization on 1 July 2011 during a plasma sheet crossing. Red dots in each figure represent the measurements with only one proton count. Blue lines are Kappa distributions fitting the hot plasma components. Green lines are Maxwellian fits to the cold components. The black line in (a) is the sum of blue and green lines. The figure is adapted from Sun et al. (2017b).

### 5.3   Ion cross-tail energization

Particle energization without any dynamic magnetospheric phenomena (e.g., dipolarization, magnetic reconnections) in the magnetotail is also possible (so-called cross-tail particle motion). When the radius of curvature of the magnetic field is comparable to the gyro-radius of a particle, non-adiabatic energization can occur as a result of impulsive centrifugal forces. The magnetic moment changes of the particle are organized according to a three-branch pattern (Delcourt and Martin Jr, 1994). In that regime, acceleration or non-adiabatic motion causes several interesting phenomena related to ion dynamics and plasma environment. For example, Ashour-Abdalla et al. (1993) reported enhanced particle trapping with a large energy gain during particles' drift motion toward the duskside. This may also cause phase bunching, a process that thins the magnetotail current sheet (e.g., Delcourt et al., 1995, 1996a, 1996b). In addition, the aurora caused by ion precipitation, energy-dispersed ion structures, and plasma sheet phenomena has been reported at Earth (West Jr et al., 1978a, 1978b; Wagner et al., 1979; Lyons and Speiser, 1982; Ashour-Abdalla et al., 1991, 1992; Keiling et al., 2004). The large magnetic moment of the particle indicates a change in its mirror point, suggesting that non-adiabatic motion causes ion precipitation.

At Mercury, Delcourt et al. (2003) examined the circulation of heavy ions of planetary origin within Mercury's magnetosphere using a simple particle tracing technique and made a statistical map of ion distributions (see Figure 28). In their calculation, sodium ions from the exosphere can be non-adiabatically accelerated and move towards the dusk magnetopause (see also, Ip, 1987). Due to their large gyro radii, those particles cannot achieve sufficient drift motion around the planet to hit the dusk magnetopause. On the other hand, protons are not non-adiabatically energized due to their small gyro radii. Due to the electric field in the magnetotail, ions move from dawn to dusk, resulting in an asymmetry in the distribution of sodium ions, which is consistent with MESSENGER observations (Zurbuchen et al., 2011; Raines et al., 2013; Gershman et al., 2014). Poh et al. (2018) have examined the transport of mass and energy in Mercury's plasma sheet and found an average polytropic index of 0.687. The predicted polytropic index assuming adiabatic behavior is 5/3, which is larger than the observed value, which suggests particles are behaving non-adiabatically. Interestingly, the polytropic index with magnetospheric activities shows a lower value (~0.58) than the index during quiet times (~0.64).

### 5.4   Open questions regarding the particle accelerations

Four years of MESSENGER observations have brought a wealth of information on the plasma environment and magnetospheric activity at Mercury. Analysis of particle acceleration combined with simulation studies has shown us many features in the Mercury magnetosphere are analogous to those observed at Earth but on a smaller scale. However, due to mission and instrument constraints, some information is still missing. For example, non-adiabatic energization of ions with a few eV has not yet been confirmed due to the mission constraints. Also, wave-particle interactions in Mercury's magnetosphere have not yet been observed directly.

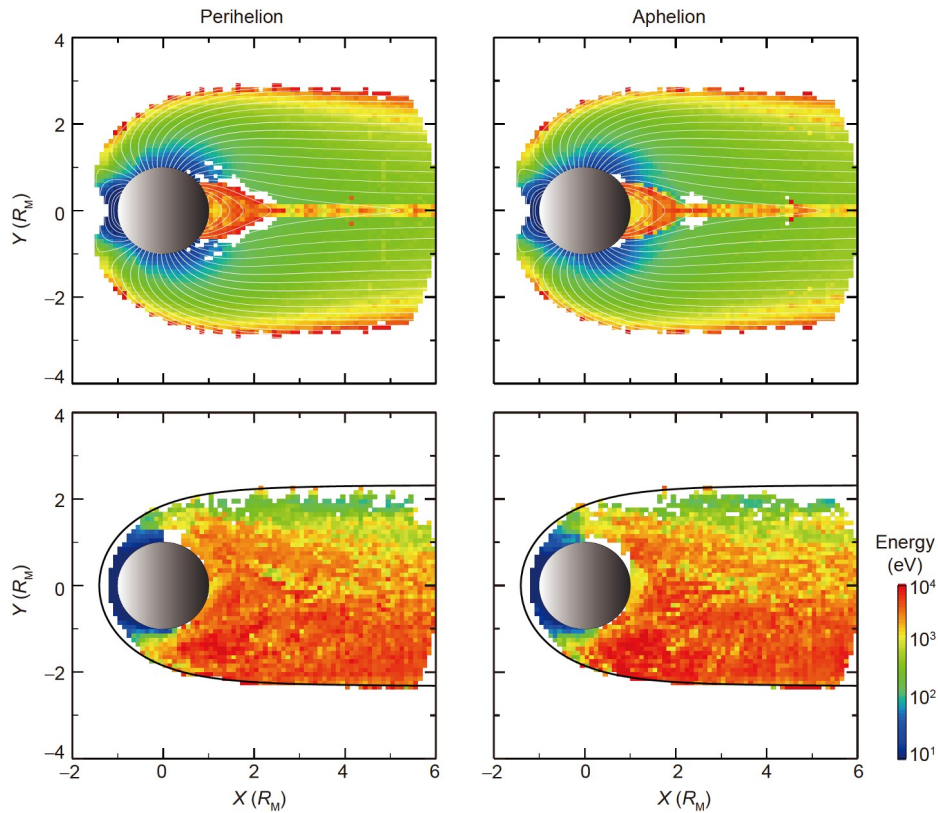

**Figure 28**    Modeled energy of Na$^+$ at (left) perihelion and (right) aphelion. The top and bottom panels show cross-sections in the noon-midnight plane and the equatorial plane, respectively. The color bar at the right represents the intensity of energies. The figure is adapted from Delcourt et al. (2003).

This chapter has reviewed the particle energizations on a large scale, including the K-H vortices, dipolarization fronts, and cross-tail energization. We know that the K-H waves are an essential source for ultra-low frequency (ULF) waves. ULF waves can be a means of energy transport (Dungey and Southwood, 1970; Pu and Kivelson, 1983; Glassmeier et al., 2004), and can energize particles through wave-particle interactions. A better understanding of ULF waves can result in a better understanding of the transfer of energy and momentum in planetary magnetospheres. ULF waves have been frequently observed on Mercury's dayside magnetosphere, and are possibly initiated by the K-H waves (Liljeblad et al., 2016; Liljeblad and Karlsson, 2017), particle anisotropy (e. g., Schriver et al., 2011; Boardsen et al., 2015), and magnetic structures in the plasma sheet (Sun et al., 2015a). However, observations are limited to single-point magnetic field measurement. Moreover, ULF waves in Earth's magnetosphere can be initiated by many processes, such as variations in the solar wind dynamic pressure, including interplanetary shocks (e.g., Kepko et al., 2002; Zong et al., 2007), and high-speed plasma flows, and magnetic structures that are accompanied by pressure variations (Keiling et al., 2014). Another important source for ULF waves is ion cyclotron waves (ICWs). At Mercury, the newly ionized H$^+$, O$^+$, or Na$^+$ can be picked up by the convection electric field and generate ICWs. A survey of Mariner 10 magnetic field measurements does not show evidence of H$^+$ or Na$^+$ ICWs in Mercury's magnetosphere (Boardsen and Slavin, 2007), which is likely because Mariner 10 crossed Mercury's magnetosphere quickly and could not take measurements over a sufficient number of Na$^+$ cyclotron periods. The latest MESSENGER measurements have provided the evidence for pick-up planetary ions (Na$^+$-group) in the solar wind ahead of Mercury's magnetosphere (Jasinski et al., 2020) but are not accompanied by ICWs. Huang et al. (2020) have done a statistical study of the magnetic field fluctuations around Mercury's magnetosphere. In the regions near the magnetopause, they reported a peak in the spectra power density of the magnetic field fluctuations. The peak has a frequency that is close to the Na$^+$ local gyro-frequency, which might be evidence of Na$^+$ cyclotron waves. Schmid et al. (2021) have done a comprehensive investigation of the ICWs generated by pick-up protons. The investigations focus on the solar wind and magnetosheath around Mercury's magnetosphere and have shown that ICWs found in the solar wind and magnetosheath exhibit different properties.

Most of the above sources have not been confirmed in Mercury's magnetosphere. Furthermore, the scale of Mercury's magnetosphere is only a few thousand kilometers, which is comparable to the wavelength of several types of ULF waves, and Mercury only has a tenuous exosphere with a very low conductivity, which can influence the propagation and dissipation of ULF waves in Mercury's magnetosphere. Although some MESSENGER observations (James et al., 2019) and simulations (Kim et al., 2016) have suggested the existence of field-line resonance in Mercury's magnetosphere, whether a large portion of the plasma waves was reflected or absorbed near Mercury's surface is still an open question.

Moreover, our knowledge of very low frequency (VLF) waves in Mercury's magnetosphere is incomplete. VLF waves are shown to be closely related to energetic electrons (from tens of keV to MeV). For example, chorus waves, whose frequency is between the lower hybrid frequency and the electron cyclotron frequency, have not been investigated in Mercury's magnetosphere. Chorus waves are widely observed at planetary magnetospheres, such as Earth (Tsurutani and Smith, 1974; Burtis and Helliwell, 1976; Tao et al., 2020), Saturn (Gurnett et al., 1981), and Jupiter (Menietti et al., 2012). Chorus waves can significantly influence electron dynamics. Chorus waves can accelerate electrons from tens of keV to hundreds of keV and even to MeV (e.g., Horne and Thorne, 1998; Thorne et al., 2013). Furthermore, chorus waves can scatter the pitch angles of low energy electrons (hundreds eV to keVs), causing some electrons to be scattered into loss cones and precipitate into the upper atmosphere (Thorne et al., 2010; Kasahara et al., 2018).

## 6. Future BepiColombo opportunities

The BepiColombo project is a joint mission between ESA and JAXA currently en route to Mercury (Benkhoff et al., 2010). The BepiColombo mission was launched on 20 October 2018 and is on an approximately seven-year journey to the planet Mercury. During its journey, BepiColombo is scheduled to perform nine planetary flybys. To date, they have flown by Earth and Venus once and will fly by Venus one more time and a total of Mercury six times. BepiColombo consists of two spacecraft: the Mercury Magnetospheric Orbiter (MMO, also called Mio) (Murakami et al., 2020), and the Mercury Planetary Orbiter (MPO). Both Mio and MPO have groups of dedicated instruments that will detect the neutral, plasma, and magnetic field environments at Mercury. BepiColombo will also enable simultaneous two-point observations of Mercury's magnetosphere for the first time. At times, one spacecraft will measure solar wind conditions upstream of the bow shock and serve as a solar wind monitor for the other spacecraft inside the magnetosphere. The recently published instrumental papers outline a variety of discussions over how the BepiColombo measurements are expected to improve the understanding of the planet Mercury. Here, it is pointed out that the dual-spacecraft measurements of BepiColombo will be beneficial to the following topics:

(1) Magnetosphere-planet's surface coupling. MPO/Search for Exosphere Refilling and Emitted Neutral Abundances (SERENA) (Orsini et al., 2021) provide comprehensive measurements of low-energy (~eV) and high-energy (up to 5 keV) neutrals and their elemental compositions in the exosphere near the planet. Comprehensive measurements of temperature profiles and spatial distributions can help to identify the different releasing processes of exospheric particles and the dynamics of planetary exospheres. MPO/Solar Intensity X-ray and particle Spectrometer (SIXS) (Huovelin et al., 2020) can measure the energetic neutral particles (ENA). Since the high-energy solar wind particles impacting on the surface can eject energetic neutrals out of the surface, the SIXS can monitor the solar wind precipitation on the planet's surface. Mio/Mercury Plasma Particle Experiment (MPPE) (Saito et al., 2010) covers wider energy ranges and can provide three-dimensional velocity distributions of ion species. With MPPE, the dynamics of the ionized particles can be investigated in detail.

(2) The properties and the drivers of the magnetospheric response modes. Firstly, the features and properties of Mercury's magnetosphere under the different magnetospheric modes can be studied with BepiColombo dual-spacecraft. Mio/MPPE can provide 3-dimensional measurements of the plasma and will be able to provide the plasma flow information (Saito et al., 2010). The convection feature in the plasma sheet can be investigated and compared among the response modes. Secondly, the solar wind monitor allows for analysis of the solar wind conditions that drive different magnetospheric response modes, including substorms, steady magnetospheric convection, and sawtooth events. Thirdly, comparative magnetospheres between Mercury and Earth might help interpret the driving conditions of different magnetospheric modes.

(3) The dawn-dusk properties of the plasma sheet. The BepiColombo dual-spacecraft measurements will be able to mitigate the limitations of single-point measurement. One spacecraft can measure solar wind conditions while the other spacecraft samples inside the magnetosphere. These upstream solar wind observations will allow for improved aberration considerations and analysis of dawn-dusk asymmetries as functions of solar wind forcing. Furthermore, the orbital planes of BepiColombo's spacecraft will be orthogonal to that of MESSENGER. While MESSENGER orbited along with the terminators at perihelion, the two BepiColombo spacecraft will orbit noon-midnight, increasing the seasonal magnetospheric coverage of in situ measurements at Mercury. With an improved understanding of cross-tail asymmetries at Mercury, we can continue to explore the fundamental aspects of planetary magnetotails.

(4) The particle energization in Mercury's magnetosphere. The BepiColombo dual-spacecraft will have a north-south symmetric orbit and will allow for many coordinated observations (Milillo et al., 2020). The dual-spacecraft will at times provide simultaneous measurements of the upstream solar wind conditions and the magnetosphere. The spacecraft Mio and MPO carry a full package of sophisticated instruments to observe both electromagnetic fields and the plasma environment. Mio/MPPE will provide information on plasma flows, especially the electrons, and the properties of energetic particles. The MPO/SERENA will provide measurements of low-energy neutral atoms and ions and their compositions (up to several keV) (Orsini et al., 2021), which again offers a great opportunity to investigate the particle dynamics of Mercury's magnetosphere and exosphere. At the same time, the spacecraft carries a variety of magnetic field and electric field instruments, including the magnetometer, i.e., MPO/MAG (Glassmeier et al., 2010; Heyner et al., 2021) and Mio/MGF (Baumjohann et al., 2020). In particular, the Plasma/Plasma wave experiment (PWI) will provide the first electric field measurements from DC to 10 MHz (Karlsson et al., 2020; Kasaba et al., 2020), and search coil magnetometers with frequency up to 20 kHz (Yagitani et al., 2020). With these high sampling rate for magnetic field and electric field measurements, we can better understand the electromagnetic field waves and plasma waves in Mercury's magnetosphere and further our investigations of wave-particle interactions and the energetic particle environment of Mercury.

**Acknowledgements**    *In this review, Weijie SUN designs the whole project. Weijie SUN drafts Chapter one, Chapter two, Chapter three, and Chapter six. Ryan DEWEY drafts Chapter four. Sae AIZAWA drafts Chapter five. Jia HUANG provides Figures 2 and 3 and writes the paragraphs in Chapter 1.1 about the Parker Solar Probe (PSP) and the Wind spacecraft. James SLAVIN provides comments and suggestions on the structure of the paper. All authors discussed and improved the manuscript. The authors are grateful to the scientific editor for the invitation to write this review. We acknowledge the NASA Parker Solar Probe Mission and the SWEAP team led by Dr. Justin KASPER and the FIELDS team led by Dr. Stuart D. BALE for use of data. We appreciate Dr. Justin KASPER and Dr. Ben ALTERMAN for processing the Wind/SWE data, and Dr. R. LEPPING for providing the Wind/MFI data. Weijie SUN thanks Dr. Jim RAINES (University of Michigan, Ann Arbor), Dr. Xianzhe JIA (University of Michigan, Ann Arbor), Dr. Gangkai POH (NASA/GSFC), Mr. Jiu-Tong ZHAO (Peking University), Dr. Jun ZHONG (Institute of Geology and Geophysics, Chinese Academy of Sciences), Dr. Yi-Hsin LIU (Dartmouth College) and Dr. Dominique DELCOURT (Ecole Polytechnique-CNRS-UPMC, France) for providing and agreeing for using the high-resolution figures in their papers. Weijie SUN and James A. SLAVIN were supported by the National Aeronautics and Space Administration (Grant Nos. 80NSSC18K1137, 80NSSC21K0052). S AIZAWA acknowledges the support of CNES for the BepiColombo mission.*

## References

Aizawa S, Delcourt D, Terada N. 2018. Sodium ion dynamics in the magnetospheric flanks of Mercury. Geophys Res Lett, 45: 595–601

Aizawa S, Delcourt D, Terada N, André N. 2020a. Statistical study of non-adiabatic energization and transport in Kelvin-Helmholtz vortices at Mercury. Planet Space Sci, 193: 105079

Aizawa S, Raines J M, Delcourt D, Terada N, André N. 2020b. MESSENGER observations of planetary ion characteristics in the vicinity of Kelvin-Helmholtz vortices at Mercury. J Geophys Res-Space Phys, 125: e27871

Akasofu S I. 1964. The development of the auroral substorm. Planet Space Sci, 12: 273–282

Akasofu S I. 1981. Energy coupling between the solar wind and the magnetosphere. Space Sci Rev, 28: 121–190

Akhavan-Tafti M, Palmroth M, Slavin J A, Battarbee M, Ganse U, Grandin M, Le G, Gershman D J, Eastwood J P, Stawarz J E. 2020. Comparative analysis of the vlasiator simulations and MMS observations of multiple X-line reconnection and flux transfer events. J Geophys Res-Space Phys, 125: e27410

Alfè D, Gillan M J, Price G D. 2007. Temperature and composition of the Earth's core. Contemp Phys, 48: 63–80

Anderson B J, Acuña M H, Korth H, Slavin J A, Uno H, Johnson C L, Purucker M E, Solomon S C, Raines J M, Zurbuchen T H, Gloeckler G, McNutt R L. 2010. The magnetic field of Mercury. Space Sci Rev, 152: 307–339

Anderson B J, Acuña M H, Lohr D A, Scheifele J, Raval A, Korth H, Slavin J A. 2007. The magnetometer instrument on MESSENGER. Space Sci Rev, 131: 417–450

Anderson B J, Johnson C L, Korth H, Slavin J A, Winslow R M, Phillips R J, McNutt Jr R L, Solomon S C. 2014. Steady-state field-aligned currents at Mercury. Geophys Res Lett, 41: 7444–7452

Anderson B J, Johnson C L, Korth H, Winslow R M, Borovsky J E, Purucker M E, Slavin J A, Solomon S C, Zuber M T, McNutt Jr R J. 2012. Low-degree structure in Mercury's planetary magnetic field. J Geophys Res, 117: E00L12

Andrews G B, Zurbuchen T H, Mauk B H, Malcom H, Fisk L A, Gloeckler G, Ho G C, Kelley J S, Koehn P L, Lefevere T W, Livi S S, Lundgren R A, Raines J M. 2007. The energetic particle and plasma spectrometer instrument on the MESSENGER spacecraft. Space Sci Rev, 131: 523–556

Artemyev A V, Clark G, Mauk B, Vogt M F, Zhang X J. 2020. Juno observations of heavy ion energization during transient dipolarizations in Jupiter magnetotail. J Geophys Res-Space Phys, 125: e27933

Artemyev A V, Petrukovich A A, Nakamura R, Zelenyi L M. 2011. Cluster statistics of thin current sheets in the Earth magnetotail: Specifics of the dawn flank, proton temperature profiles and electrostatic effects. J Geophys Res, 116: A09233

Ashour-Abdalla M, Berchem J P, Buechner J, Zelenyi L M. 1993. Shaping of the magnetotail from the mantle: Global and local structuring. J Geophys Res, 98: 5651–5676

Ashour-Abdalla M, Berchem J, Büchner J, Zelenyi L M. 1991. Large and small scale structures in the plasma sheet: A signature of chaotic motion and resonance effects. Geophys Res Lett, 18: 1603–1606

Ashour-Abdalla M, Zelenyi L M, Bosqued J M, Peroomian V, Wang Z, Schriver D, Richard R L. 1992. The formation of the wall region: Consequences in the near Earth magnetotail. Geophys Res Lett, 19:

1739–1742

Badman S V, Cowley S W H. 2007. Significance of Dungey-cycle flows in Jupiter's and Saturn's magnetospheres, and their identification on closed equatorial field lines. Ann Geophys, 25: 941–951

Badman S V, Jackman C M, Nichols J D, Clarke J T, Gérard J C. 2014. Open flux in Saturn's magnetosphere. Icarus, 231: 137–145

Baker D N, Dewey R M, Lawrence D J, Goldsten J O, Peplowski P N, Korth H, Slavin J A, Krimigis S M, Anderson B J, Ho G C, McNutt Jr R L, Raines J M, Schriver D, Solomon S C. 2016. Intense energetic electron flux enhancements in Mercury's magnetosphere: An integrated view with high-resolution observations from MESSENGER. J Geophys Res-Space Phys, 121: 2171–2184

Bale S D, Goetz K, Harvey P R, Turin P, Bonnell J W, Dudok de Wit T, Ergun R E, MacDowall R J, Pulupa M, Andre M, Bolton M, Bougeret J L, Bowen T A, Burgess D, Cattell C A, Chandran B D G, Chaston C C, Chen C H K, Choi M K, Connerney J E, Cranmer S, Diaz-Aguado M, Donakowski W, Drake J F, Farrell W M, Fergeau P, Fermin J, Fischer J, Fox N, Glaser D, Goldstein M, Gordon D, Hanson E, Harris S E, Hayes L M, Hinze J J, Hollweg J V, Horbury T S, Howard R A, Hoxie V, Jannet G, Karlsson M, Kasper J C, Kellogg P J, Kien M, Klimchuk J A, Krasnoselskikh V V, Krucker S, Lynch J J, Maksimovic M, Malaspina D M, Marker S, Martin P, Martinez-Oliveros J, McCauley J, McComas D J, McDonald T, Meyer-Vernet N, Moncuquet M, Monson S J, Mozer F S, Murphy S D, Odom J, Oliverson R, Olson J, Parker E N, Pankow D, Phan T, Quataert E, Quinn T, Ruplin S W, Salem C, Seitz D, Sheppard D A, Siy A, Stevens K, Summers D, Szabo A, Timofeeva M, Vaivads A, Velli M, Yehle A, Werthimer D, Wygant J R. 2016. The FIELDS instrument suite for solar probe plus. Space Sci Rev, 204: 49–82

Baumjohann W, Haerendel G. 1985. Magnetospheric convection observed between 0600 and 2100 LT: Solar wind and IMF dependence. J Geophys Res, 90: 6370–6378

Baumjohann W, Matsuoka A, Narita Y, Magnes W, Heyner D, Glassmeier K H, Nakamura R, Fischer D, Plaschke F, Volwerk M, Zhang T L, Auster H U, Richter I, Balogh A, Carr C M, Dougherty M, Horbury T S, Tsunakawa H, Matsushima M, Shinohara M, Shibuya H, Nakagawa T, Hoshino M, Tanaka Y, Anderson B J, Russell C T, Motschmann U, Takahashi F, Fujimoto A. 2020. The BepiColombo-Mio magnetometer en route to Mercury. Space Sci Rev, 216: 125

Benkhoff J, van Casteren J, Hayakawa H, Fujimoto M, Laakso H, Novara M, Ferri P, Middleton H R, Ziethe R. 2010. BepiColombo—Comprehensive exploration of Mercury: Mission overview and science goals. Planet Space Sci, 58: 2–20

Benninghoven A. 1975. Developments in secondary ion mass spectroscopy and applications to surface studies. Surf Sci, 53: 596–625

Bida T A, Killen R M, Morgan T H. 2000. Discovery of calcium in Mercury's atmosphere. Nature, 404: 159–161

Birn J, Nakamura R, Panov E V, Hesse M. 2011. Bursty bulk flows and dipolarization in MHD simulations of magnetotail reconnection. J Geophys Res, 116: A01210

Biskamp D, Welter H. 1980. Coalescence of magnetic islands. Phys Rev Lett, 44: 1069–1072

Boardsen S A, Kim E H, Raines J M, Slavin J A, Gershman D J, Anderson B J, Korth H, Sundberg T, Schriver D, Travnicek P. 2015. Interpreting ~1 Hz magnetic compressional waves in Mercury's inner magnetosphere in terms of propagating ion-Bernstein waves. J Geophys Res-Space Phys, 120: 4213–4228

Boardsen S A, Slavin J A. 2007. Search for pick-up ion generated Na$^+$ cyclotron waves at Mercury. Geophys Res Lett, 34: L22106

Bowers C F, Slavin J A, DiBraccio G A, Poh G, Hara T, Xu S, Brain D A. 2021. MAVEN survey of magnetic flux rope properties in the Martian ionosphere: Comparison with three types of formation mechanisms. Geophys Res Lett, 48: e93296

Boyle C B, Reiff P H, Hairston M R. 1997. Empirical polar cap potentials. J Geophys Res, 102: 111–125

Brambles O J, Lotko W, Zhang B, Wiltberger M, Lyon J, Strangeway R J. 2011. Magnetosphere sawtooth oscillations induced by ionospheric outflow. Science, 332: 1183–1186

Broadfoot A L, Shemansky D E, Kumar S. 1976. Mariner 10: Mercury atmosphere. Geophys Res Lett, 3: 577–580

Büchner J, Kilian P, Muñoz P A, Spanier F, Widmer F, Zhou X, Jain N. 2018. Kinetic simulations of electron acceleration at Mercury. In: Lühr H, Wicht J, Gilder S A, Holschneider M, eds. Magnetic Fields in the Solar System: Planets, Moons and Solar Wind Interactions. Cham: Springer International Publishing. 201–240

Burch J L, Moore T E, Torbert R B, Giles B L. 2016. Magnetospheric multiscale overview and science objectives. Space Sci Rev, 199: 5–21

Burkholder B L, Delamere P A, Johnson J R, Ng C S. 2020. Identifying active Kelvin-Helmholtz vortices on Saturn's magnetopause boundary. Geophys Res Lett, 47: e84206

Burlaga L F. 2001. Magnetic fields and plasmas in the inner heliosphere: Helios results. Planet Space Sci, 49: 1619–1627

Burtis W J, Helliwell R A. 1976. Magnetospheric chorus: Occurrence patterns and normalized frequency. Planet Space Sci, 24: 1007–1024

Chen C, Sun T R, Wang C, Huang Z H, Tang B B, Guo X C. 2019. The effect of solar wind mach numbers on the occurrence rate of flux transfer events at the dayside magnetopause. Geophys Res Lett, 46: 4106–4113

Chen Y, Tóth G, Jia X, Slavin J A, Sun W, Markidis S, Gombosi T I, Raines J M. 2019. Studying dawn-dusk asymmetries of Mercury's magnetotail using MHD-EPIC simulations. J Geophys Res-Space Phys, 124: 8954–8973

Cheng A F, Johnson R E, Krimigis S M, Lanzerotti L J. 1987. Magnetosphere, exosphere, and surface of Mercury. Icarus, 71: 430–440

Cheng X, Guo Y, Ding M D. 2017. Origin and structures of solar eruptions I: Magnetic flux rope. Sci China Earth Sci, 60: 1383–1407

Christensen U R. 2006. A deep dynamo generating Mercury's magnetic field. Nature, 444: 1056–1058

Christon S P. 1987. A comparison of the Mercury and Earth magnetospheres: Electron measurements and substorm time scales. Icarus, 71: 448–471

Cooling B M A, Owen C J, Schwartz S J. 2001. Role of the magnetosheath flow in determining the motion of open flux tubes. J Geophys Res, 106: 18763–18775

Coroniti F V, Kennel C F. 1972. Changes in magnetospheric configuration during the substorm growth phase. J Geophys Res, 77: 3361–3370

Crooker N U. 1992. Reverse convection. J Geophys Res, 97: 19363–19372

Dahlin J T, Drake J F, Swisdak M. 2014. The mechanisms of electron heating and acceleration during magnetic reconnection. Phys Plasmas, 21: 092304

Daughton W, Scudder J, Karimabadi H. 2006. Fully kinetic simulations of undriven magnetic reconnection with open boundary conditions. Phys Plasmas, 13: 072101

Delamere P A, Burkholder B, Ma X. 2018. Three-dimensional hybrid simulation of viscous-like processes at Saturn's magnetopause boundary. Geophys Res Lett, 45: 7901–7908

Delcourt D C, Belmont G, Sauvaud J A, Moore T E, Martin Jr R F. 1996a. Centrifugally driven phase bunching and related current sheet structure in the near-Earth magnetotail. J Geophys Res, 101: 19839–19847

Delcourt D C, Grimald S, Leblanc F, Berthelier J J, Millilo A, Mura A, Orsini S, Moore T E. 2003. A quantitative model of the planetary Na$^+$ contribution to Mercury's magnetosphere. Ann Geophys, 21: 1723–1736

Delcourt D C, Leblanc F, Seki K, Terada N, Moore T E, Fok M C. 2007. Ion energization during substorms at Mercury. Planet Space Sci, 55: 1502–1508

Delcourt D C, Malova H V, Zelenyi L M. 2017. On the response of quasiadiabatic particles to magnetotail reconfigurations. Ann Geophys, 35: 11–23

Delcourt D C, Martin Jr R F. 1994. Application of the centrifugal impulse model to particle motion in the near-Earth magnetotail. J Geophys Res, 99: 23583–23590

Delcourt D C, Moore T E, Orsini S, Millilo A, Sauvaud J A. 2002. Centrifugal acceleration of ions near Mercury. Geophys Res Lett, 29: 32

Delcourt D C, Sauvaud J A, Martin Jr R F, Moore T E. 1995. Gyrophase effects in the centrifugal impulse model of particle motion in the magnetotail. J Geophys Res, 100: 17211–17220

Delcourt D C, Sauvaud J A, Martin Jr R F, Moore T E. 1996b. On the nonadiabatic precipitation of ions from the near-Earth plasma sheet. J Geophys Res, 101: 17409–17418

Delcourt D C. 2013. On the supply of heavy planetary material to the magnetotail of Mercury. Ann Geophys, 31: 1673–1679

Dewey R M, Raines J M, Sun W, Slavin J A, Poh G. 2018. MESSENGER observations of fast plasma flows in Mercury's magnetotail. Geophys Res Lett, 45: 10,110–10,118

Dewey R M, Slavin J A, Raines J M, Azari A R, Sun W. 2020. MES-SENGER observations of flow braking and flux pileup of dipolariza-tions in Mercury's magnetotail: evidence for current wedge formation. J Geophys Res-Space Phys, 125: e2020JA028112

Dewey R M, Slavin J A, Raines J M, Baker D N, Lawrence D J. 2017. Energetic Electron Acceleration and Injection During Dipolarization Events in Mercury's Magnetotail. J Geophys Res-Space Phys, 122: 12,170–12,188

DiBraccio G A, Slavin J A, Imber S M, Gershman D J, Raines J M, Jackman C M, Boardsen S A, Anderson B J, Korth H, Zurbuchen T H, McNutt Jr R L, Solomon S C. 2015b. MESSENGER observations of flux ropes in Mercury's magnetotail. Planet Space Sci, 115: 77–89

DiBraccio G A, Slavin J A, Raines J M, Gershman D J, Tracy P J, Boardsen S A, Zurbuchen T H, Anderson B J, Korth H, McNutt Jr R L, Solomon S C. 2015a. First observations of Mercury's plasma mantle by MES-SENGER. Geophys Res Lett, 42: 9666–9675

Ding D Q, Lee L C, Kennel C F. 1992. The beta dependence of the collisionless tearing instability at the dayside magnetopause. J Geophys Res, 97: 8257–8267

Dominique D L, Koehn P L, Killen R M, Sprague A L, Sarantos M, Cheng A F, Bradley E T, McClintock W E. 2007. Mercury's Atmosphere: A Surface-Bounded Exosphere. In: Dominique D L, Russell C T, eds. The Messenger Mission to Mercury. New York, NY: Springer New York. 161–186

Dong C, Wang L, Hakim A, Bhattacharjee A, Slavin J A, DiBraccio G A, Germaschewski K. 2019. Global ten-moment multifluid simulations of the Solar wind interaction with Mercury: From the planetary conducting core to the dynamic magnetosphere. Geophys Res Lett, 46: 11584–11596

Dong Y, Fang X, Brain D A, McFadden J P, Halekas J S, Connerney J E, Curry S M, Harada Y, Luhmann J G, Jakosky B M. 2015. Strong plume fluxes at Mars observed by MAVEN: An important planetary ion escape channel. Geophys Res Lett, 42: 8942–8950

Dorelli J C, Bhattacharjee A. 2009. On the generation and topology of flux transfer events. J Geophys Res, 114: A06213

Drake J F, Swisdak M, Che H, Shay M A. 2006. Electron acceleration from contracting magnetic islands during reconnection. Nature, 443: 553–556

Dubinin E, Fraenz M, Fedorov A, Lundin R, Edberg N, Duru F, Vaisberg O. 2011. Ion energization and escape on Mars and Venus. Space Sci Rev, 162: 173–211

Dungey J W, Southwood D J. 1970. Ultra low frequency waves in the magnetosphere. Space Sci Rev, 10: 672–688

Dungey J W. 1961. Interplanetary magnetic field and the auroral zones. Phys Rev Lett, 6: 47–48

Eastwood J P, Videira J J H, Brain D A, Halekas J S. 2012. A chain of magnetic flux ropes in the magnetotail of Mars. Geophys Res Lett, 39: L03104

Elphic R C, Funsten III H O, Hervig R L. 1993. Solar wind-induced sec-ondary ions and their relation to lunar surface composition. In: the 24th Lunar and Planetary Science Conference. Houston. 439

Eraker J H, Simpson J A. 1986. Acceleration of charged particles in Mercury's magnetosphere. J Geophys Res, 91: 9973–9994

Eriksson S, Lavraud B, Wilder F D, Stawarz J E, Giles B L, Burch J L, Baumjohann W, Ergun R E, Lindqvist P A, Magnes W, Pollock C J, Russell C T, Saito Y, Strangeway R J, Torbert R B, Gershman D J, Khotyaintsev Y V, Dorelli J C, Schwartz S J, Avanov L, Grimes E, Vernisse Y, Sturner A P, Phan T D, Marklund G T, Moore T E, Paterson W R, Goodrich K A. 2016. Magnetospheric Multiscale observations of magnetic reconnection associated with Kelvin-Helmholtz waves. Geo-phys Res Lett, 43: 5606–5615

Espinoza C M, Stepanova M, Moya P S, Antonova E E, Valdivia J A. 2018. Ion and electron κ distribution functions along the plasma sheet. Geo-phys Res Lett, 45: 6362–6370

Exner W, Heyner D, Liuzzo L, Motschmann U, Shiota D, Kusano K, Shibayama T. 2018. Coronal mass ejection hits Mercury: A.I.K.E.F. hybrid-code results compared to MESSENGER data. Planet Space Sci, 153: 89–99

Exner W, Simon S, Heyner D, Motschmann U. 2020. Influence of Mer-cury's exosphere on the structure of the magnetosphere. J Geophys Res-Space Phys, 125: e2019JA027691

Fairfield D H. 1986. The magnetic field of the equatorial magnetotail from 10 to 40 $R_E$. J Geophys Res, 91: 4238–4244

Fear R C, Coxon J C, Jackman C M. 2019. The contribution of flux transfer events to Mercury's dungey cycle. Geophys Res Lett, 46: 14239–14246

Fear R C, Milan S E, Fazakerley A N, Owen C J, Asikainen T, Taylor M G G T, Lucek E A, Rème H, Dandouras I, Daly P W. 2007. Motion of flux transfer events: A test of the cooling model. Ann Geophys, 25: 1669–1690

Fear R C, Trenchi L, Coxon J C, Milan S E. 2017. How much flux does a flux transfer event transfer? J Geophys Res-Space Phys, 122: 12,310–12,327

Fejer B G, Gonzales C A, Farley D T, Kelley M C, Woodman R F. 1979. Equatorial electric fields during magnetically disturbed conditions 1. The effect of the interplanetary magnetic field. J Geophys Res, 84: 5797–5802

Fermo R L, Drake J F, Swisdak M, Hwang K J. 2011. Comparison of a statistical model for magnetic islands in large current layers with Hall MHD simulations and Cluster FTE observations. J Geophys Res, 116: A09226

Ferraro V C A. 1960. An approximate method of estimating the size and shape of the stationary hollow carved out in a neutral ionized stream of corpuscles impinging on the geomagnetic field. J Geophys Res, 65: 3951–3953

Fox N J, Velli M C, Bale S D, Decker R, Driesman A, Howard R A, Kasper J C, Kinnison J, Kusterer M, Lario D, Lockwood M K, McComas D J, Raouafi N E, Szabo A. 2016. The Solar probe plus mission: Humanity's first visit to our star. Space Sci Rev, 204: 7–48

Freeman J W, Ibrahim M. 1975. Lunar electric fields, surface potential and associated plasma sheaths. Moon, 14: 103–114

Frey H U, Phan T D, Fuselier S A, Mende S B. 2003. Continuous magnetic reconnection at Earth's magnetopause. Nature, 426: 533–537

Fu H S, Cao J B, Khotyaintsev Y V, Sitnov M I, Runov A, Fu S Y, Hamrin M, André M, Retinò A, Ma Y D, Lu H Y, Wei X H, Huang S Y. 2013. Dipolarization fronts as a consequence of transient reconnection: *In situ* evidence. Geophys Res Lett, 40: 6023–6027

Fu H, Grigorenko E E, Gabrielse C, Liu C, Lu S, Hwang K J, Zhou X, Wang Z, Chen F. 2020. Magnetotail dipolarization fronts and particle acceleration: A review. Sci China Earth Sci, 63: 235–256

Fu S Y, Shi Q Q, Wang C, Parks G, Zheng L, Zheng H, Sun W J. 2011. High-speed flowing plasmas in the Earth's plasma sheet. Chin Sci Bull, 56: 1182–1187

Fu Z F, Lee L C, Shi Y. 1990. A three-dimensional MHD simulation of the multiple X line reconnection process. In: Russell C T, Priest E R, Lee L C, eds. Physics of Magnetic Flux Ropes. Geophysical Monograph Series. 58. Washington D C: American Geophysical Union (AGU). 515–519

Gabrielse C, Angelopoulos V, Runov A, Turner D L. 2014. Statistical characteristics of particle injections throughout the equatorial magne-totail. J Geophys Res-Space Phys, 119: 2512–2535

Genestreti K J, Fuselier S A, Goldstein J, Nagai T, Eastwood J P. 2014. The location and rate of occurrence of near-Earth magnetotail reconnection as observed by Cluster and Geotail. J Atmos Sol-Terr Phys, 121: 98–

109

Genova A, Goossens S, Mazarico E, Lemoine F G, Neumann G A, Kuang W, Sabaka T J, Hauck II S A, Smith D E, Solomon S C, Zuber M T. 2019. Geodetic evidence that Mercury has a solid inner core. Geophys Res Lett, 46: 3625–3633

Gershman D J, DiBraccio G A. 2020. Solar cycle dependence of Solar wind coupling with giant planet magnetospheres. Geophys Res Lett, 47: e2020GL089315

Gershman D J, Raines J M, Slavin J A, Zurbuchen T H, Sundberg T, Boardsen S A, Anderson B J, Korth H, Solomon S C. 2015. MESSENGER observations of multiscale Kelvin-Helmholtz vortices at Mercury. J Geophys Res-Space Phys, 120: 4354–4368

Gershman D J, Slavin J A, Raines J M, Zurbuchen T H, Anderson B J, Korth H, Baker D N, Solomon S C. 2013. Magnetic flux pileup and plasma depletion in Mercury's subsolar magnetosheath. J Geophys Res-Space Phys, 118: 7181–7199

Gershman D J, Slavin J A, Raines J M, Zurbuchen T H, Anderson B J, Korth H, Baker D N, Solomon S C. 2014. Ion kinetic properties in Mercury's pre-midnight plasma sheet. Geophys Res Lett, 41: 5740–5747

Gilbert J A, Lepri S T, Landi E, Zurbuchen T H. 2012. First measurements of the complete heavy-ion charge state distributions of C, O, and Fe associated with interplanetary coronal mass ejections. Astrophys J, 751: 20

Glassmeier K H. 2000. Currents in Mercury's magnetosphere. In: Obtani S, Fujii R, Hesse M, Lysak R L, eds. Magnetospheric Current Systems. Geophys Monogr Ser. 118. Washington D C: American Geophysical Union (AGU). 371–380

Glassmeier K H, Auster H U, Heyner D, Okrafka K, Carr C, Berghofer G, Anderson B J, Balogh A, Baumjohann W, Cargill P, Christensen U, Delva M, Dougherty M, Fornaçon K H, Horbury T S, Lucek E A, Magnes W, Mandea M, Matsuoka A, Matsushima M, Motschmann U, Nakamura R, Narita Y, O'Brien H, Richter I, Schwingenschuh K, Shibuya H, Slavin J A, Sotin C, Stoll B, Tsunakawa H, Vennerstrom S, Vogt J, Zhang T. 2010. The fluxgate magnetometer of the BepiColombo Mercury Planetary Orbiter. Planet Space Sci, 58: 287–299

Glassmeier K H, Grosser J, Auster U, Constantinescu D, Narita Y, Stellmach S. 2007. Electromagnetic induction effects and dynamo action in the hermean system. Space Sci Rev, 132: 511–527

Glassmeier K H, Klimushkin D, Othmer C, Mager P. 2004. ULF waves at Mercury: Earth, the giants, and their little brother compared. Adv Space Res, 33: 1875–1883

Gloeckler G, Galvin A B, Ipavich F M, Geiss J, Balsiger H, von Steiger R, Fisk L A, Ogilvie K W, Wilken B. 1993. Detection of interstellar pick-up hydrogen in the Solar System. Science, 261: 70–73

Goertz C K, Nielsen E, Korth A, Haldoupis C, Hoeg P, Hayward D, Glassmeier K H. 1985. Observations of a possible ground signature of flux transfer events. J Geophys Res, 90: 4069–4078

Goertz C K. 1980. Io's interaction with the plasma torus. J Geophys Res, 85: 2949–2956

Goldstein B E, Suess S T, Walker R J. 1981. Mercury: Magnetospheric processes and the atmospheric supply and loss rates. J Geophys Res, 86: 5485–5499

Grasset O, Sotin C, Deschamps F. 2000. On the internal structure and dynamics of Titan. Planet Space Sci, 48: 617–636

Gurnett D A, Kurth W S, Scarf F L. 1981. Plasma waves near Saturn: Initial results from voyager 1. Science, 212: 235–239

Haerendel G, Paschmann G, Sckopke N, Rosenbauer H, Hedgecock P C. 1978. The frontside boundary layer of the magnetosphere and the problem of reconnection. J Geophys Res, 83: 3195–3216

Halekas J S, Delory G T, Lin R P, Stubbs T J, Farrell W M. 2008. Lunar Prospector observations of the electrostatic potential of the lunar surface and its response to incident currents. J Geophys Res, 113: A09102

Hara T, Harada Y, Mitchell D L, DiBraccio G A, Espley J R, Brain D A, Halekas J S, Seki K, Luhmann J G, McFadden J P, Mazelle C, Jakosky B M. 2017. On the origins of magnetic flux ropes in near-Mars magnetotail current sheets. Geophys Res Lett, 44: 7653–7662

Hasegawa H, Fujimoto M, Phan T D, Rème H, Balogh A, Dunlop M W, Hashimoto C, Tandokoro R. 2004. Transport of solar wind into Earth's magnetosphere through rolled-up Kelvin-Helmholtz vortices. Nature, 430: 755–758

Hasegawa H, Fujimoto M, Takagi K, Saito Y, Mukai T, Rème H. 2006. Single-spacecraft detection of rolled-up Kelvin-Helmholtz vortices at the flank magnetopause. J Geophys Res, 111: A09203

Hasegawa H, Wang J, Dunlop M W, Pu Z Y, Zhang Q H, Lavraud B, Taylor M G G T, Constantinescu O D, Berchem J, Angelopoulos V, McFadden J P, Frey H U, Panov E V, Volwerk M, Bogdanova Y V. 2010. Evidence for a flux transfer event generated by multiple X-line reconnection at the magnetopause. Geophys Res Lett, 37: L16101

Hauck II S A, Margot J L, Solomon S C, Phillips R J, Johnson C L, Lemoine F G, Mazarico E, McCoy T J, Padovan S, Peale S J, Perry M E, Smith D E, Zuber M T. 2013. The curious case of Mercury's internal structure. J Geophys Res-Planets, 118: 1204–1220

Henderson M G, Reeves G D, Skoug R, Thomsen M F, Denton M H, Mende S B, Immel T J, Brandt P C, Singer H J. 2006. Magnetospheric and auroral activity during the 18 April 2002 sawtooth event. J Geophys Res, 111: A01S90

Heyner D, Auster H U, Fornaçon K H, Carr C, Richter I, Mieth J Z D, Kolhey P, Exner W, Motschmann U, Baumjohann W, Matsuoka A, Magnes W, Berghofer G, Fischer D, Plaschke F, Nakamura R, Narita Y, Delva M, Volwerk M, Balogh A, Dougherty M, Horbury T, Langlais B, Mandea M, Masters A, Oliveira J S, Sánchez-Cano B, Slavin J A, Vennerstrøm S, Vogt J, Wicht J, Glassmeier K H. 2021. The BepiColombo planetary magnetometer MPO-MAG: What can we learn from the hermean magnetic field? Space Sci Rev, 217: 52

Heyner D, Nabert C, Liebert E, Glassmeier K H. 2016. Concerning reconnection-induction balance at the magnetopause of Mercury. J Geophys Res-Space Phys, 121: 2935–2961

Ho G C, Starr R D, Krimigis S M, Vandegriff J D, Baker D N, Gold R E, Anderson B J, Korth H, Schriver D, McNutt Jr R L, Solomon S C. 2016. MESSENGER observations of suprathermal electrons in Mercury's magnetosphere. Geophys Res Lett, 43: 550–555

Hodges Jr R R. 1975. Formation of the lunar atmosphere. Moon, 14: 139–157

Hofer W O. 1991. Angular, energy, and mass distribution of sputtered particles. In: Behrisch R, Wittmaack K, eds. Sputtering by Particle Bombardment III: Characteristics of Sputtered Particles, Technical Applications. Berlin, Heidelberg: Springer Berlin Heidelberg. 15–90

Hoilijoki S, Ganse U, Pfau-Kempf Y, Cassak P A, Walsh B M, Hietala H, von Alfthan S, Palmroth M. 2017. Reconnection rates and X line motion at the magnetopause: Global 2D-3V hybrid-Vlasov simulation results. J Geophys Res-Space Phys, 122: 2877–2888

Hood L L, Schubert G. 1979. Inhibition of solar wind impingement on Mercury by planetary induction currents. J Geophys Res, 84: 2641–2647

Horne R B, Thorne R M. 1998. Potential waves for relativistic electron scattering and stochastic acceleration during magnetic storms. Geophys Res Lett, 25: 3011–3014

Hu S Q. 2017. The Grad-Shafranov reconstruction in twenty years: 1996–2016. Sci China Earth Sci, 60: 1466–1494

Huang C S, Foster J C, Kelley M C. 2005. Long-duration penetration of the interplanetary electric field to the low-latitude ionosphere during the main phase of magnetic storms. J Geophys Res, 110: A11309

Huang C S, Foster J C, Reeves G D, Le G, Frey H U, Pollock C J, Jahn J M. 2003. Periodic magnetospheric substorms: Multiple space-based and ground-based instrumental observations. J Geophys Res, 108: 1411

Huang S Y, Wang Q Y, Sahraoui F, Yuan Z G, Liu Y J, Deng X H, Sun W J, Jiang K, Xu S B, Yu X D, Wei Y Y, Zhang J. 2020. Analysis of turbulence properties in the Mercury plasma environment using MESSENGER observations. Astrophys J, 891: 159

Hunten D M, Morgan T H, Shemansky D E. 1988. The Mercury atmosphere. Mercury: University of Arizona Press. 562–612

Huovelin J, Vainio R, Kilpua E, Lehtolainen A, Korpela S, Esko E, Muinonen K, Bunce E, Martindale A, Grande M, Andersson H, Nenonen S,

Lehti J, Schmidt W, Genzer M, Vihavainen T, Saari J, Peltonen J, Valtonen E, Talvioja M, Portin P, Narendranath S, Jarvinen R, Okada T, Mililo A, Laurenza M, Heino E, Oleynik P. 2020. Solar intensity x-ray and particle spectrometer SIXS: Instrument design and first results. Space Sci Rev, 216: 94

Hwang K J, Kuznetsova M M, Sahraoui F, Goldstein M L, Lee E, Parks G K. 2011. Kelvin-Helmholtz waves under southward interplanetary magnetic field. J Geophys Res, 116: A08210

Iess L, Jacobson R A, Ducci M, Stevenson D J, Lunine J I, Armstrong J W, Asmar S W, Racioppa P, Rappaport N J, Tortora P. 2012. The tides of Titan. Science, 337: 457–459

Imber S M, Slavin J A, Auster H U, Angelopoulos V. 2011. A THEMIS survey of flux ropes and traveling compression regions: Location of the near-Earth reconnection site during solar minimum. J Geophys Res, 116: A02201

Imber S M, Slavin J A, Boardsen S A, Anderson B J, Korth H, McNutt Jr R L, Solomon S C. 2014. MESSENGER observations of large dayside flux transfer events: Do they drive Mercury's substorm cycle? J Geophys Res-Space Phys, 119: 5613–5623

Imber S M, Slavin J A. 2017. MESSENGER observations of magnetotail loading and unloading: Implications for substorms at Mercury. J Geophys Res-Space Phys, 122: 11,402–11,412

Ip W H, Axford W I. 1980. A weak interaction model for Io and the jovian magnetosphere. Nature, 283: 180–183

Ip W H. 1987. Dynamics of electrons and heavy ions in Mercury's magnetosphere. Icarus, 71: 441–447

Jackman C M, Achilleos N, Bunce E J, Cowley S W H, Dougherty M K, Jones G H, Milan S E, Smith E J. 2004. Interplanetary magnetic field at ~9 AU during the declining phase of the solar cycle and its implications for Saturn's magnetospheric dynamics. J Geophys Res, 109: A11203

Jackman C M, Arridge C S. 2011. Solar cycle effects on the dynamics of Jupiter's and Saturn's magnetospheres. Sol Phys, 274: 481–502

James M K, Imber S M, Bunce E J, Yeoman T K, Lockwood M, Owens M J, Slavin J A. 2017. Interplanetary magnetic field properties and variability near Mercury's orbit. J Geophys Res-Space Phys, 122: 7907–7924

James M K, Imber S M, Yeoman T K, Bunce E J. 2019. Field line resonance in the hermean magnetosphere: Structure and implications for plasma distribution. J Geophys Res-Space Phys, 124: 211–228

Janhunen P, Kallio E. 2004. Surface conductivity of Mercury provides current closure and may affect magnetospheric symmetry. Ann Geophys, 22: 1829–1837

Jasinski J M, Akhavan-Tafti M, Sun W, Slavin J A, Coates A J, Fuselier S A, Sergis N, Murphy N. 2021. Flux transfer events at a reconnection-suppressed magnetopause: Cassini observations at Saturn. J Geophys Res-Space Phys, 126: e2020JA028786

Jasinski J M, Regoli L H, Cassidy T A, Dewey R M, Raines J M, Slavin J A, Coates A J, Gershman D J, Nordheim T A, Murphy N. 2020. A transient enhancement of Mercury's exosphere at extremely high altitudes inferred from pickup ions. Nat Commun, 11: 4350

Jasinski J M, Slavin J A, Arridge C S, Poh G, Jia X, Sergis N, Coates A J, Jones G H, Waite Jr J H. 2016. Flux transfer event observation at Saturn's dayside magnetopause by the Cassini spacecraft. Geophys Res Lett, 43: 6713–6723

Jasinski J M, Slavin J A, Raines J M, DiBraccio G A. 2017. Mercury's solar wind interaction as characterized by magnetospheric plasma mantle observations With MESSENGER. J Geophys Res-Space Phys, 122: 12,153–12,169

Jia X, Slavin J A, Gombosi T I, Daldorff L K S, Toth G, Holst B. 2015. Global MHD simulations of Mercury's magnetosphere with coupled planetary interior: Induction effect of the planetary conducting core on the global interaction. J Geophys Res-Space Phys, 120: 4763–4775

Jia X, Slavin J A, Poh G, DiBraccio G A, Toth G, Chen Y, Raines J M, Gombosi T I. 2019. MESSENGER observations and global simulations of highly compressed magnetosphere events at Mercury. J Geophys Res-Space Phys, 124: 229–247

Johnson C L, Philpott L C, Anderson B J, Korth H, Hauck Ii S A, Heyner D, Phillips R J, Winslow R M, Solomon S C. 2016. MESSENGER observations of induced magnetic fields in Mercury's core. Geophys Res Lett, 43: 2436–2444

Johnson J R, Cheng C Z. 1997. Kinetic Alfvén waves and plasma transport at the magnetopause. Geophys Res Lett, 24: 1423–1426

Johnson J R, Cheng C Z. 2001. Stochastic ion heating at the magnetopause due to kinetic Alfvén waves. Geophys Res Lett, 28: 4421–4424

Johnson R E, Leblanc F, Yakshinskiy B V, Madey T E. 2002. Energy distributions for desorption of sodium and potassium from ice: The Na/K ratio at Europa. Icarus, 156: 136–142

Juusola L, Østgaard N, Tanskanen E, Partamies N, Snekvik K. 2011b. Earthward plasma sheet flows during substorm phases. J Geophys Res, 116: A10228

Juusola L, Østgaard N, Tanskanen E. 2011a. Statistics of plasma sheet convection. J Geophys Res, 116: A08201

Kan J R, Lee L C. 1979. Energy coupling function and solar wind-magnetosphere dynamo. Geophys Res Lett, 6: 577–580

Karlsson T, Kasaba Y, Wahlund J E, Henri P, Bylander L, Puccio W, Jansson S E, Åhlen L, Kallio E, Kojima H, Kumamoto A, Lappalainen K, Lybekk B, Ishisaka K, Eriksson A, Morooka M. 2020. The ME-FISTO and WPT electric field sensors of the plasma wave investigation on the BepiColombo Mio spacecraft. Space Sci Rev, 216: 132

Kasaba Y, Takashima T, Matsuda S, Eguchi S, Endo M, Miyabara T, Taeda M, Kuroda Y, Kasahara Y, Imachi T, Kojima H, Yagitani S, Moncuquet M, Wahlund J E, Kumamoto A, Matsuoka A, Baumjohann W, Yokota S, Asamura K, Saito Y, Delcourt D, Hirahara M, Barabash S, Andre N, Kobayashi M, Yoshikawa I, Murakami G, Hayakawa H. 2020. Mission data processor aboard the BepiColombo Mio spacecraft: Design and scientific operation concept. Space Sci Rev, 216: 34

Kasahara S, Miyoshi Y, Yokota S, Mitani T, Kasahara Y, Matsuda S, Kumamoto A, Matsuoka A, Kazama Y, Frey H U, Angelopoulos V, Kurita S, Keika K, Seki K, Shinohara I. 2018. Pulsating aurora from electron scattering by chorus waves. Nature, 554: 337–340

Kasper J C, Abiad R, Austin G, Balat-Pichelin M, Bale S D, Belcher J W, Berg P, Bergner H, Berthomier M, Bookbinder J, Brodu E, Caldwell D, Case A W, Chandran B D G, Cheimets P, Cirtain J W, Cranmer S R, Curtis D W, Daigneau P, Dalton G, Dasgupta B, DeTomaso D, Diaz-Aguado M, Djordjevic B, Donaskowski B, Effinger M, Florinski V, Fox N, Freeman M, Gallagher D, Gary S P, Gauron T, Gates R, Goldstein M, Golub L, Gordon D A, Gurnee R, Guth G, Halekas J, Hatch K, Heerikuisen J, Ho G, Hu Q, Johnson G, Jordan S P, Korreck K E, Larson D, Lazarus A J, Li G, Livi R, Ludlam M, Maksimovic M, McFadden J P, Marchant W, Maruca B A, McComas D J, Messina L, Mercer T, Park S, Peddie A M, Pogorelov N, Reinhart M J, Richardson J D, Robinson M, Rosen I, Skoug R M, Slagle A, Steinberg J T, Stevens M L, Szabo A, Taylor E R, Tiu C, Turin P, Velli M, Webb G, Whittlesey P, Wright K, Wu S T, Zank G. 2016. Solar wind electrons alphas and protons (SWEAP) investigation: Design of the solar wind and coronal plasma instrument suite for solar probe plus. Space Sci Rev, 204: 131–186

Katsura T, Shimizu H, Momoki N, Toh H. 2021. Electromagnetic induction revealed by MESSENGER's vector magnetic data: The size of Mercury's core. Icarus, 354: 114112

Kavosi S, Raeder J. 2015. Ubiquity of Kelvin-Helmholtz waves at Earth's magnetopause. Nat Commun, 6: 7019

Keesee A M, Buzulukova N, Goldstein J, McComas D J, Scime E E, Spence H, Fok M C, Tallaksen K. 2011. Remote observations of ion temperatures in the quiet time magnetosphere. Geophys Res Lett, 38: L03104

Keiling A, Marghitu O, Vogt J, Amm O, Bunescu C, Constantinescu V, Frey H, Hamrin M, Karlsson T, Nakamura R, Nilsson H, Semeter J, Sorbalo E. 2014. Magnetosphere-ionosphere coupling of global Pi2 pulsations. J Geophys Res-Space Phys, 119: 2717–2739

Keiling A, Rème H, Dandouras I, Bosqued J M, Sergeev V, Sauvaud J A, Jacquey C, Lavraud B, Louarn P, Moreau T, Vallat C, Escoubet C P, Parks G K, McCarthy M, Möbius E, Amata E, Klecker B, Korth A, Lundin R, Daly P, Zong Q G. 2004. New properties of energy-dispersed ions in the plasma sheet boundary layer observed by Cluster. J Geophys Res, 109: A05215

Kepko L, Glassmeier K H, Slavin J A, Sundberg T. 2015. Substorm Current Wedge at Earth and Mercury. In: Keiling A, Jackman C M, Delamere P A, eds. Magnetotails in the Solar System. Hoboken, NJ: John Wiley & Sons, Inc. 361–372

Kepko L, Spence H E, Singer H J. 2002. ULF waves in the solar wind as direct drivers of magnetospheric pulsations. Geophys Res Lett, 29: 39-1–39-4

Killen R M, Burger M H, Vervack Jr R J, Cassidy T A. 2018. Understanding Mercury's exosphere: Models derived from MESSENGER observations In: Anderson B J, Nittler L R, Solomon S C, eds. Mercury: The View after MESSENGER. Cambridge Planetary Science. Cambridge: Cambridge University Press. 407–429

Killen R M, Morgan T H. 1993. Diffusion of Na and K in the uppermost regolith of Mercury. J Geophys Res, 98: 23589–23601

Killen R M, Potter A, Fitzsimmons A, Morgan T H. 1999. Sodium D2 line profiles: Clues to the temperature structure of Mercury's exosphere. Planet Space Sci, 47: 1449–1458

Killen R, Cremonese G, Lammer H, Orsini S, Potter A E, Sprague A L, Wurz P, Khodachenko M L, Lichtenegger H I M, Milillo A, Mura A. 2007. Processes that promote and deplete the exosphere of Mercury. Space Sci Rev, 132: 433–509

Kim E-H, Boardsen S A, Johnson J R, Slavin J A. 2016. ULF Waves at Mercury. In: Keiling A, Lee D H, Nakariakov V, eds. Low-Frequency Waves in Space Plasmas. Washington D C: American Geophysical Union. 323–341

Kivelson M G, Ridley A J. 2008. Saturation of the polar cap potential: Inference from Alfvén wing arguments. J Geophys Res, 113: A05214

Korth H, Anderson B J, Gershman D J, Raines J M, Slavin J A, Zurbuchen T H, Solomon S C, McNutt Jr R L. 2014. Plasma distribution in Mercury's magnetosphere derived from MESSENGER magnetometer and fast imaging plasma spectrometer observations. J Geophys Res-Space Phys, 119: 2917–2932

Korth H, Anderson B J, Johnson C L, Slavin J A, Raines J M, Zurbuchen T H. 2018. Structure and configuration of Mercury's magnetosphere. In: Anderson B J, Nittler L R, Solomon S C, eds. Mercury: The View after MESSENGER. Cambridge Planetary Science. Cambridge: Cambridge University Press. 430–460

Kuo H, Russell C T, Le G. 1995. Statistical studies of flux transfer events. J Geophys Res, 100: 3513–3519

Lawrence D J, Anderson B J, Baker D N, Feldman W C, Ho G C, Korth H, McNutt Jr R L, Peplowski P N, Solomon S C, Starr R D, Vandegriff J D, Winslow R M. 2015. Comprehensive survey of energetic electron events in Mercury's magnetosphere with data from the MESSENGER Gamma-Ray and neutron spectrometer. J Geophys Res-Space Phys, 120: 2851–2876

Leblanc F, Doressoundiram A, Schneider N, Massetti S, Wedlund M, López Ariste A, Barbieri C, Mangano V, Cremonese G. 2009. Short-term variations of Mercury's Na exosphere observed with very high spectral resolution. Geophys Res Lett, 36: L07201

Leblanc F, Johnson R E. 2003. Mercury's sodium exosphere. Icarus, 164: 261–281

Lee L C, Fu Z F. 1985. A theory of magnetic flux transfer at the Earth's magnetopause. Geophys Res Lett, 12: 105–108

Lei W, Gendrin R, Higel B, Berchem J. 1981. Relationships between the solar wind electric field and the magnetospheric convection electric field. Geophys Res Lett, 8: 1099–1102

Lepping R P, Acũna M H, Burlaga L F, Farrell W M, Slavin J A, Schatten K H, Mariani F, Ness N F, Neubauer F M, Whang Y C, Byrnes J B, Kennon R S, Panetta P V, Scheifele J, Worley E M. 1995. The WIND magnetic field investigation. Space Sci Rev, 71: 207–229

Lepri S T, Zurbuchen T H. 2010. Direct observational evidence of filament material within interplanetary coronal mass ejections. Astrophys J, 723: L22–L27

Leyser R P, Imber S M, Milan S E, Slavin J A. 2017. The influence of IMF clock angle on dayside flux transfer events at Mercury. Geophys Res Lett, 44: 10,829–10,837

Liljeblad E, Karlsson T, Raines J M, Slavin J A, Kullen A, Sundberg T, Zurbuchen T H. 2015. MESSENGER observations of the dayside low-latitude boundary layer in Mercury's magnetosphere. J Geophys Res-Space Phys, 120: 8387–8400

Liljeblad E, Karlsson T, Sundberg T, Kullen A. 2016. Observations of magnetospheric ULF waves in connection with the Kelvin-Helmholtz instability at Mercury. J Geophys Res-Space Phys, 121: 8576–8588

Liljeblad E, Karlsson T. 2017. Investigation of ~ 20–40 mHz ULF waves and their driving mechanisms in Mercury's dayside magnetosphere. Ann Geophys, 35: 879–884

Liljeblad E, Sundberg T, Karlsson T, Kullen A. 2014. Statistical investigation of Kelvin-Helmholtz waves at the magnetopause of Mercury. J Geophys Res-Space Phys, 119: 9670–9683

Lindsay S T, James M K, Bunce E J, Imber S M, Korth H, Martindale A, Yeoman T K. 2016. MESSENGER X-ray observations of magnetosphere-surface interaction on the nightside of Mercury. Planet Space Sci, 125: 72–79

Liu J, Angelopoulos V, Runov A, Zhou X Z. 2013. On the current sheets surrounding dipolarizing flux bundles in the magnetotail: The case for wedgelets. J Geophys Res-Space Phys, 118: 2000–2020

Liu Y H, Li T C, Hesse M, Sun W J, Liu J, Burch J, Slavin J A, Huang K. 2019. Three-dimensional magnetic reconnection with a spatially confined X-line extent: Implications for dipolarizing flux bundles and the dawn-dusk asymmetry. J Geophys Res-Space Phys, 124: 2819–2830

Lockwood M, Cowley S W H, Sandholt P E, Lepping R P. 1990. The ionospheric signatures of flux transfer events and solar wind dynamic pressure changes. J Geophys Res, 95: 17113–17135

Lockwood M, Cowley S W H, Smith M F, Rijnbeek R P, Elphic R C. 1995. The contribution of flux transfer events to convection. Geophys Res Lett, 22: 1185–1188

Loureiro N F, Schekochihin A A, Cowley S C. 2007. Instability of current sheets and formation of plasmoid chains. Phys Plasmas, 14: 100703

Lu S, Lu Q, Lin Y, Wang X, Ge Y, Wang R, Zhou M, Fu H, Huang C, Wu M, Wang S. 2015. Dipolarization fronts as earthward propagating flux ropes: A three-dimensional global hybrid simulation. J Geophys Res-Space Phys, 120: 6286–6300

Luhmann J G, Kozyra J U. 1991. Dayside pickup oxygen ion precipitation at Venus and Mars: Spatial distributions, energy deposition and consequences. J Geophys Res, 96: 5457–5467

Lv X, Liu W L. 2018. Measurements of convection electric field in the inner magnetosphere. Sci China Tech Sci, 61: 1866–1871

Lyons L R, Speiser T W. 1982. Evidence for current sheet acceleration in the geomagnetic tail. J Geophys Res, 87: 2276–2286

Mangano V, Massetti S, Milillo A, Plainaki C, Orsini S, Rispoli R, Leblanc F. 2015. THEMIS Na exosphere observations of Mercury and their correlation with in-situ magnetic field measurements by MESSENGER. Planet Space Sci, 115: 102–109

Mangano V, Milillo A, Mura A, Orsini S, De Angelis E, Di Lellis A M, Wurz P. 2007. The contribution of impulsive meteoritic impact vapourization to the Hermean exosphere. Planet Space Sci, 55: 1541–1556

Manka R H. 1973. Plasma and potential at the Lunar surface. In: Grard R J L, ed. Photon and Particle Interactions with Surfaces in Space. 37. Dordrecht: Springer Netherlands. 347–361

Masson A, Nykyri K. 2018. Kelvin-Helmholtz instability: Lessons learned and ways forward. Space Sci Rev, 214: 71

Masters A. 2015. The dayside reconnection voltage applied to Saturn's magnetosphere. Geophys Res Lett, 42: 2577–2585

McClintock W E, Lankton M R. 2007. The Mercury atmospheric and surface composition spectrometer for the MESSENGER mission. Space Sci Rev, 131: 481–521

McClintock W E, Vervack R J, Bradley E T, Killen R M, Mouawad N, Sprague A L, Burger M H, Solomon S C, Izenberg N R. 2009. MESSENGER Observations of Mercury's exosphere: Detection of magnesium and distribution of constituents. Science, 324: 610

McGrath M A, Johnson R E, Lanzerotti L J. 1986. Sputtering of sodium on the planet Mercury. Nature, 323: 694–696

McLain J L, Sprague A L, Grieves G A, Schriver D, Travinicek P, Orlando T M. 2011. Electron-stimulated desorption of silicates: A potential source for ions in Mercury's space environment. J Geophys Res, 116: E03007

Menietti J D, Shprits Y Y, Horne R B, Woodfield E E, Hospodarsky G B, Gurnett D A. 2012. Chorus, ECH, and Z mode emissions observed at Jupiter and Saturn and possible electron acceleration. J Geophys Res, 117: A12214

Milan S E, Cowley S W H, Lester M, Wright D M, Slavin J A, Fillingim M, Carlson C W, Singer H J. 2004. Response of the magnetotail to changes in the open flux content of the magnetosphere. J Geophys Res, 109: A04220

Milan S, Provan G, Hubert B. 2006. Magnetic flux transport in the dungey cycle: The role of sub-storms in flux closure. In: Syrjäsuo, Donovan, ed. the Eighth International Conference on Substorms (ICS-8), Alberta, Canada: University of Calgary. 187–190

Milillo A, Fujimoto M, Murakami G, Benkhoff J, Zender J, Aizawa S, Dósa M, Griton L, Heyner D, Ho G, Imber S M, Jia X, Karlsson T, Killen R M, Laurenza M, Lindsay S T, McKenna-Lawlor S, Mura A, Raines J M, Rothery D A, André N, Baumjohann W, Berezhnoy A, Bourdin P A, Bunce E J, Califano F, Deca J, de la Fuente S, Dong C, Grava C, Fatemi S, Henri P, Ivanovski S L, Jackson B V, James M, Kallio E, Kasaba Y, Kilpua E, Kobayashi M, Langlais B, Leblanc F, Lhotka C, Mangano V, Martindale A, Massetti S, Masters A, Morooka M, Narita Y, Oliveira J S, Odstrcil D, Orsini S, Pelizzo M G, Plainaki C, Plaschke F, Sahraoui F, Seki K, Slavin J A, Vainio R, Wurz P, Barabash S, Carr C M, Delcourt D, Glassmeier K H, Grande M, Hirahara M, Huovelin J, Korablev O, Kojima H, Lichtenegger H, Livi S, Matsuoka A, Moissl R, Moncuquet M, Muinonen K, Quèmerais E, Saito Y, Yagitani S, Yoshikawa I, Wahlund J E. 2020. Investigating Mercury's environment with the two-spacecraft BepiColombo mission. Space Sci Rev, 216: 93

Möbius E, Hovestadt D, Klecker B, Scholer M, Gloeckler G, Ipavich F M. 1985. Direct observation of He⁺ pick-up ions of interstellar origin in the solar wind. Nature, 318: 426–429

Moore T W, Nykyri K, Dimmock A P. 2016. Cross-scale energy transport in space plasmas. Nat Phys, 12: 1164–1169

Moore T W, Nykyri K, Dimmock A P. 2017. Ion-scale wave properties and enhanced ion heating across the low-latitude boundary layer during Kelvin-Helmholtz instability. J Geophys Res-Space Phys, 122: 11,128–11,153

Morgan T H, Zook H A, Potter A E. 1988. Impact-driven supply of sodium and potassium to the atmosphere of Mercury. Icarus, 75: 156–170

Mura A, Milillo A, Orsini S, Massetti S. 2007. Numerical and analytical model of Mercury's exosphere: Dependence on surface and external conditions. Planet Space Sci, 55: 1569–1583

Murakami G, Hayakawa H, Ogawa H, Matsuda S, Seki T, Kasaba Y, Saito Y, Yoshikawa I, Kobayashi M, Baumjohann W, Matsuoka A, Kojima H, Yagitani S, Moncuquet M, Wahlund J E, Delcourt D, Hirahara M, Barabash S, Korablev O, Fujimoto M. 2020. Mio—First comprehensive exploration of Mercury's space environment: Mission overview. Space Sci Rev, 216: 113

Nagai T, Shinohara I, Zenitani S, Nakamura R, Nakamura T K M, Fujimoto M, Saito Y, Mukai T. 2013. Three-dimensional structure of magnetic reconnection in the magnetotail from Geotail observations. J Geophys Res-Space Phys, 118: 1667–1678

Ness N F, Behannon K W, Lepping R P, Whang Y C, Schatten K H. 1974. Magnetic field observations near Mercury: Preliminary results from Mariner 10. Science, 185: 151–160

Ness N F, Behannon K W, Lepping R P, Whang Y C. 1976. Observations of Mercury's magnetic field. Icarus, 28: 479–488

Nichols J D, Cowley S W H, McComas D J. 2006. Magnetopause reconnection rate estimates for Jupiter's magnetosphere based on interplanetary measurements at ~5 AU. Ann Geophys, 24: 393–406

Nikoukar R, Lawrence D J, Peplowski P N, Dewey R M, Korth H, Baker D N, McNutt Jr R L. 2018. Statistical Study of Mercury's energetic

electron events as observed by the Gamma-Ray and neutron spectrometer instrument onboard MESSENGER. J Geophys Res-Space Phys, 123: 4961–4978

Nishida A. 1966. Formation of plasmapause, or magnetospheric plasma knee, by the combined action of magnetospheric convection and plasma escape from the tail. J Geophys Res, 71: 5669–5679

Nykyri K, Otto A, Lavraud B, Mouikis C, Kistler L M, Balogh A, Rème H. 2006. Cluster observations of reconnection due to the Kelvin-Helmholtz instability at the dawnside magnetospheric flank. Ann Geophys, 24: 2619–2643

Nykyri K, Otto A. 2001. Plasma transport at the magnetospheric boundary due to reconnection in Kelvin-Helmholtz vortices. Geophys Res Lett, 28: 3565–3568

O'Brien T P, Thompson S M, McPherron R L. 2002. Steady magnetospheric convection: Statistical signatures in the solar wind and AE. Geophys Res Lett, 29: 1130

Odstrcil D, Pizzo V J, Linker J A, Riley P, Lionello R, Mikic Z. 2004. Initial coupling of coronal and heliospheric numerical magnetohydrodynamic codes. J Atmos Sol-Terr Phys, 66: 1311–1320

Ogilvie K W, Chornay D J, Fritzenreiter R J, Hunsaker F, Keller J, Lobell J, Miller G, Scudder J D, Sittler E.~C J, Torbert R B, Bodet D, Needell G, Lazarus A J, Steinberg J T, Tappan J H, Mavretic A, Gergin E. 1995. SWE, a comprehensive plasma instrument for the WIND spacecraft. Space Sci Rev, 71: 55–77

Orsini S, Livi S A, Lichtenegger H, Barabash S, Milillo A, De Angelis E, Phillips M, Laky G, Wieser M, Olivieri A, Plainaki C, Ho G, Killen R M, Slavin J A, Wurz P, Berthelier J J, Dandouras I, Kallio E, McKenna-Lawlor S, Szalai S, Torkar K, Vaisberg O, Allegrini F, Daglis I A, Dong C, Escoubet C P, Fatemi S, Fränz M, Ivanovski S, Krupp N, Lammer H, Leblanc F, Mangano V, Mura A, Nilsson H, Raines J M, Rispoli R, Sarantos M, Smith H T, Szego K, Aronica A, Camozzi F, Di Lellis A M, Fremuth G, Giner F, Gurnee R, Hayes J, Jeszenszky H, Tominetti F, Trantham B, Balaz J, Baumjohann W, Brienza D, Bührke U, Bush M D, Cantatore M, Cibella S, Colasanti L, Cremonese G, Cremonesi L, D'Alessandro M, Delcourt D, Delva M, Desai M, Fama M, Ferris M, Fischer H, Gaggero D, Gamborino D, Garnier P, Gibson W C, Goldstein R, Grande M, Grishin V, Haggerty D, Holmström M, Horvath I, Hsieh K C, Jacques A, Johnson R E, Kazakov A, Kecskemety K, Krüger H, Kürbisch C, Lazzarotto F, Leblanc F, Leichtfried M, Leoni R, Loose A, Maschietti D, Massetti S, Mattioli F, Miller G, Moissenko D, Morbidini A, Noschese R, Nuccilli F, Nunez C, Paschalidis N, Persyn S, Piazza D, Oja M, Ryno J, Schmidt W, Scheer J A, Shestakov A, Shuvalov S, Seki K, Selci S, Smith K, Sordini R, Svensson J, Szalai L, Toublanc D, Urdiales C, Varsani A, Vertolli N, Wallner R, Wahlstroem P, Wilson P, Zampieri S. 2021. SERENA: Particle instrument suite for determining the Sun-Mercury interaction from BepiColombo. Space Sci Rev, 217: 11

Orsini S, Mangano V, Milillo A, Plainaki C, Mura A, Raines J M, De Angelis E, Rispoli R, Lazzarotto F, Aronica A. 2018. Mercury sodium exospheric emission as a proxy for solar perturbations transit. Sci Rep, 8: 928

Pan D X, Sun W J, Shi Q Q, Tian A M, Yao Z H, Fu S Y, Zong Q G, Zhou X Z, Pu Z Y. 2016. THEMIS statistical study on the plasma properties of high-speed flows in Earth's magnetotail. Sci China Earth Sci, 59: 548–555

Pan D X, Khotyaintsev Y V, Graham D B, Vaivads A, Zhou X Z, André M, Lindqvist P A, Ergun R E, Le C O, Russell C T, Torbert R B, Giles B, Burch J J. 2018. Rippled electron-scale structure of a dipolarization front. Geophys Res Lett, 45: 12,116–12,124

Park R S, Folkner W M, Williams J G, Boggs D H. 2021. The JPL planetary and lunar ephemerides DE440 and DE441. Astron J, 161: 105

Partamies N, Pulkkinen T I, McPherron R L, McWilliams K, Bryant C R, Tanskanen E, Singer H J, Reeves G D, Thomsen M F. 2009. Statistical survey on sawtooth events, SMCs and isolated substorms. Adv Space Res, 44: 376–384

Paschmann G, Haerendel G, Papamastorakis I, Sckopke N, Bame S J, Gosling J T, Russell C T. 1982. Plasma and magnetic field character-

istics of magnetic flux transfer events. J Geophys Res, 87: 2159–2168

Paschmann G, Papamastorakis I, Baumjohann W, Sckopke N, Carlson C W, Sonnerup B U Ö, Lühr H. 1986. The magnetopause for large magnetic shear: AMPTE/IRM observations. J Geophys Res, 91: 11099–11115

Peale S J. 1976. Does Mercury have a molten core? Nature, 262: 765–766

Perrault P, Akasofu S I. 1978. A study of geomagnetic storms. Geophys J Int, 54: 547–573

Pilcher C B, Ridgway S T, McCord T B. 1972. Galilean satellites: Identification of water frost. Science, 178: 1087–1089

Poh G, Slavin J A, Jia X, DiBraccio G A, Raines J M, Imber S M, Gershman D J, Sun W J, Anderson B J, Korth H, Zurbuchen T H, McNutt Jr R L, Solomon S C. 2016. MESSENGER observations of cusp plasma filaments at Mercury. J Geophys Res-Space Phys, 121: 8260–8285

Poh G, Slavin J A, Jia X, Raines J M, Imber S M, Sun W J, Gershman D J, DiBraccio G A, Genestreti K J, Smith A W. 2017b. Mercury's cross-tail current sheet: Structure, X-line location and stress balance. Geophys Res Lett, 44: 678–686

Poh G, Slavin J A, Jia X, Raines J M, Imber S M, Sun W J, Gershman D J, DiBraccio G A, Genestreti K J, Smith A W. 2017a. Coupling between Mercury and its nightside magnetosphere: Cross-tail current sheet asymmetry and substorm current wedge formation. J Geophys Res-Space Phys, 122: 8419–8433

Poh G, Slavin J A, Jia X, Sun W J, Raines J M, Imber S M, DiBraccio G A, Gershman D J. 2018. Transport of mass and energy in Mercury's plasma sheet. Geophys Res Lett, 45: 12,163–12,170

Poh G, Slavin J A, Lu S, Le G, Ozturk D S, Sun W J, Zou S, Eastwood J P, Nakamura R, Baumjohann W, Russell C T, Gershman D J, Giles B L, Pollock C J, Moore T E, Torbert R B, Burch J L. 2019. Dissipation of earthward propagating flux rope through Re-reconnection with geomagnetic field: An MMS case study. J Geophys Res-Space Phys, 124: 7477–7493

Pokorný P, Sarantos M, Janches D. 2017. Reconciling the dawn-dusk asymmetry in Mercury's exosphere with the micrometeoroid impact directionality. Astrophys J, 842: L17

Potter A E, Killen R M, Sarantos M. 2006. Spatial distribution of sodium on Mercury. Icarus, 181: 1–12

Potter A E, Morgan T H. 1986. Potassium in the atmosphere of Mercury. Icarus, 67: 336–340

Potter A, Morgan T. 1985. Discovery of sodium in the atmosphere of Mercury. Science, 229: 651–653

Pritchett P L, Coroniti F V. 2010. A kinetic ballooning/interchange instability in the magnetotail. J Geophys Res, 115: A06301

Pritchett P L, Coroniti F V. 2013. Structure and consequences of the kinetic ballooning/interchange instability in the magnetotail. J Geophys Res-Space Phys, 118: 146–159

Provan G, Yeoman T K, Milan S E. 1998. CUTLASS Finland radar observations of the ionospheric signatures of flux transfer events and the resulting plasma flows. Ann Geophys, 16: 1411–1422

Pu Z Y, Raeder J, Zhong J, Bogdanova Y V, Dunlop M, Xiao C J, Wang X G, Fazakerley A. 2013. Magnetic topologies of an in vivo FTE observed by Double Star/TC-1 at Earth's magnetopause. Geophys Res Lett, 40: 3502–3506

Pu Z Y, Kivelson M G. 1983. Kelvin-Helmholtz instability at the magnetopause: Energy flux into the magnetosphere. J Geophys Res, 88: 853–862

Raeder J. 2006. Flux Transfer Events: 1. generation mechanism for strong southward IMF. Ann Geophys, 24: 381–392

Raines J M, DiBraccio G A, Cassidy T A, Delcourt D C, Fujimoto M, Jia X, Mangano V, Milillo A, Sarantos M, Slavin J A, Wurz P. 2015. Plasma sources in planetary magnetospheres: Mercury. Space Sci Rev, 192: 91–144

Raines J M, Gershman D J, Zurbuchen T H, Sarantos M, Slavin J A, Gilbert J A, Korth H, Anderson B J, Gloeckler G, Krimigis S M, Baker D N, McNutt Jr R L, Solomon S C. 2013. Distribution and compositional variations of plasma ions in Mercury's space environment: The first three Mercury years of MESSENGER observations. J Geophys Res-Space Phys, 118: 1604–1619

Raines J M, Slavin J A, Zurbuchen T H, Gloeckler G, Anderson B J, Baker D N, Korth H, Krimigis S M, McNutt Jr R L. 2011. MESSENGER observations of the plasma environment near Mercury. Planet Space Sci, 59: 2004–2015

Reiff P H, Spiro R W, Hill T W. 1981. Dependence of polar cap potential drop on interplanetary parameters. J Geophys Res, 86: 7639–7648

Rijnbeek R P, Cowley S W H, Southwood D J, Russell C T. 1984. A survey of dayside flux transfer events observed by ISEE 1 and 2 magnetometers. J Geophys Res, 89: 786–800

Rivoldini A, Van Hoolst T. 2013. The interior structure of Mercury constrained by the low-degree gravity field and the rotation of Mercury. Earth Planet Sci Lett, 377-378: 62–72

Rong Z J, Ding Y, Slavin J A, Zhong J, Poh G, Sun W J, Wei Y, Chai L H, Wan W X, Shen C. 2018. The magnetic field structure of Mercury's magnetotail. J Geophys Res-Space Phys, 123: 548–566

Rong Z J, Wan W X, Shen C, Li X, Dunlop M W, Petrukovich A A, Zhang T L, Lucek E. 2011. Statistical survey on the magnetic structure in magnetotail current sheets. J Geophys Res, 116: A09218

Rostoker G, Akasofu S I, Foster J, Greenwald R A, Kamide Y, Kawasaki K, Lui A T Y, McPherron R L, Russell C T. 1980. Magnetospheric substorms—Definition and signatures. J Geophys Res, 85: 1663–1668

Runov A, Angelopoulos V, Zhou X Z. 2012. Multipoint observations of dipolarization front formation by magnetotail reconnection. J Geophys Res, 117: A05230

Russell C T, Baker D N, Slavin J A. 1988. The magnetosphere of Mercury. In: Vilas F, Chapman C R, Matthews M S, eds. Mercury. Tucson: University of Arizona Press. 514–561

Russell C T, Elphic R C. 1978. Initial ISEE magnetometer results: Magnetopause observations. Space Sci Rev, 22: 681–715

Russell C T, Elphic R C. 1979. Observation of magnetic flux ropes in the Venus ionosphere. Nature, 279: 616–618

Russell C T, Walker R J. 1985. Flux transfer events at Mercury. J Geophys Res, 90: 11067–11074

Ruzmaikin A, Sokoloff D, Shukurov A. 1989. The dynamo origin of magnetic fields in galaxy clusters. Mon Not R Astron Soc, 241: 1–14

Saito Y, Sauvaud J A, Hirahara M, Barabash S, Delcourt D, Takashima T, Asamura K. 2010. Scientific objectives and instrumentation of Mercury plasma particle experiment (MPPE) onboard MMO. Planet Space Sci, 58: 182–200

Salvail J R, Fanale F P. 1994. Near-surface ice on Mercury and the Moon: A topographic thermal model. Icarus, 111: 441–455

Sarantos M, Killen R M, Kim D. 2007. Predicting the long-term solar wind ion-sputtering source at Mercury. Planet Space Sci, 55: 1584–1595

Schmid D, Narita Y, Plaschke F, Volwerk M, Nakamura R, Baumjohann W. 2021. Pick-up ion cyclotron waves around Mercury. Geophys Res Lett, 48: e2021GL092606

Schmidt C A, Baumgardner J, Mendillo M, Wilson J K. 2012. Escape rates and variability constraints for high-energy sodium sources at Mercury. J Geophys Res, 117: A03301

Schriver D, Trávníček P M, Anderson B J, Ashour-Abdalla M, Baker D N, Benna M, Boardsen S A, Gold R E, Hellinger P, Ho G C, Korth H, Krimigis S M, McNutt Jr R L, Raines J M, Richard R L, Slavin J A, Solomon S C, Starr R D, Zurbuchen T H. 2011. Quasi-trapped ion and electron populations at Mercury. Geophys Res Lett, 38: L23103

Scurry L, Russell C T, Gosling J T. 1994. Geomagnetic activity and the beta dependence of the dayside reconnection rate. J Geophys Res, 99: 14811–14814

Shabansky V P. 1971. Some processes in the magnetosphere. Space Sci Rev, 12: 299–418

Shiokawa K, Baumjohann W, Haerendel G. 1997. Braking of high-speed flows in the near-Earth tail. Geophys Res Lett, 24: 1179–1182

Shue J H, Song P, Russell C T, Steinberg J T, Chao J K, Zastenker G, Vaisberg O I, Kokubun S, Singer H J, Detman T R, Kawano H. 1998. Magnetopause location under extreme solar wind conditions. J Geophys Res, 103: 17691–17700

Sibeck D G, Lopez R E, Roelof E C. 1991. Solar wind control of the magnetopause shape, location, and motion. J Geophys Res, 96: 5489–5495

Sigmund P. 1969. Theory of sputtering. I. Sputtering yield of amorphous and polycrystalline targets. Phys Rev, 184: 383–416

Siscoe G L, Huang T S. 1985. Polar cap inflation and deflation. J Geophys Res, 90: 543–547

Siscoe G L, Ness N F, Yeates C M. 1975. Substorms on Mercury? J Geophys Res, 80: 4359–4363

Sitnov M I, Swisdak M, Divin A V. 2009. Dipolarization fronts as a signature of transient reconnection in the magnetotail. J Geophys Res, 114: A04202

Skoug R M, Bame S J, Feldman W C, Gosling J T, McComas D J, Steinberg J T, Tokar R L, Riley P, Burlaga L F, Ness N F, Smith C W. 1999. A prolonged He$^+$ enhancement within a coronal mass ejection in the solar wind. Geophys Res Lett, 26: 161–164

Slavin J A, Acuña M H, Anderson B J, Baker D N, Benna M, Boardsen S A, Gloeckler G, Gold R E, Ho G C, Korth H, Krimigis S M, McNutt R L, Raines J M, Sarantos M, Schriver D, Solomon S C, Trávníček P, Zurbuchen T H. 2009. MESSENGER observations of magnetic reconnection in Mercury's magnetosphere. Science, 324: 606–610

Slavin J A, Acuña M H, Anderson B J, Baker D N, Benna M, Gloeckler G, Gold R E, Ho G C, Killen R M, Korth H, Krimigis S M, McNutt R L, Nittler L R, Raines J M, Schriver D, Solomon S C, Starr R D, Trávníček P, Zurbuchen T H. 2008. Mercury's magnetosphere after MESSENGER's first flyby. Science, 321: 85–89

Slavin J A, Anderson B J, Baker D N, Benna M, Boardsen S A, Gloeckler G, Gold R E, Ho G C, Korth H, Krimigis S M, McNutt R L, Nittler L R, Raines J M, Sarantos M, Schriver D, Solomon S C, Starr R D, Trávníček P M, Zurbuchen T H. 2010a. MESSENGER observations of extreme loading and unloading of Mercury's magnetic tail. Science, 329: 665–668

Slavin J A, Anderson B J, Baker D N, Benna M, Boardsen S A, Gold R E, Ho G C, Imber S M, Korth H, Krimigis S M, McNutt Jr R L, Raines J M, Sarantos M, Schriver D, Solomon S C, Trávníček P, Zurbuchen T H. 2012a. MESSENGER and Mariner 10 flyby observations of magnetotail structure and dynamics at Mercury. J Geophys Res, 117: A01215

Slavin J A, Baker D N, Gershman D J, Ho G C, Imber S M, Krimigis S M, Sundberg T. 2018. Mercury's dynamic magnetosphere. In: Anderson B J, Nittler L R, Solomon S C, eds. Mercury: The View after MESSENGER. Cambridge Planetary Science. Cambridge: Cambridge University Press. 461–496

Slavin J A, DiBraccio G A, Gershman D J, Imber S M, Poh G K, Raines J M, Zurbuchen T H, Jia X, Baker D N, Glassmeier K H, Livi S A, Boardsen S A, Cassidy T A, Sarantos M, Sundberg T, Masters A, Johnson C L, Winslow R M, Anderson B J, Korth H, McNutt Jr R L, Solomon S C. 2014. MESSENGER observations of Mercury's dayside magnetosphere under extreme solar wind conditions. J Geophys Res-Space Phys, 119: 8087–8116

Slavin J A, Holzer R E. 1979. The effect of erosion on the solar wind stand-off distance at Mercury. J Geophys Res, 84: 2076–2082

Slavin J A, Holzer R E. 1981. Solar wind flow about the terrestrial planets 1. Modeling bow shock position and shape. J Geophys Res, 86: 11401–11418

Slavin J A, Imber S M, Boardsen S A, DiBraccio G A, Sundberg T, Sarantos M, Nieves-Chinchilla T, Szabo A, Anderson B J, Korth H, Zurbuchen T H, Raines J M, Johnson C L, Winslow R M, Killen R M, McNutt Ralph L. J, Solomon S C. 2012b. MESSENGER observations of a flux-transfer-event shower at Mercury. J Geophys Res, 117: A00M06

Slavin J A, Imber S M, Raines J M. 2021. A dungey cycle in the life of Mercury's magnetosphere. In: Maggiolo R N A, Hasegawa H, Welling D T, Zhang Y, Paxton L J, eds. Magnetospheres in the Solar System. Hoboken, N J: The American Geophysical Union and John Wiley and Sons, Inc. 535–556

Slavin J A, Lepping R P, Gjerloev J, Fairfield D H, Hesse M, Owen C J, Moldwin M B, Nagai T, Ieda A, Mukai T. 2003. Geotail observations of magnetic flux ropes in the plasma sheet. J Geophys Res, 108: 10

Slavin J A, Lepping R P, Wu C C, Anderson B J, Baker D N, Benna M, Boardsen S A, Killen R M, Korth H, Krimigis S M, McClintock W E, McNutt Jr R L, Sarantos M, Schriver D, Solomon S C, Trávníček P, Zurbuchen T H. 2010b. MESSENGER observations of large flux transfer events at Mercury. Geophys Res Lett, 37: L02105

Slavin J A, Middleton H R, Raines J M, Jia X, Zhong J, Sun W J, Livi S, Imber S M, Poh G K, Akhavan-Tafti M, Jasinski J M, DiBraccio G A, Dong C, Dewey R M, Mays M L. 2019. MESSENGER observations of disappearing dayside magnetosphere events at Mercury. J Geophys Res-Space Phys, 124: 6613–6635

Slavin J A, Owen J C J, Connerney J E P, Christon S P. 1997. Mariner 10 observations of field-aligned currents at Mercury. Planet Space Sci, 45: 133–141

Slavin J A, Tanskanen E I, Hesse M, Owen C J, Dunlop M W, Imber S, Lucek E A, Balogh A, Glassmeier K H. 2005. Cluster observations of traveling compression regions in the near-tail. J Geophys Res, 110: A06207

Slavin J A. 2004. Mercury's magnetosphere. Adv Space Res, 33: 1859–1874

Smith A W, Jackman C M, Frohmaier C M, Coxon J C, Slavin J A, Fear R C. 2018b. Evaluating single spacecraft observations of planetary magnetotails with simple monte carlo simulations: 1. Spatial distributions of the neutral line. J Geophys Res-Space Phys, 123: 10,109–10,123

Smith A W, Jackman C M, Frohmaier C M, Fear R C, Slavin J A, Coxon J C. 2018a. Evaluating single spacecraft observations of planetary magnetotails with simple monte carlo simulations: 2. Magnetic flux rope signature selection effects. J Geophys Res-Space Phys, 123: 10,124–10,138

Smith A W, Slavin J A, Jackman C M, Fear R C, Poh G K, DiBraccio G A, Jasinski J M, Trenchi L. 2017a. Automated force-free flux rope identification. J Geophys Res-Space Phys, 122: 780–791

Smith A W, Slavin J A, Jackman C M, Poh G K, Fear R C. 2017b. Flux ropes in the Hermean magnetotail: Distribution, properties, and formation. J Geophys Res-Space Phys, 122: 8136–8153

Solomon S C, Anderson B J. 2018. The MESSENGER mission: Science and implementation overview. In: Anderson B J, Nittler L R, Solomon S C, eds. Mercury: The View after MESSENGER. Cambridge Planetary Science. Cambridge: Cambridge University Press. 1–29

Sonnerup B U O, Paschmann G, Papamastorakis I, Sckopke N, Haerendel G, Bame S J, Asbridge J R, Gosling J T, Russell C T. 1981. Evidence for magnetic field reconnection at the Earth's magnetopause. J Geophys Res, 86: 10049–10067

Sonnerup B U Ö. 1974. Magnetopause reconnection rate. J Geophys Res, 79: 1546–1549

Sonnerup B U Ö. 1979. Magnetic Field Reconnection, Space Plasma Physics: The Study of Solar System Plasmas. Washington D C: National Academy of Sciences. 879–972

Spence H E, Kivelson M G. 1993. Contributions of the low-latitude boundary layer to the finite width magnetotail convection model. J Geophys Res, 98: 15487–15496

Sprague A L. 1990. A diffusion source for sodium and potassium in the atmospheres of Mercury and the Moon. Icarus, 84: 93–105

Starr R D, Schriver D, Nittler L R, Weider S Z, Byrne P K, Ho G C, Rhodes E A, Schlemm II C E, Solomon S C, Trávníček P M. 2012. MESSENGER detection of electron-induced X-ray fluorescence from Mercury's surface. J Geohys Res-Planets, 117: E00L02, doi: 10.1029/2012JE004118

Suess S T, Goldstein B E. 1979. Compression of the Hermaean magnetosphere by the solar wind. J Geophys Res, 84: 3306–3312

Sun W J, Fu S Y, Slavin J A, Raines J M, Zong Q G, Poh G K, Zurbuchen T H. 2016. Spatial distribution of Mercury's flux ropes and reconnection fronts: MESSENGER observations. J Geophys Res-Space Phys, 121: 7590–7607

Sun W J, Fu S Y, Wei Y, Yao Z H, Rong Z J, Zhou X Z, Slavin J A, Wan W X, Zong Q G, Pu Z Y, Shi Q Q, Shen X C. 2017a. Plasma sheet pressure variations in the near-earth magnetotail during substorm growth phase: THEMIS observations. J Geophys Res-Space Phys, 122: 12,212–12,228

Sun W J, Raines J M, Fu S Y, Slavin J A, Wei Y, Poh G K, Pu Z Y, Yao Z H, Zong Q G, Wan W X. 2017b. MESSENGER observations of the energization and heating of protons in the near-Mercury magnetotail. Geophys Res Lett, 44: 8149–8158

Sun W J, Slavin J A, Dewey R M, Chen Y, DiBraccio G A, Raines J M, Jasinski J M, Jia X, Akhavan-Tafti M. 2020b. MESSENGER observations of Mercury's nightside magnetosphere under extreme solar wind conditions: Reconnection-generated structures and steady convection. J Geophys Res-Space Phys, 125: e27490

Sun W J, Slavin J A, Dewey R M, Raines J M, Fu S Y, Wei Y, Karlsson T, Poh G K, Jia X, Gershman D J, Zong Q G, Wan W X, Shi Q Q, Pu Z Y, Zhao D. 2018. A comparative study of the proton properties of magnetospheric substorms at Earth and Mercury in the near magnetotail. Geophys Res Lett, 45: 7933–7941

Sun W J, Slavin J A, Fu S, Raines J M, Zong Q G, Imber S M, Shi Q, Yao Z, Poh G, Gershman D J, Pu Z, Sundberg T, Anderson B J, Korth H, Baker D N. 2015b. MESSENGER observations of magnetospheric substorm activity in Mercury's near magnetotail. Geophys Res Lett, 42: 3692–3699

Sun W J, Slavin J A, Smith A W, Dewey R M, Poh G K, Jia X, Raines J M, Livi S, Saito Y, Gershman D J, DiBraccio G A, Imber S M, Guo J P, Fu S Y, Zong Q G, Zhao J T. 2020a. Flux transfer event showers at Mercury: Dependence on plasma $\beta$ and magnetic shear and their contribution to the dungey cycle. Geophys Res Lett, 47: e89784

Sun W J, Slavin J A, Tian A M, Bai S C, Poh G K, Akhavan-Tafti M, Lu S, Yao S T, Le G, Nakamura R, Giles B L, Burch J L. 2019. MMS study of the structure of ion-scale flux ropes in the Earth's cross-tail current sheet. Geophys Res Lett, 46: 6168–6177

Sun W J, Slavin J A, Fu S, Raines J M, Sundberg T, Zong Q G, Jia X, Shi Q, Shen X, Poh G, Pu Z, Zurbuchen T H. 2015a. MESSENGER observations of Alfvénic and compressional waves during Mercury's substorms. Geophys Res Lett, 42: 6189–6198

Sundberg T, Boardsen S A, Slavin J A, Anderson B J, Korth H, Zurbuchen T H, Raines J M, Solomon S C. 2012. MESSENGER orbital observations of large-amplitude Kelvin-Helmholtz waves at Mercury's magnetopause. J Geophys Res-Space Phys, 117(A4), doi: 10.1029/2011JA017268

Sundberg T, Boardsen S A, Slavin J A, Blomberg L G, Cumnock J A, Solomon S C, Anderson B J, Korth H. 2011. Reconstruction of propagating Kelvin-Helmholtz vortices at Mercury's magnetopause. Planet Space Sci, 59: 2051–2057

Sundberg T, Slavin J A. 2015. Mercury's magnetotail. In: Keiling A, Jackman C M, Delamere P A, eds. Magnetotails in the Solar System. Washington D C: The American Geophysical Union and John Wiley and Sons, Inc. 21–42

Swisdak M, Opher M, Drake J F, Alouani Bibi F. 2010. The vector direction of the interstellar magnetic field outside the heliosphere. Astrophys J, 710: 1769–1775

Takahashi F, Matsushima M. 2006. Dipolar and non-dipolar dynamos in a thin shell geometry with implications for the magnetic field of Mercury. Geophys Res Lett, 33: L10202

Tao X, Zonca F, Chen L, Wu Y. 2020. Theoretical and numerical studies of chorus waves: A review. Sci China Earth Sci, 63: 78–92

Taylor G J, Scott E R D. 2005. Mercury. In: Davis A M, ed. Meteorites, Comets, and Planets. Amsterdam: Elsevier Science. 477–486

Thorne R M, Li W, Ni B, Ma Q, Bortnik J, Chen L, Baker D N, Spence H E, Reeves G D, Henderson M G, Kletzing C A, Kurth W S, Hospodarsky G B, Blake J B, Fennell J F, Claudepierre S G, Kanekal S G. 2013. Rapid local acceleration of relativistic radiation-belt electrons by magnetospheric chorus. Nature, 504: 411–414

Thorne R M, Ni B, Tao X, Horne R B, Meredith N P. 2010. Scattering by chorus waves as the dominant cause of diffuse auroral precipitation. Nature, 467: 943–946

Tóth G, Odstrčil D. 1996. Comparison of some flux corrected transport and total variation diminishing numerical schemes for hydrodynamic and magnetohydrodynamic problems. J Comput Phys, 128: 82–100

Trávníček P M, Schriver D, Hellinger P, Herčík D, Anderson B J, Sarantos M, Slavin J A. 2010. Mercury's magnetosphere-solar wind interaction for northward and southward interplanetary magnetic field: Hybrid simulation results. Icarus, 209: 11–22

Tsurutani B T, Smith E J. 1974. Postmidnight chorus: A substorm phenomenon. J Geophys Res, 79: 118–127

van Eyken A P, Rishbeth H, Willis D M, Cowley S W H. 1984. Initial EISCAT observations of plasma convection at invariant latitudes 70°–77°. J Atmos Terrestrial Phys, 46: 635–641

Verhoeven O, Tarits P, Vacher P, Rivoldini A, Van Hoolst T. 2009. Composition and formation of Mercury: Constraints from future electrical conductivity measurements. Planet Space Sci, 57: 296–305

Vervack Jr R J, Killen R M, McClintock W E, Merkel A W, Burger M H, Cassidy T A, Sarantos M. 2016. New discoveries from MESSENGER and insights into Mercury's exosphere. Geophys Res Lett, 43: 11,545–11,551

Vignes D, Acuña M H, Connerney J E P, Crider D H, Rème H, Mazelle C. 2004. Magnetic flux ropes in the martian atmosphere: Global characteristics. In: Winterhalter D, Acuña M, Zakharov A, eds. Mars' Magnetism and Its Interaction with the Solar Wind. 18. Dordrecht: Springer Netherlands. 223–231

Wagner J S, Kan J R, Akasofu S I. 1979. Particle dynamics in the plasma sheet. J Geophys Res, 84: 891–897

Walker R J, Russell C T. 1985. Flux transfer events at the Jovian magnetopause. J Geophys Res, 90: 7397–7404

Walsh A P, Haaland S, Forsyth C, Keesee A M, Kissinger J, Li K, Runov A, Soucek J, Walsh B M, Wing S, Taylor M G G T. 2014. Dawn-dusk asymmetries in the coupled solar wind-magnetosphere-ionosphere system: A review. Ann Geophys, 32: 705–737

Walsh B M, Ryou A S, Sibeck D G, Alexeev I I. 2013. Energetic particle dynamics in Mercury's magnetosphere. J Geophys Res-Space Phys, 118: 1992–1999

Wang C P, Lyons L R, Weygand J M, Nagai T, McEntire R W. 2006. Equatorial distributions of the plasma sheet ions, their electric and magnetic drifts, and magnetic fields under different interplanetary magnetic $B_z$ conditions. J Geophys Res, 111: A04215

Wardinski I, Langlais B, Thébault E. 2019. Correlated time-varying magnetic fields and the core size of Mercury. J Geophys Res-Planets, 124: 2178–2197

Wei Y, Pu Z, Hong M, Zong Q, Ren Z, Fu S, Xie L, Alex S, Cao X, Wang J, Chu X. 2009. Westward ionospheric electric field perturbations on the dayside associated with substorm processes. J Geophys Res, 114: A12209

Wei Y, Zhao B, Li G, Wan W. 2015. Electric field penetration into Earth's ionosphere: A brief review for 2000–2013. Sci Bull, 60: 748–761

West Jr H I, Buck R M, Kivelson M G. 1978a. On the configuration of the magnetotail near midnight during quiet and weakly disturbed periods: State of the magnetosphere. J Geophys Res, 83: 3805–3818

West Jr H I, Buck R M, Kivelson M G. 1978b. On the configuration of the magnetotail near midnight during quiet and weakly disturbed periods: Magnetic field modeling. J Geophys Res, 83: 3819–3831

Wing S, Johnson J R, Newell P T, Meng C I. 2005. Dawn-dusk asymmetries, ion spectra, and sources in the northward interplanetary magnetic field plasma sheet. J Geophys Res, 110: A08205

Winslow R M, Anderson B J, Johnson C L, Slavin J A, Korth H, Purucker M E, Baker D N, Solomon S C. 2013. Mercury's magnetopause and bow shock from MESSENGER Magnetometer observations. J Geophys Res-Space Phys, 118: 2213–2227

Winslow R M, Johnson C L, Anderson B J, Gershman D J, Raines J M, Lillis R J, Korth H, Slavin J A, Solomon S C, Zurbuchen T H, Zuber M T. 2014. Mercury's surface magnetic field determined from proton-reflection magnetometry. Geophys Res Lett, 41: 4463–4470

Winslow R M, Johnson C L, Anderson B J, Korth H, Slavin J A, Purucker M E, Solomon S C. 2012. Observations of Mercury's northern cusp region with MESSENGER's Magnetometer. Geophys Res Lett, 39: L08112, doi: 10.1029/2012GL051472

Winslow R M, Lugaz N, Philpott L, Farrugia C J, Johnson C L, Anderson B J, Paty C S, Schwadron N A, Asad M A. 2020. Observations of extreme ICME ram pressure compressing Mercury's dayside magnetosphere to the surface. Astrophys J, 889: 184

Wurz P, Whitby J A, Rohner U, Martín-Fernández J A, Lammer H, Kolb C. 2010. Self-consistent modelling of Mercury's exosphere by sputtering, micro-meteorite impact and photon-stimulated desorption. Planet Space Sci, 58: 1599–1616

Wygant J R, Torbert R B, Mozer F S. 1983. Comparison of S3-3 polar cap potential drops with the interplanetary magnetic field and models of magnetopause reconnection. J Geophys Res, 88: 5727–5735

Xie L, Lee L C. 2019. A new mechanism for the field line twisting in the ionospheric magnetic flux rope. J Geophys Res-Space Phys, 124: 3266–3275

Yagi M, Seki K, Matsumoto Y, Delcourt D C, Leblanc F. 2017. Global structure and sodium ion dynamics in Mercury's magnetosphere with the offset dipole. J Geophys Res-Space Phys, 122: 10,990–11,002

Yagitani S, Ozaki M, Sahraoui F, Mirioni L, Mansour M, Chanteur G, Coillot C, Ruocco S, Leray V, Hikishima M, Alison D, Le Contel O, Kojima H, Kasahara Y, Kasaba Y, Sasaki T, Yumoto T, Takeuchi Y. 2020. Measurements of magnetic field fluctuations for plasma wave investigation by the search coil magnetometers (SCM) onboard Bepicolombo Mio (Mercury Magnetospheric Orbiter). Space Sci Rev, 216: 111

Yakshinskiy B V, Madey T E. 1999. Photon-stimulated desorption as a substantial source of sodium in the lunar atmosphere. Nature, 400: 642–644

Yakshinskiy B V, Madey T E. 2000. Desorption induced by electronic transitions of Na from SiO₂: Relevance to tenuous planetary atmospheres. Surf Sci, 451: 160–165

Yakshinskiy B V, Madey T E. 2004. Photon-stimulated desorption of Na from a lunar sample: Temperature-dependent effects. Icarus, 168: 53–59

Yan G Q, Mozer F S, Shen C, Chen T, Parks G K, Cai C L, McFadden J P. 2014. Kelvin-Helmholtz vortices observed by THEMIS at the duskside of the magnetopause under southward interplanetary magnetic field. Geophys Res Lett, 41: 4427–4434

Zebker H A, Stiles B, Hensley S, Lorenz R, Kirk R L, Lunine J. 2009. Size and shape of Saturn's Moon Titan. Science, 324: 921–923

Zelenyi L, Oka M, Malova H, Fujimoto M, Delcourt D, Baumjohann W. 2007. Particle acceleration in Mercury's magnetosphere. Space Sci Rev, 132: 593–609

Zhang B, Brambles O J, Lotko W, Lyon J G. 2020. Is nightside outflow required to induce magnetospheric sawtooth oscillations. Geophys Res Lett, 47: e86419

Zhang H, Khurana K K, Kivelson M G, Angelopoulos V, Pu Z Y, Zong Q G, Liu J, Zhou X Z. 2008. Modeling a force-free flux transfer event probed by multiple Time History of Events and Macroscale Interactions during Substorms (THEMIS) spacecraft. J Geophys Res, 113: A00C05

Zhang Q H, Lockwood M, Foster J C, Zhang S R, Zhang B C, McCrea I W, Moen J, Lester M, Ruohoniemi J M. 2015. Direct observations of the full Dungey convection cycle in the polar ionosphere for southward interplanetary magnetic field conditions. J Geophys Res-Space Phys, 120: 4519–4530

Zhang T L, Lu Q M, Baumjohann W, Russell C T, Fedorov A, Barabash S, Coates A J, Du A M, Cao J B, Nakamura R, Teh W L, Wang R S, Dou X K, Wang S, Glassmeier K H, Auster H U, Balikhin M. 2012. Magnetic reconnection in the near venusian magnetotail. Science, 336: 567–570

Zhao J T, Sun W J, Zong Q G, Slavin J A, Zhou X Z, Dewey R M, Poh G K, Raines J M. 2019. A statistical study of the force balance and structure in the flux ropes in Mercury's magnetotail. J Geophys Res-Space Phys, 124: 5143–5157

Zhao J T, Zong Q G, Slavin J A, Sun W J, Zhou X Z, Yue C, Raines J M, Ip W H. 2020. Proton properties in Mercury's magnetotail: A statistical study. Geophys Res Lett, 47: e88075

Zhong J, Lee L C, Wang X G, Pu Z Y, He J S, Wei Y, Wan W X. 2020a. Multiple X-line reconnection observed in Mercury's magnetotail driven by an interplanetary coronal mass ejection. Astrophys J, 893: L11

Zhong J, Pu Z Y, Dunlop M W, Bogdanova Y V, Wang X G, Xiao C J, Guo R L, Hasegawa H, Raeder J, Zhou X Z, Angelopoulos V, Zong Q G, Fu S Y, Xie L, Taylor M G G T, Shen C, Berchem J, Zhang Q H, Volwerk M, Eastwood J P. 2013. Three-dimensional magnetic flux rope structure formed by multiple sequential X-line reconnection at the magnetopause. J Geophys Res-Space Phys, 118: 1904–1911

Zhong J, Wan W X, Slavin J A, Wei Y, Lin R L, Chai L H, Raines J M, Rong Z J, Han X H. 2015a. Mercury's three-dimensional asymmetric magnetopause. J Geophys Res-Space Phys, 120: 7658–7671

Zhong J, Wan W X, Wei Y, Slavin J A, Raines J M, Rong Z J, Chai L H, Han X H. 2015b. Compressibility of Mercury's dayside magnetosphere. Geophys Res Lett, 42: 10,135–10,139

Zhong J, Wei Y, Lee L C, He J S, Slavin J A, Pu Z Y, Zhang H, Wang X G, Wan W X. 2020b. Formation of macroscale flux transfer events at Mercury. Astrophys J, 893: L18

Zhou X, Büchner J, Widmer F, Muñoz P A. 2018. Electron acceleration by turbulent plasmoid reconnection. Phys Plasmas, 25: 042904

Zong Q G, Zhou X Z, Li X, Song P, Fu S Y, Baker D N, Pu Z Y, Fritz T A, Daly P, Balogh A, Réme H. 2007. Ultralow frequency modulation of energetic particles in the dayside magnetosphere. Geophys Res Lett, 34: L12105

Zurbuchen T H, Raines J M, Gloeckler G, Krimigis S M, Slavin J A, Koehn P L, Killen R M, Sprague A L, McNutt R L, Solomon S C. 2008. MESSENGER observations of the composition of Mercury's ionized exosphere and plasma environment. Science, 321: 90–92

Zurbuchen T H, Raines J M, Slavin J A, Gershman D J, Gilbert J A, Gloeckler G, Anderson B J, Baker D N, Korth H, Krimigis S M, Sarantos M, Schriver D, McNutt R L, Solomon S C. 2011. MESSENGER observations of the spatial distribution of planetary ions near Mercury. Science, 333: 1862–1865